\documentclass[fleqn,usenatbib,useAMS]{mnras}
\usepackage{amsmath,fleqn,graphicx,txfonts}
\usepackage{times}
\usepackage{gensymb}
\usepackage{epsf}
\usepackage[dvips]{epsfig}
\usepackage{xspace}
\usepackage{amssymb}
\usepackage{multicol, blindtext} 
\usepackage{bm}
\usepackage{pdflscape}
\usepackage{natbib}
\usepackage{enumitem}
\usepackage{tikz}
\usepackage{verbatim}
\usetikzlibrary{arrows,shapes}
\usepackage{etoolbox}

\newcommand{\ite}{\ensuremath{\,{\rm \left [it\right]}}\xspace}
\newcommand{\e}{\ensuremath{\,{\!=\!}\,}\xspace}
\newcommand{\sm}{\ensuremath{\,{\rm M_{\odot}}}\xspace}
\newcommand{\slu}{\ensuremath{\,{\rm L_{\odot}}}\xspace}
\newcommand{\ml}{\ensuremath{\,{\rm M_{\odot}\,L_{\odot}^{-1}}}\xspace}
\newcommand{\mum}{\ensuremath{{\rm \mu m}}\xspace}
\newcommand{\si}{\ensuremath{\,{\!\sim\!}\,}\xspace}
\newcommand{\ie}{{\it i.e.}\xspace}
\newcommand{\eg}{{\it e.g.}\xspace}

\newcommand{\kpc}{\ensuremath{\,{\rm kpc}}\xspace}
\newcommand{\pc}{\ensuremath{\,{\rm pc}}\xspace}
\newcommand{\Myr}{\ensuremath{\,{\rm Myr}}\xspace}
\newcommand{\Gyr}{\ensuremath{\,{\rm Gyr}}\xspace}
\newcommand{\kms}{\ensuremath{\,{\rm km}\,{\rm s}^{\rm -1}}\xspace}
\newcommand{\psu}{\ensuremath{\,{\rm km}\,{\rm s}^{\rm -1}\,{\rm kpc}^{-1}}\xspace}
\newcommand{\zh}{\ensuremath{\,{\rm Z/H}}\xspace}
\newcommand{\dex}{\ensuremath{\,{\rm dex}}\xspace}

\newcommand{\magasq}{\ensuremath{\,{\rm mag \, {\rm arcsec^{-2}}}}\xspace}
\newcommand{\mags}{\ensuremath{\,{\rm mag}}\xspace}

\newcommand{\HI}{{\textnormal{H}$\,$}{\small \textnormal{I}}\xspace}
\newcommand{\pa}{\ensuremath{{\rm PA}}\xspace}
\newcommand{\degreep}{\ensuremath{{\degree\!\!.}}\xspace}
\newcommand{\mone}{\ensuremath{^{ -1}}\xspace}

\newcommand{\as}{\ensuremath{\,{\rm arcsec}}\xspace}
\newcommand{\de}{\ensuremath{\right)}\xspace}
\newcommand{\iz}{\ensuremath{\left(}\xspace}

\newcommand{\setAo}{\ensuremath{{{\chi}^2}\,^{\rm CBR}_{\mu}}\xspace}
\newcommand{\setBo}{\ensuremath{{{\chi}^2}\,^{\rm CBR}_{\sigma}}\xspace}
\newcommand{\setEo}{\ensuremath{{{\chi}^2}\,^{\rm BPR}_{\upsilon}}\xspace}
\newcommand{\setCo}{\ensuremath{{{\chi}^2}\,^{\rm BPR}_{\mu}}\xspace}
\newcommand{\setDo}{\ensuremath{{{\chi}^2}\,^{\rm BPR}_{\sigma}}\xspace}

\newcommand{\setAomin}{\ensuremath{{{\chi}^2}\,^{\rm CBR}_{\mu\,{\rm min}}}\xspace}
\newcommand{\setBomin}{\ensuremath{{{\chi}^2}\,^{\rm CBR}_{\sigma\,{\rm min}}}\xspace}
\newcommand{\setEomin}{\ensuremath{{{\chi}^2}\,^{\rm BPR}_{\upsilon\,{\rm min}}}\xspace}
\newcommand{\setComin}{\ensuremath{{{\chi}^2}\,^{\rm BPR}_{\mu\,{\rm min}}}\xspace}
\newcommand{\setDomin}{\ensuremath{{{\chi}^2}\,^{\rm BPR}_{\sigma\,{\rm min}}}\xspace}

\newcommand{\setAos}{\ensuremath{{s}\,^{\rm CBR}_{\mu}}\xspace}
\newcommand{\setBos}{\ensuremath{{s}\,^{\rm CBR}_{\sigma}}\xspace}
\newcommand{\setEos}{\ensuremath{{s}\,^{\rm BPR}_{\upsilon}}\xspace}
\newcommand{\setCos}{\ensuremath{{s}\,^{\rm BPR}_{\mu}}\xspace}
\newcommand{\setDos}{\ensuremath{{s}\,^{\rm BPR}_{\sigma}}\xspace}

\newcommand{\setA}{\ensuremath{{\Delta\hat{\chi}^2}\,^{\rm CBR}_{\mu}}\xspace}
\newcommand{\setB}{\ensuremath{{\Delta\hat{\chi}^2}\,^{\rm CBR}_{\sigma}}\xspace}
\newcommand{\setE}{\ensuremath{{\Delta\hat{\chi}^2}\,^{\rm BPR}_{\upsilon}}\xspace}
\newcommand{\setC}{\ensuremath{{\Delta\hat{\chi}^2}\,^{\rm BPR}_{\mu}}\xspace}
\newcommand{\setD}{\ensuremath{{\Delta\hat{\chi}^2}\,^{\rm BPR}_{\sigma}}\xspace}

\newcommand{\setsum}{\ensuremath{\hat{\chi}^2_{\rm sum}}\xspace}
\newcommand{\setdsum}{\ensuremath{\Delta\hat{\chi}^2_{\rm sum}}\xspace}

\newcommand{\pml}{\ensuremath{{\Upsilon_{3.6}\,\,}\xspace}}
\newcommand{\pdm}{\ensuremath{{M^{\rm B}_{\rm DM}\,\,}\xspace}}
\newcommand{\pps}{\ensuremath{{\Omega_{\rm p}\,\,}\xspace}}

\newcommand{\bmml}{\ensuremath{{\Upsilon_{3.6}\!=\!0.72\,{\rm M_{\odot}\,L_{\odot}^{-1}}\,}\xspace}}
\newcommand{\bmdm}{\ensuremath{{M^{\rm B}_{\rm DM}\!=\!1.2\times10^{10}\,{\rm M_{\odot}}\,}\xspace}}
\newcommand{\bmps}{\ensuremath{{\Omega_{\rm p}\!=\!40\,{\rm km}\,{\rm s}^{\rm -1}\,{\rm kpc}^{-1}}\xspace}}

\newcommand{\bml}{\ensuremath{{\Upsilon_{3.6}\!=\!0.72\!\pm\! 0.02\,{\rm M_{\odot}\,L_{\odot}^{-1}}\,}\xspace}}
\newcommand{\bdm}{\ensuremath{{M^{\rm B}_{\rm DM}\!=\!1.2^{+0.2}_{-0.4}\times10^{10}\,{\rm M_{\odot}}\,}\xspace}}
\newcommand{\bps}{\ensuremath{{\Omega_{\rm p}\!=\!40\pm5\,{\rm km}\,{\rm s}^{\rm -1}\,{\rm kpc}^{-1}}\xspace}}

\newcommand{\bnfwml}{\ensuremath{{\Upsilon_{3.6}\!=\!0.70^{+0.02}_{-0.04}\,{\rm M_{\odot}\,L_{\odot}^{-1}}\,}\xspace}}
\newcommand{\bnfwdm}{\ensuremath{{M^{\rm B}_{\rm DM}\!=\!1.0^{+0.4}_{-0.2}\times10^{10}\,{\rm M_{\odot}}\,}\xspace}}
\newcommand{\bnfwps}{\ensuremath{{\Omega_{\rm p}\!=\!40\!\pm\!5\,{\rm km}\,{\rm s}^{\rm -1}\,{\rm kpc}^{-1}}\xspace}}

\newcommand{\bmein}{{\ensuremath{\vec{\rm M}_{\rm AM}^{\rm EIN}\,}\xspace}}
\newcommand{\bmnfw}{{\ensuremath{\vec{\rm M}_{\rm AM}^{\rm NFW}\,}\xspace}}

\newcommand{\bein}{{\ensuremath{\vec{\rm M}_{\rm BM}^{\rm EIN}}\xspace}}
\newcommand{\bnfw}{{\ensuremath{\vec{\rm M}_{\rm BM}^{\rm NFW}}\xspace}}

\newcommand{\CB}{\ensuremath{\rm CB}\xspace}
\newcommand{\BPB}{\ensuremath{\rm BPB}\xspace}

\newcommand{\bfun}{}

\defcitealias{Blana2017}{B17}
\defcitealias{Syer1996}{ST96}
\defcitealias{Opitsch2016}{O16}
\defcitealias{Opitsch2017}{O17}

\graphicspath{{Figures/}}

\newtoggle{paper}
\toggletrue{paper}

\begin{document}

\title[M2M models for M31's bar and composite bulge]{Sculpting Andromeda -- 
made-to-measure models for M31's bar and composite bulge: dynamics, stellar and dark matter mass} 

\author[Bla\~{n}a et al.]
{Mat\'ias Bla\~{n}a D\'iaz$^{1}$\thanks{E-mail: mblana@mpe.mpg.de},
Ortwin Gerhard$^{1}$,
Christopher Wegg$^{1}$,
Matthieu Portail$^{1}$,
\newauthor
Michael Opitsch$^{1,2}$,
Roberto Saglia$^{1,2}$,
Maximilian Fabricius$^{1,2}$,
Peter Erwin$^{1}$,
Ralf Bender$^{1,2}$\\
$^{1}$ Max-Planck-Institut f\"ur extraterrestrische Physik, Gie\ss enbachstra\ss e 1, 85748 Garching bei M\"unchen, Germany\\
$^{2}$ Universit\"ats-Sternwarte M\"unchen, Scheinerstr. 1, D-81679 M\"unchen, Germany}

\date{Accepted 2018 August 17. Received 2018 August 14; in original form 2018 April 23}

\pagerange{\pageref{firstpage}--\pageref{lastpage}} \pubyear{2018}
\maketitle
\label{firstpage}

\begin{abstract}
The Andromeda galaxy (M31) contains a box/peanut bulge (BPB) entangled with a classical bulge (CB) requiring 
a triaxial modelling to determine the dynamics, stellar and dark matter mass.
We construct made-to-measure models fitting new VIRUS-W IFU bulge stellar kinematic observations, 
the IRAC-3.6\mum photometry, and the disc's \HI rotation curve. 
We explore the parameter space for the 3.6\mum mass-to-light ratio $\left(\pml\!\!\right)$, 
the bar pattern speed ($\pps\!\!$),
and the dark matter mass in the composite bulge ($\pdm\!\!$) within $3.2\kpc$.
Considering Einasto dark matter profiles, 
we find the best models for \bml, \bdm and \bps. These models have a dynamical bulge mass 
of $M_{\rm dyn}^{\rm B}\e4.25^{+0.10}_{-0.29}\times\!10^{10}\sm$ including a stellar mass of 
$M_{\star}^{\rm B}\e3.09^{+0.10}_{-0.12}\times\!10^{10}\sm$(73\%),
of which the CB has $M_{\star}^{\rm CB}\e1.18^{+0.06}_{-0.07}\times\!10^{10}\sm$(28\%)
and the BPB $M_{\star}^{\rm BPB}\e1.91\pm0.06\!\times\!10^{10}\sm$(45\%).
We also explore models with NFW haloes finding that, while the Einasto models 
better fit the stellar kinematics, the obtained parameters agree within the errors.
The \pdm values agree with adiabatically contracted cosmological NFW haloes with M31's virial mass and radius.
The best model has two bulge components with completely different kinematics
that only together successfully reproduce the observations 
($\mu_{3.6}$, $\upsilon_{\rm los}, \sigma_{\rm los}$, $h3$, $h4$).
The modelling includes dust absorption which reproduces the observed kinematic asymmetries. Our results provide 
new constraints for the early formation of M31 given the lower mass found for the classical bulge and the shallow 
dark matter profile, as well as the secular evolution of M31 implied by the bar and its resonant interactions with the 
classical bulge, stellar halo and disc.
\end{abstract}


\section{Introduction}
\label{sec:intro}
The Andromeda galaxy (M31, NGC224) is the closest neighbouring massive spiral galaxy, presenting us a unique opportunity to study in depth the 
dynamics of disc galaxy substructures, such as classical bulges and bars, the latter found in approximately 70 per cent of the disc galaxies in the 
local Universe \citep{MenendezDelmestre2007, Erwin2018}. In addition, our external perspective more easily proves a global view of M31 in comparison 
to the Milky Way, while as a similar mass disk galaxy, it allows us to place our home galaxy in context.

Historically, M31's triaxial bulge has been mostly addressed as a classical bulge, while generally the bar component has been only qualitatively 
considered in the modelling of its stellar dynamics. However, an accurate dynamical estimation of the mass distribution of the stellar and the 
dark matter in the bulge must take into account the barred nature of M31's central regions \citep{Lindblad1956}. More recent observations better quantify the 
triaxiality of the bulge which is produced by its box/peanut bulge (\BPB) component \citep{Beaton2007, Opitsch2017}, a situation similar in 
many aspects to the Milky Way's box/peanut bulge \citep{Shen2010,Wegg2013,Bland-Hawthorn2016}. The M31 \BPB is in addition entangled with 
a classical bulge (\CB) component \citep{Athanassoula2006}. The \CB is much more concentrated than the \BPB, with the two components 
contributing with \si1/3 and \si2/3  of the total stellar mass of the bulge respectively, as shown by \citet[hereafter \citetalias{Blana2017}]{Blana2017}.

Each substructure in M31 can potentially teach us about the different mechanisms involved in the formation and the evolution of the whole galaxy. 
In particular, the properties of the \CB component of M31 can give us information about the early formation epoch. Current galaxy 
formation theories consider classical bulges as remnants of a very early formation process, such as a protogalactic collapse, and/or as remnants 
of mergers of galaxies that occurred during the first gigayears of violent hierarchical formation \citep{Toomre1977, Naab2003, Bournaud2005}.
On the other hand, the massive \BPB of M31 provides us information about the evolution of the disc, as box/peanut bulges are formed later from the disc 
material. 
Box/peanut bulges in N-body models are triaxial structures formed through the buckling instability of the bar, which typically lasts for $\lesssim 1\Gyr$, 
generating a vertically thick structure \citep{Combes1990, Raha1991}. Recent observations of two barred galaxies also show evidence of their bars in the buckling
process \citep{Erwin2016}. Box/peanut bulges are frequent being found in 79 per cent of massive barred local galaxies \citep[$M_{\star}\gtrsim 10^{10.4}\sm $,][]{Erwin2017}. 
Note that box/peanut bulges are sometimes referred as box/peanut pseudobulges, however not to be confused with discy pseudobulges, which are 
formed by gas accreted into the centres of disc galaxies \citep{Kormendy2013}. 

Moreover, on even longer time-scales, box/peanut bulges and bars can interact through resonances with the disc and thereby redistribute its material, generating 
for example surface brightness breaks, as well as ring-like substructures \citep{Buta1991, Debattista2006b, Erwin2008, Buta2017b}. 
Bars also transfer their angular momentum to the spheroid components, such as classical bulges \citep{Saha2012, Saha2016}, stellar haloes 
\citep{Perez-Villegas2017a} and dark matter haloes \citep{Athanassoula2002}, changing their dynamical properties. 
Furthermore, \citet{Erwin2016} show also with observations that classical bulges can coexist with discy 
pseudobulges and box/peanut bulges building composite bulges, a scenario that has also been reproduced in galaxy formation simulations \citep{Athanassoula2016}.
This makes M31 a convenient laboratory to test formation theories of composite bulges and to better understand their dynamics.

To understand the formation and the evolution of Andromeda, and to accurately compare it with galaxy formation simulations, it is imperative to 
first determine the contribution and the properties of each of the substructures, such as their masses and sizes, as well as the dark matter 
distribution. In the outer disc region the gas kinematics constrain the dark matter distribution \citep{Chemin2009, Corbelli2010}. However,
in the centre, the gas may not be in equilibrium due to the triaxial potential generated by the bar. Therefore, we model the stellar kinematics taking into 
consideration the triaxial structure of the \BPB. \citet[][hereafter \citetalias{Opitsch2016}]{Opitsch2016} and 
\citet[][hereafter \citetalias{Opitsch2017}]{Opitsch2017} obtained kinematic observations of exquisite detail using the integral field unit (IFU) 
VIRUS-W \citep{Fabricius2012}, completely covering the classical bulge, the \BPB and most of the projected thin or planar 
bar. In this paper we use these kinematic observations to fit a series of made-to-measure models that allow us to find constraints for the stellar 
and dark matter mass within the bulge region, as well as other dynamical parameters such as the pattern speed of the \BPB and the thin 
bar.

This paper is ordered as follows: Section \ref{sec:mod} describes the observational data, its implementation, and the made-to-measure modelling of M31.
Section \ref{sec:res} shows the results of the models that are separated in two main parts. 
In the first, Section \ref{sec:res:param}, we present the main results of the parameter search exploration.  
In the second part, in Section \ref{sec:res:bm}, we present the properties of the best model and we compare it 
with the M31 observations. 
In Section \ref{sec:conc} we conclude with a summary and a discussion of the implications of our findings.

\section{Modelling the bulge of M31}
\label{sec:mod}
Most dynamical models for the bulge of M31 assume a spherical or an oblate geometry for the bulge 
\citep{Ruiz1976, Kent1989, Widrow2003, Widrow2005, Block2006, Hammer2010}, making the mass 
estimations in the centre less accurate due to the barred nature of this galaxy. N-body barred galaxy models 
can represent the bulge and the bar of M31 much better. However, finding an N-body model that exactly reproduces all 
the properties of the M31 substructures is very difficult, because N-body models depend on their initial conditions and 
on the bar formation and buckling instabilities, evolving with some degree of stochasticity. 
Therefore, here we use the Made-to-measure (M2M) method to model the bulge of M31 
\citep[][hereafter ST96]{Syer1996}. 
This method can model triaxial systems and therefore it is the most suitable approach to model M31's bar.

In the following sections we describe our technique 
that implements the M2M method to fit the kinematic and the photometric observations, which allows us
to determine the main dynamical properties of the M31 composite bulge: the pattern speed of the bar ($\pps$), 
the stellar mass-to-light ratio of the bulge in the 3.6\mum band ($\pml$) and the dark matter mass within the 
bulge ($\pdm$). 

\subsection{Made-to-measure method}
\label{sec:mod:m2m}
We use the program {\sc nmagic} that implements the M2M method to fit N-body models to observations \citep{DeLorenzi2007a, DeLorenzi2008, Morganti2012, Portail2015, Portail2017a}.
In the original implementation of the M2M method \citepalias{Syer1996} the potential and the model observables are calculated from the initial mass 
distribution of the particles, where their masses are then optimised to match observations, requiring a mass distribution of the particles that is close to the final 
model. In the {\sc nmagic} implementation the potential is periodically recomputed to generate a system that is gravitationally self-consistent.

A discrete model observable is defined for a system with $N$ particles with phase-space time ($t$) depending coordinates $\vec{z}_i\iz t\de\e\iz \vec{r}_i,\vec{v}_i\de$ as:
\begin{align}
\label{eq:obsd}
& y\iz t\de = \sum_{i\e1}^{N}\,K_i\iz \vec{z}_i\iz t\de\de \,w_i
\end{align}
where $K_i$ is a known kernel that is used to calculate the distribution moments, $w_i$ is the weight of each particle that contributes to the observable,
corresponding here to the particle's mass. We increase the effective number of particles implementing an
exponential temporal smoothing with timescale $\tau_{\rm s}$, obtaining the smoothed observable $y_{\tau}$.

The observational data is composed by $j$ observations (\eg $j$ number of pixels in an image), 
and by $k$ different sets of observations; here we work with one set of photometric observations and four sets of 
kinematic observations. Therefore, we generalise to $Y_j^k$ observations with $Y_{\rm err}\,_j^k$ errors, and by observing the model similarly 
we have $y_{\tau}\,_j^k$ temporally smoothed model observables and $K_{i\,j}^k$ kernels . 
The deviation between the model observables and the observations is defined by the delta
\begin{align}
\label{eq:deltas}
&\Delta_j^k\iz t\de =\frac{y_{\tau}\,_j^k\iz t\de-Y_j^k}{Y_{\rm err}\,_j^k} 
\end{align}
and therefore the sum in time of $\iz\Delta_j^k\de^2$ is the chi-square $\chi^2\,_j^k$ of the 
temporal smoothed model observables and the observations.

The heart of the M2M method is the algorithm that determines how the weights of the particles change in time 
during the iterative fit to the observations. Here we use the ``force-of-change'' (FOC) defined by \citetalias{Syer1996} as:
\begin{align}
\label{eq:FOC}
&\frac{dw_i}{dt} =\epsilon\, w_i \,\partial_{w_i} F
\end{align}
where $\epsilon$ is a constant adjusting the strength of the FOC. This relation is a gradient ascent algorithm 
that maximises $F$ in the space of the weights, defined in {\sc NMAGIC} as
\begin{align}
\label{eq:F}
&F = -\frac{1}{2}\chi^2_{\rm tot} + \mu S
\end{align}
Here the first term is just the total chi-square
\begin{align}
\label{eq:X2}
& \chi^2_{\rm tot} = \sum_{k,j}\lambda_k\, \chi^2\,_j^k
\end{align}
where $\chi^2 \,_j^k=\iz\Delta_j^k\de^2$, and $\lambda_k$ are $k$ constants that balance the contributions between 
different $k$ sets of observables \citep{Long2010, Portail2015}. 
The term $S$ is an ``entropy'' introduced by \citetalias{Syer1996} that forces the weights of the particle distribution to remain close to their 
initial distribution, defined here as in \citet{Morganti2013,Portail2017a}. 
\begin{align}
\label{eq:S}
& S = \sum_i w_i \iz 1- \ln \frac {w_i}{\hat{w}_i}\de
\end{align}
where the ``priors" $\hat{w}_i$ are the averages of the weights of each of the stellar particle types.
The entropy term also forces the model to slowly change its initial 3D mass density distribution. 
The factor $\mu$ balances the contribution between the entropy term and the chi-square term \citep{DeLorenzi2007a}. 
Introducing the previous terms in equation \ref{eq:FOC} we have now the FOC equation
\begin{align}
\label{eq:FOC2} 
&\frac{dw_i}{dt} =-\epsilon\, w_i \left[\mu\ln\iz \frac {w_i}{\hat{w}_i} \de +
\sum_k \lambda_k \sum_j \iz K_{i\,j}^k + w_i\,\partial_{w_i} K_{i\,j}^k\de \frac{\Delta_j^k}{Y_{\rm err }\,_j^k}\right] 
\end{align}
With the observables that we define later the differential term becomes zero ($\partial_{w_i} K_{i\,j}^k  \e0$).

\subsection{Inputs to the M2M modelling from \citetalias{Blana2017}: initial N-body model and projection angles}
\label{sec:mod:input}
The M2M modelling requires an initial input particle model that contains the orbits required to construct a new model that 
successfully matches the observations. Therefore, we use the best matching particle model for the M31 bulge from 
\citetalias{Blana2017}, \ie Model 1, which comes from a set of 72 N-body models built with a box/peanut bulge (\BPB) component 
and a classical bulge (\CB) component with different masses and scale lengths. These models evolved from a Hernquist density profile for the classical 
bulge and another for the dark matter halo, where none of these components have initial rotation. 
During these simulations the initial disc forms a bar that later buckles 
forming a \BPB, but leaving bar material in the plane which is the thin bar. The thin bar 
is aligned with the \BPB extending beyond this. We reserve the term ``bar'' for whole structure of the thin bar and the \BPB together. 
The bar and disc particles have the same label, as the bar evolved from the initial disc. 
The bar is entangled with the \CB, where both structures evolve due to the transfer 
of angular momentum from the bar to the \CB and the dark matter halo as well, gaining both rotation. The light of the \CB
bulge and the \BPB dominate in the centre, and therefore no stellar halo component is included. The number of particles 
used for the \CB, bar and disc and the dark matter halo are $N_{\rm CB}\e10^6$, $N_{\rm bar + disc}\e10^6$ and $N_{\rm halo}\e2\times10^6$.

Model 1 (see \citetalias{Blana2017}) has a concentrated \CB with {\bfun a 3D half-mass radius $r_{\rm half}^{\rm CB}\e0.53\kpc$ $(140\as)$} and a \BPB with a 3D semimajor axis of $r^{\rm BPB}\e3.2\kpc\, (840\as)$ {\bfun and a half-mass radius of $r_{\rm half}^{\rm BPB}\e1.3\kpc$ $(340\as)$}.
 Within the radius $r^{\rm BPB}$, \citetalias{Blana2017} measure a stellar mass of the composite bulge of $M^{\rm B}_{\star}\e3.3\times10^{10}\sm$, 
 where the \CB and the \BPB have $\si 1/3$ and $\si 2/3$ of the bulge total stellar mass, respectively.
They estimate a stellar mass-to-light ratio in the 3.6\mum of $\pml_{\rm B17}\e0.813\ml$.
{\bfun The initial dark matter halo mass within 51\kpc is $3.8\times10^{11}\sm$ and within $r^{\rm BPB}$ is $\pdm\e0.7\times 10^{10}\sm$.}
This model has a bar pattern speed of $\pps\e38\psu$. 

We tested our final results using another model from \citetalias{Blana2017} {\bfun as the input N-body model for the M2M fits. 
This model had the same initial conditions as Model 1, except for the classical bulge mass being 30 per cent higher}. We found only small differences 
in the final fitted M2M model.

We also need to project the M2M models on the sky to calculate the model observables defined later in Section \ref{sec:mod:obs},
requiring the distance to M31 $d_{\rm M31}$, the disc inclination angle $i$, the disc major axis position angle $\pa_{\rm disk}$, and the bar angle $\theta_{\rm bar}$.
For this we use the same quantities adopted as in \citetalias{Blana2017}:
$d_{\rm M31}\e785\pm25\kpc$ \citep{McConnachie2005} (at this distance $3.8\pc \e 1 \as$, $1\kpc \e 260 \as$ and $13.7\kpc \e 1\degree$ on the sky),
$i\e77\degree$ \citep{Corbelli2010}, 
$\pa_{\rm disk}\e38\degree$ \citep{DeVaucouleurs1958}, and the bar angle 
$\theta_{\rm bar}\e54\degreep7\pm3\degreep8$ measured in \citetalias{Blana2017}. 
The bar angle is defined in the plane of the disc (where the bar major axis would be aligned with the disc projected major axis for 
$\theta_{\rm bar}\e0\degree$, see \citetalias{Blana2017} Figure 1). Projecting $\theta_{\rm bar}$ into the sky results in an angle of $\theta_{\rm proj}\e17\degreep7\pm2\degreep5$ 
measured from the line of nodes of the disc major axis, corresponding to a position angle of $\pa_{\rm bar}\e55\degreep7\pm2\degreep5$.
We corroborate later in Section \ref{sec:res:param:ba} that $\theta_{\rm bar}\e54\degreep7\pm3\degreep8$ 
is the bar angle that best matches the photometry of the bulge, reproducing the bulge isophotal twist.

\subsection{Fitting the photometry and IFU kinematics}
\label{sec:mod:obs}
In this section we describe how we prepare M31's photometric and kinematic observational data to use as constraints for 
the M2M fitting with {\sc nmagic}. 
The photometric data consist of an image of M31 from the Infrared Array Camera 1 (IRAC 1) .
The kinematic data correspond to IFU observations of the bulge region of M31, and to \HI rotation curves in the disc region.
Consistently with the observations, we build model observables that measure the same quantities in the model and are used to fit to the equivalent data values.
However, as we explain later in Section \ref{sec:mod:tech}, to find our range of the best matching models we select a 
subsample of the fitted observations to compare them with the models. 
All the model observables $y_j^k$ defined here are temporally smoothed to $y_{\tau}\,_j^k$.

\subsubsection{Photometry I: IRAC 3.6\mum observations}
\label{sec:mod:obs:phot1}
The imaging data that we use come from the large-scale IRAC mosaic images of M31 of the 
\textit{Spitzer Space Telescope} \citep{Barmby2006} kindly made available to us by Pauline Barmby.
We use the IRAC 1 band that at 3.6\mum wavelength for two reasons: 
i) it traces well the old stars (bulk of the population) where the light is dominated by giant stars that 
populate the red giant branch (RGB), and ii) this band has the advantage of being only weakly affected 
by the dust emission or absorption \citep{Meidt2014}.
The IRAC1 mosaic of \citet{Barmby2006} has pixels with size of 0.863\as
and covers a region of $3\degreep7 \times 1\degreep6$. 
We are interested in covering the inner bulge region, both the region where the \CB dominates within 
$\si100\as\,(0.4\kpc)$ in the projected radius, 
also where the \BPB is at $\si700\as\,(2.7\kpc)$ in projection.
We use a resolution of 8.63\as ($32.8\pc$) per pixel for the image,  
which is a convenient scale that faithfully shows the light gradients in the central region 
where the transition between the \CB and the \BPB is.
As we are interested in the scenario where the 10\kpc-ring could be 
connected to the outer Lindblad resonance, we include the region of the stellar disc out to $3950\as\,(15\kpc)$. 
We define an ellipse with this projected semimajor axis by fitting to the isophotes with the \texttt{ellipse} task in iraf.
We mask the pixels of the image that are outside this 15\kpc ellipse and proceed to fit the image.
We also mask hot pixels in the image, foreground stars, and the dwarf galaxy M32.
At the end of the filtering, the total number of photometric observable (pixels) used for the M2M fit is 170651. 

The original image pixel values are in intensity $I\,[{\rm MJy}\,{\rm sr}\mone]$.
The surface-brightness figures in the paper that are in \magasq are in the Vega system, 
and they are transformed from the original units using the 3.6\mum zero-point calibration 280.9 ${\rm Jy}$ \citep{Reach2005}. 
The conversion between the SB in \magasq and the luminosity \slu is done 
using the absolute solar magnitude value $M_{\odot}^{3.6}\e3.24\mags$ \citep{Oh2008}, and 
multiplying $I$ by the pixel area $A_{\rm pixel}\e8.63\times8.63\as^2\e32.8\times32.8\pc^2$.

We also require photometric error maps for the M2M modelling. 
Given that the M2M models are a representation of M31 in dynamical equilibrium, they cannot reproduce the observed 
substructures in M31 that are produced by perturbations such as spiral arms.
Therefore, we include these smaller scale deviations between M31 and the models in the errors.
For this we combined three types of error maps: the observational error $L_{\rm err}^{\rm obs}$,
the variability between pixels $L_{\rm err}^{\rm stdv}$ and the asymmetry error $L_{\rm err}^{\rm asym}$.
The first error ($L_{\rm err}^{\rm obs}$) is calculated from the square root of the sum in quadrature of the pixel error
and the standard deviation for each pixel that comes from the original 0.863\as pixels.
The typical $L_{\rm err}^{\rm obs}$ errors are between one and 5 per cent of the intensity depending on 
the pixel location in the image. 
This error is smaller than the variability observed between 
contiguous pixels and so we therefore include a second error that takes into account the pixel-to-pixel scatter. 
The surface-brightness image of our M2M models is smoother 
than the observations. We take into account this variability by including in the photometric error 
the standard deviation within a radius of one 8.63\as-pixel around each pixel of the image, obtaining 
the error $L_{\rm err}^{\rm stdv}$.
Finally we also include the variability observed at kiloparsec scales due to substructures 
like the spiral arms beyond the bar region, and the 10\kpc-ring. For this we 
subtract the image with the same image, but rotated 180\degree around the centre of the 
bulge, obtaining $L_{\rm err}^{\rm asym}$.  
The bulge is roughly symmetric making this term smaller in the bulge than in the disc region.
The combined photometric error per pixel $Y_{{\rm err}}\,^k_j$ with $k\e0$ is then:
\begin{align}
\label{eq:phot:err}
&Y_{{\rm err}}\,^{k\e0}_j=L_{\rm err}\,_j\e\left[ \iz L_{\rm err}^{\rm obs}\,_j\de^2 + 
\iz L_{\rm err}^{\rm stdv}\,_j\de^2 + \iz L_{\rm err}^{\rm asym}\,_j\de^2 \right]^{1/2}
\end{align}

\subsubsection{Photometry II: model observables and the mass-to-light ratio ($\pml$)}
\label{sec:mod:obs:phot2}
The photometric model observables consist of an array of pixels that extends from the bulge centre
out to the disc until 15\kpc along the disc major axis, where each model pixel uniquely corresponds to 
each observed pixel, with the same pixel size ($8.63\as$).
Each $j$th pixel measures the stellar masses $m_i$ of $N_{j}$ particles that pass through each pixel, 
which are converted to light in the 3.6\mum band using the stellar mass-to-light ratio $\pml$.
The total light per pixel $L_{j}$ is the photometric model observable $y_j^k$ with $k\e0$ :
\begin{align}
\label{eq:obs:phot}
&y_j^k=L_{j}\e \sum_{i}^{N_{j}} \,l_i \e \sum_{i}^{N_{j}} \Upsilon_i\mone\,m_i 
\end{align}
where the light per particle ($l_i$) is just $\Upsilon_i\mone \,m_i$. 
We define three mass-to-light ratio parameters in the 3.6\mum band: 
$\Upsilon^{\rm CB}$ for the classical bulge, $\Upsilon^{\rm BPB}$ for the \BPB 
and $\Upsilon^{\rm d}$ for the outer disc, which are assigned to the particles according to the relation:
\begin{align}
\label{eq:ML}
\Upsilon_i = 
\begin{cases} 
\Upsilon^{\rm CB}  & \text{if } i \in \text{CB} \\
\Upsilon^{\rm BPB}   & \text{if } i \notin \text{CB} \wedge  \ R_i\leq R_{\rm t}\\
\iz\Upsilon^{\rm BPB}-\Upsilon^{\rm d}\de e^{ \frac{-\iz R_i-R_{\rm t}\de^2}{2\,R_{\rm s}^2}} +\Upsilon^{\rm d}   & \text{if } i \notin \text{CB}  \wedge\ R_i>R_{\rm t}\\
\end{cases}
\end{align}
where the \CB particles are assigned $\Upsilon^{\rm CB}$ everywhere, and the bar and 
disc particles at the cylindrical radius $R_{i}$ are assigned $\Upsilon^{\rm BPB}$ within $R_{\rm t}$, and $\Upsilon^{\rm d}$ if they are outside this radius. 
The last Gaussian term provides a smooth transition of $\Upsilon_i$ from the value of $\Upsilon^{\rm BPB}$ to the value in the disc $\Upsilon^{\rm d}$, 
where $R_{\rm t}$ is the transition radius between the end of the thin bar and the disc $R_{\rm t}\e4\kpc$ \citepalias{Blana2017}, 
and $R_{\rm s}$ is the scale of the transition ($R_{\rm s}\e1.5\kpc$). 

In Section \ref{sec:mod:tech} we explain in more detail the different 
mass-to-light values that we explored, however in our fiducial M2M fits we assumed $\pml\e\Upsilon^{\rm CB}\e\Upsilon^{\rm BPB}\e\Upsilon^{\rm d}$. 
In Sections \ref{sec:res:param:ML} and \ref{sec:res:param:mass} we explore further different values for each component, finding only small differences compared with 
our range of best models.
From equation \ref{eq:obs:phot} we have that the photometric kernel ($k\e0$) is
\begin{align}
\label{eq:obs:kernphot}
&K^{k\e0}_{i\,j}\e\Upsilon_i\mone
\end{align}

\subsubsection{Kinematics I: M31 Bulge IFU observations}
\label{sec:mod:obs:kin}
\citetalias{Opitsch2016} and \citetalias{Opitsch2017} obtained kinematic IFU observations of the central region 
of M31 using the McDonald Observatory's 2.7-meter Harlan J. Smith Telescope and the VIRUS-W Spectrograph 
\citep{Fabricius2012}. They cover the whole bulge and bar region and also sample the disc out to one disc scale 
length along six different directions, obtaining line-of-sight velocity distribution profiles 
(LOSVDs). From this they calculate the four Gauss-Hermite expansion coefficient moments \citep{Gerhard1993,Bender1994},
and obtain kinematic maps for the velocity $\upsilon_{\rm los}$, the velocity dispersion $\sigma_{\rm los}$
and the kinematic moments $h3$ and $h4$. The velocity 
maps are corrected for the systemic velocity of $-300\kms$ \citep{DeVaucouleurs1991}. Note that the 
light weighted mean line-of-sight velocity $\langle \upsilon \rangle_{\rm los}$ and the light weighted velocity 
standard deviation (or dispersion) 
$\langle \sigma \rangle_{\rm los} \e\sqrt{\langle \upsilon^2 \rangle_{\rm los} - \langle \upsilon \rangle^2_{\rm los}}$, differ slightly from $\upsilon_{\rm los}$ and $\sigma_{\rm los}$ when the LOSVDs deviate from a Gaussian distribution ($h3\!\neq\!0$ or $h4\!\neq\!0$ or non-zero higher moments). 
This is because $\upsilon_{\rm los}$ and $\sigma_{\rm los}$ are instead chosen so that the lower order Gauss-Hermite terms, $h1$ and $h2$, are zero.

We re-grid the kinematic observations into new maps with the same spatial resolution of the photometric data. The new values 
of $\upsilon_{\rm los}$, $\sigma_{\rm los}$, $h3$ and $h4$ are calculated from the error weighted average of the 
original values, leaving 13400 measurements for each kinematic variable, and therefore 53600 kinematic values in total. The re-gridded 
observational kinematic errors ($Y_{\rm err}^{\rm obs}\,_j^k$, with $k\e1,2,3,4$) are calculated from the standard 
deviation of the error weighted average. Similarly to the photometry, we combined the new observational error 
and the error due to the variability between different kinematic pixels within one pixel radius ($Y_{\rm err}^{\rm stdv}\,^k_j$), 
obtaining a total kinematic error per observable and per set of:
\begin{align}
\label{eq:kinerr}
Y_{{\rm err}}\,^k_j\e \left[\iz Y_{\rm err}^{\rm obs}\,_j^k\de^2 + \iz Y_{\rm err}^{\rm stdv}\,^k_j\de^2 \right]^{1/2}.
\end{align}

\subsubsection{Kinematics II: model observables}
\label{sec:mod:obs:kin2}
We now proceed to build the kinematic model observables.
Because the kinematic observations are performed in the V band, we need to include the effects of dust in our model observables.  
A further description is given later in Section \ref{sec:res:bm:dust}. Our dust absorption implementation consists of using M31 dust mass maps 
\citep{Draine2014} converted to a V band absorption map by the dust model of \citet{Draine2007}
\begin{align}
\label{eq:Av}
&A_{\rm V}^{j} = 0.74 \iz\frac{\Sigma_{\rm dust}^{j}}{10^5 \sm\,\kpc^{-2}}\de \mags~.
\end{align}
We convert this to a 3D absorption map $p^{A{\rm V}}$, deprojected as
\begin{align}
\label{eq:PAv}
p^{A{\rm V}}_{i,j} = 
  \begin{cases} 
   10^{-0.4\, A_{\rm V}^{j}}  & \text{if } z_i \leq 0\kpc \\
   1       & \text{if } z_i > 0\kpc
  \end{cases}
\end{align}
where for simplicity we assume that the dust is located in the plane of the disk, and therefore
any stellar $i$th particle that is temporarily passing behind the disc at the moment that the kinematic model 
observable is measured, is attenuated by the corresponding value of $p^{A{\rm V}}_{i,j}$ in the $j$th pixel.\\

So that the kernel of \autoref{eq:obsd} does not depend on weight, we desire kinematic model observables that are linear in the particle weights. 
Therefore, we fit the Gauss-Hermite moments of the observations, $h1\e0$ and $h2\e0$, instead of directly fitting $\sigma_{\rm los}$
and $\upsilon_{\rm los}$ \citep{DeLorenzi2007a}.
The model kinematic observables are then the light-weighted Gauss-Hermite coefficient moments, 
calculated as in \citet{DeLorenzi2007a}, but with the inclusion of dust absorption:
\begin{align}
\label{eq:GHC}
&y_j^k={\rm H}k_j = \sum^{N_j}_i\,p^{A{\rm V}}_{i,j}\, l_i\,h_{n,i}\e\sum^{N_j}_i\,p^{A{\rm V}}_{i,j}\,\Upsilon_i\mone\,m_i\,2\sqrt \pi\, u_k\iz\beta_i\de ~.
\end{align}
Here $k\e1,2,3,4$, and $u_k\iz \beta_i\de$ are the dimensionless Gauss-Hermite functions \citep{Gerhard1993},
\begin{align}
\label{eq:GH}
&u_k\iz \beta_i\de = \iz2^{n+1}\pi n!\de^{-1/2}\,H_k\iz \beta_i\de \exp\iz-\beta_i^2/2\de 
\end{align}
where $H_k$ are the standard Hermite polynomials, are
\begin{align}
\label{eq:beta}
&\beta_i = \iz \upsilon_i - \upsilon_{\rm los} \de/\sigma_{\rm los}
\end{align}
where $\upsilon_i$ is the particle's line-of-sight velocity. The expansion is performed with 
the observational values of $\sigma_{\rm los}$ and $\upsilon_{\rm los}$ so that while $h$1 and $h$2 are 
zero in the observations, they are in general non-zero when observing the model.
From this we obtain the light weighted model observables ${\rm H1}$, ${\rm H2}$, ${\rm H3}$, and ${\rm H4}$.
The corresponding kinematic kernel that changes the weights of the particles is 
\begin{align}
\label{eq:obs:kernkin}
&K^k_{i\,j}\e p^{A{\rm V}}_{i,j}\,\Upsilon_i\mone\,2\sqrt \pi\, u_k\iz\beta_i\de ~.
\end{align}

Concordantly, the observational data that we fit are the Gauss-Hermite moments $h1\e0$, $h2\e0$, $h3$ and $h4$,
which are light-weighted by the extincted light model observable 
\begin{align}
\label{eq:obs:kinlight}
&L_{j}^{A{\rm V}}\e \sum_{i}^{N_{j}}\,p^{A{\rm V}}_{i,j} \, \Upsilon_i\mone\,m_i  ~.
\end{align}
This is then used to light weight the kinematic observations \eg ${\rm H1}\e h1\,L^{A{\rm V}}$,
obtaining the observations that we fit: ${\rm H1}$, ${\rm H2}$, ${\rm H3}$ and
${\rm H4}$. 

The errors for $h1$ and $h2$ are calculated from the 
observations $\upsilon_{\rm los}$ and $\sigma_{\rm los}$ as in \citet{VanderMarel1993, Rix1997}.
\begin{align}
\label{eq:herr}
&h1_{\rm err} = \frac{1}{\sqrt{2}} \frac{\upsilon_{\rm los, err}}{\sigma_{\rm los}} ;&
h2_{\rm err} = \frac{1}{\sqrt{2}} \frac{\sigma_{\rm los, err}}{\sigma_{\rm los}}
\end{align}
Then, the kinematic errors $h1_{\rm err}$, $h2_{\rm err}$, $h3_{\rm err}$ and $h4_{\rm err}$ are also light-weighted in the form 
${\rm H1}_{\rm err}\,_j\,\e h1_{\rm err}\,_j\,\iz L_j\de^2 \, \iz L_j^{A{\rm V}}\de\mone$, 
which gives larger errors to the regions with more light extinction.
From this we obtained the light weighted errors ${\rm H1_{\rm err}}$, ${\rm H2_{\rm err}}$, ${\rm H3_{\rm err}}$, ${\rm H4_{\rm err}}$.
We also test our best model fit considering no dust absorption ($A_{\rm V}^{j}\e0 \mags$) and 
a constant value $A_{\rm V}^{j} \e 0.5 \mags$.\\

To facilitate side-by-side comparison of the model with the observations, and also for the selection of the range of best models
defined in Section \ref{sec:mod:tech}, we also compute after the M2M fitting 
the temporally smoothed $\upsilon_{\rm los}$ and $\sigma_{\rm los}$ of the model, and use these values to calculate 
$h3$ and $h4$ of the model. For this we observe the model and calculate ${\rm H1}$, ${\rm H2}$, ${\rm H3}$, 
${\rm H4}$ of the model using equation \ref{eq:GHC}, but in equation \ref{eq:beta} we replace $\upsilon_{\rm los}$ 
and $\sigma_{\rm los}$ of the observations by the mean velocity $\langle \upsilon \rangle_{\rm los}$ and the 
velocity standard deviation $\langle \sigma \rangle_{\rm los}$ of the model. 
The non-light weighted quantities are recovered dividing by $L^{A{\rm V}}_j$, \ie h1=H1/$L^{A{\rm V}}_j$ and 
similarly for $h2$, $h3$, $h4$. The parametrisation of the LOSVD with the Gauss-Hermite moments dictates 
that the variables $\sigma_{\rm los}$ and $\upsilon_{\rm los}$ are chosen such that  $h1$ and 
$h2$ are zero. If this is not the case we use again the approximation  \citep{VanderMarel1993, Rix1997} to 
correct and replace the old values of the velocity and the dispersion ($\upsilon_{\rm o} ,\sigma_{\rm o}$) with 
the new values ($\upsilon_{\rm n} ,\sigma_{\rm n}$) that result in new h$1_{\rm n}$ and h$2_{\rm n}$ values 
closer to zero:
\begin{subequations}
\label{eq:corr}
\begin{align}
& \upsilon_{\rm n}=\upsilon_{\rm o}+ \sqrt{2}\, \sigma_{\rm o} h1_{\rm o}\iz\upsilon_{\rm o} ,\sigma_{\rm o}\de\\
&\sigma_{\rm n} = \sigma_{\rm o}+ \sqrt{2}\, \sigma_{\rm o} h2_{\rm o}\iz\upsilon_{\rm o} ,\sigma_{\rm o}\de
\end{align}
\end{subequations}

We repeat the previous corrections observing the model and calculating the new $h1$, $h2$, $h3$, 
$h4$ from the new dispersion and velocity using equation \ref{eq:GHC}, repeating this iteratively until the terms 
$h1$ and $h2$ converge to zero or values smaller than the observational errors.

\subsection{Adjusting the dark matter mass within the bulge ($\pdm$), and fitting the \HI rotation curve}
\label{sec:mod:dm}
Our goal is to determine the dark matter mass within 3.2\kpc of the bulge $\pdm$,
by exploring a vast range of values given in Section \ref{sec:mod:tech}.
For this we change the initial dark matter mass distribution of the input N-body model to match
a target analytical profile. 
As we also want to explore the cusped or cored nature of the dark matter density in the central region, we consider 
different shapes for the target dark halo, making M2M models with two different target profiles. 
We consider the Einasto density profile \citep{Einasto1965} which has a central core, parametrised here as:
\begin{align}
\label{eq:haloein}
&\rho_{\rm DM}^{\rm EIN}\iz m\de=\rho_{\rm E}\,\exp  \left\lbrace  -\iz\frac{2}{\alpha}\de \, \left[ \iz\frac{m}{m_{\rm E}}\de^{\alpha}-1\right]\right\rbrace 
\end{align}
where $m\e\sqrt{x^2+y^2+(z/q)^2}$ is the elliptical radius for a flattening $q$, $m_{\rm E}$ is the scale length, $\rho_{\rm E}$ 
is the central density and $\alpha$ is the steepness of the profile.
We also comte models with a Navarro-Frenk-White (NFW) dark matter mass density profile, which has a cuspy central profile \citep{Navarro1996}, 
parametrised here as
\begin{align}
\label{eq:halonfw}
&\rho_{\rm DM}^{\rm NFW}\iz m\de=\frac{\rho_{\rm N}}{ \iz m/m_{\rm N}\de \left[1+\iz m/m_{\rm N}\de^2\right]} 
\end{align}
where $\rho_{\rm N}$ is the central density and $m_{\rm N}$ is the scale length.

The parameters of these target analytical profiles are determined during each M2M run similarly to \citet{Portail2017a},
by fitting the dark matter halo profile together with the current stellar mass distribution to match:
i) the dark matter mass enclosed within an ellipsoidal volume of the major axis of the bulge 
($r^{\rm BPB}\e m^{\rm BPB}\e3.2\kpc$) is fixed to the chosen value $\pdm$, with $\pdm=\int\!\!dv\, \rho_{\rm DM}$
(or $M_{\rm DM}^{\rm B (p)}$ from the particles);
and ii) that the total circular velocity of the model matches well the disc \HI rotation curve data \citep{Corbelli2010} described in Section \ref{sec:mod:obs:kinH}.

To adjust the particle dark matter distribution to the target analytical dark matter profile we also use the M2M method
\citep{DeLorenzi2007a}. This is done by expanding the initial dark matter density distribution of the particles and 
the target analytical dark matter density profile in spherical harmonics, which are then fitted with the M2M scheme.
The adaptation of the dark matter particles is performed while the photometric and the stellar kinematic observations are also being fitted.

A change in the dark matter mass profile may significantly change the total circular velocity, particularly in the disc region, 
affecting the orbits of the particles. This is not desirable for particles in the disc that should remain on near-circular or epicyclic orbits. 
To alleviate this we measure the circular velocity for a $i$th particle before and after the potential update, and then
re-scale the velocity of the particle living in the old potential 
$\phi_{\rm old}$ to a new velocity given by the new potential $\phi_{\rm new}$ by multiplying its velocity by the factor
$f_{V_{\rm c},i}$ that is the ratio between the new and the old circular velocities: 
\begin{align}
\label{eq:fvc}
& f_{V_{\rm c},i}=\sqrt{\vec{A}_i\cdot \vec{\nabla}_{A_i}\,\phi_{\rm new}/\vec{A}_i\cdot  \vec{\nabla}_{A_i}\,\phi_{\rm old}}
\end{align}
using the spherical radius vector $\vec{A}_i\e\vec{r}_i$ for the dark matter and \CB particles that 
have a spheroidal geometric distribution, and the cylindrical radius $\vec{A}_i\e\vec{R}_i$ for the disc particles.

\subsubsection{Kinematics III: \HI rotation curve}
\label{sec:mod:obs:kinH}
We use the de-projected azimuthally averaged \HI rotation velocity curve estimated by \citet{Corbelli2010} 
to fit the total circular velocity of our M2M models modifying the dark matter profile for a given \pml (see Section 
\ref{sec:mod:dm}) . This data extend from 8.5\kpc out to 50\kpc.  We do not fit the rotation curve beyond 20\kpc, for 
two reasons: i) the contribution of the mass of the \HI disc to the circular velocity beyond this radius becomes as 
important as the stellar disc \citep{Chemin2009}, and ii) the outer disc shows a warp ($R>27\kpc$) 
changing the inclination with respect to the inner part of the stellar and gaseous discs 
\citep{Newton1977, Henderson1979, Brinks1984, Chemin2009}. 
This region includes the 10\kpc-ring and the 15\kpc ring structures \citet{Gordon2006,Barmby2006}.
{\bfun We do not include the mass of the gas component in the potential as the gas mass and surface mass contribution within
20\kpc is estimated to be less than 10 per cent of the stellar mass ($\Sigma_{\rm gas}/\Sigma_{\star}<0.1$) \citep{Chemin2009} 
and, as we show later in Section \ref{sec:res:param:mass}, the choice of different dark matter profiles introduces variations larger than this.}

\subsection{Bar pattern speed adjustment ($\pps$)}
\label{sec:ps}
The pattern speed of the bar of the model found in \citetalias{Blana2017} is $\pps\e38\psu$. 
As we want to find constraints for this quantity, we also explore pattern speeds (see Section \ref{sec:mod:tech}).
To change the initial pattern speed, we adiabatically and linearly change its initial value to the desired final value
with a certain frequency defined in Section \ref{sec:mod:fit}
\citep[see][]{Martinez-Valpuesta2012a,Portail2017a}.
This pattern speed change is performed while the kinematic and the photometric observables are fitted and the potential is 
frequently recalculated from the new density distribution, resulting at the end of the M2M fit in a self-consistent dynamical system.

\subsection{Potential solver and orbital integration}
\label{sec:pot}
As in \citet{Portail2017a}, the {\sc NMAGIC} modelling here uses the hybrid particle-mesh code from \citet{Sellwood2003} 
to calculate the potential from the particle mass distribution. 
The potential solver uses a cylindrical mesh Fourier method to calculate 
the potential for the disc and the bulge components \citep{Sellwood1997}. 
Due to the disc geometry and our interest in resolving the vertical and 
the in plane distribution, instead of using a spherical softening, we use an oblate 
softening with 67\pc in the plane and 17\pc in the vertical direction.
The potential of the particles of the dark matter 
component is calculated using a spherical mesh with a spherical harmonics 
potential solver that extends to 42\kpc and includes terms up to the 16th order 
\citep{DeLorenzi2007a}.
The cylindrical mesh extends in the disc plane out to $R\e10\kpc$ 
and $z\pm3\kpc $ in the vertical direction, and any stellar mass particle that extends beyond the limits of this mesh
is considered during the run in the spherical mesh for the calculation of the potential.

The orbits of the particles are integrated forward in time with an adaptive leap-frog algorithm 
using the acceleration due to the gravitational potential of all the particles.
In the {\sc nmagic} M2M implementation the rotating bar is kept fix in the reference 
frame of the potential by rotating the phase-space coordinates of all the particles around the $z$-axis at the same rate of the pattern speed 
of the bar, but opposite in sign \citep{Martinez-Valpuesta2012a, Portail2015} (note that the rotated system is still in an inertial frame).

The integration time is measured in iteration units \ite, with a time step of 
$1\ite\e0.23\Myr$ \citepalias[see][]{Blana2017}.
We require that the orbits always have at least 1000 steps per orbit.

\subsection{M2M fitting procedure and parameters}
\label{sec:mod:fit}
Each M2M fitting done here with \textsc{nmagic} takes a total number of iterations of $T_{\rm tot}\e80000\ite$, 
where each fit is divided in three main phases.
The first phase uses $T_{\rm obs}\e5000\ite$, and is when the temporal smoothed 
measurements of the model observables are calculated. The temporal smoothing scale is $\tau_{\rm s}\e1600\ite$, and it is chosen to be 
larger than the period (${\rm T_{orbit}}$) of a circular orbit at 5\kpc with circular velocity $V_c$, 
which typically is $\si1000\ite$.

The second phase is when the M2M fitting is performed, and it takes $T_{\rm M2M}\e50000\ite$.
The bar pattern speed is adjusted during this phase, starting at $T_{\rm i}^{\rm ps}\e10000\ite$
and finishing at $T_{\rm f}^{\rm ps}\e40000\ite$, with an update of the new value every 
$T_{\rm up}^{\rm ps}\e3000\ite$.
During the second phase the total mass of the system may change. 
Therefore, we recalculate and update the potential from the new mass density distribution every $T_{\rm pot}\e6400\ite$.
These regular potential updates are important to build a system that is gravitationally self-consistent with its density.

The final phase is the stability check that takes $T_{\rm stab}\e25000\ite$, where the M2M fitting stops and the model 
is only observed. During this phase we recover the values of $\sigma$, $\upsilon$, $h3$ and $h4$ for the model according to equation \ref{eq:corr}
correcting them every $\tau_{\rm corr}\e3\times\tau_{\rm s}$.

\begin{figure}
\begin{center}
\includegraphics[width=8cm,trim=0.1cm 0.4cm 0.4cm 0.4cm,clip]{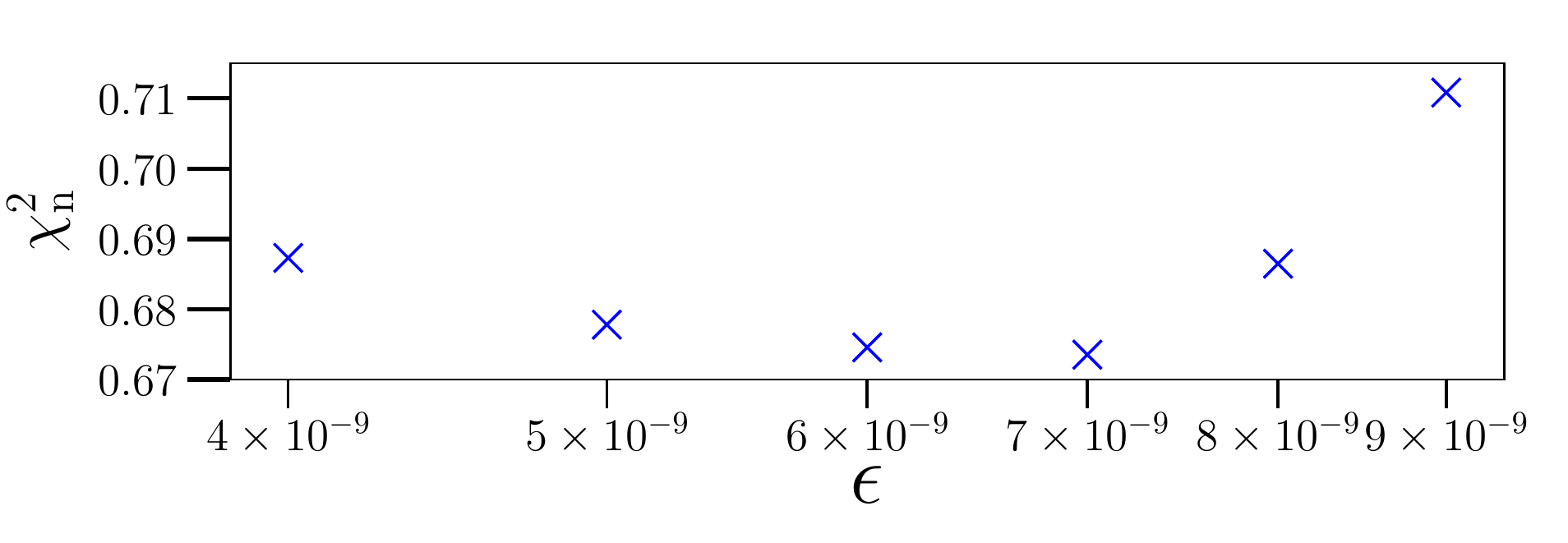}\\
\includegraphics[width=8cm,trim=0.1cm 0.35cm 0.4cm 0.4cm,clip]{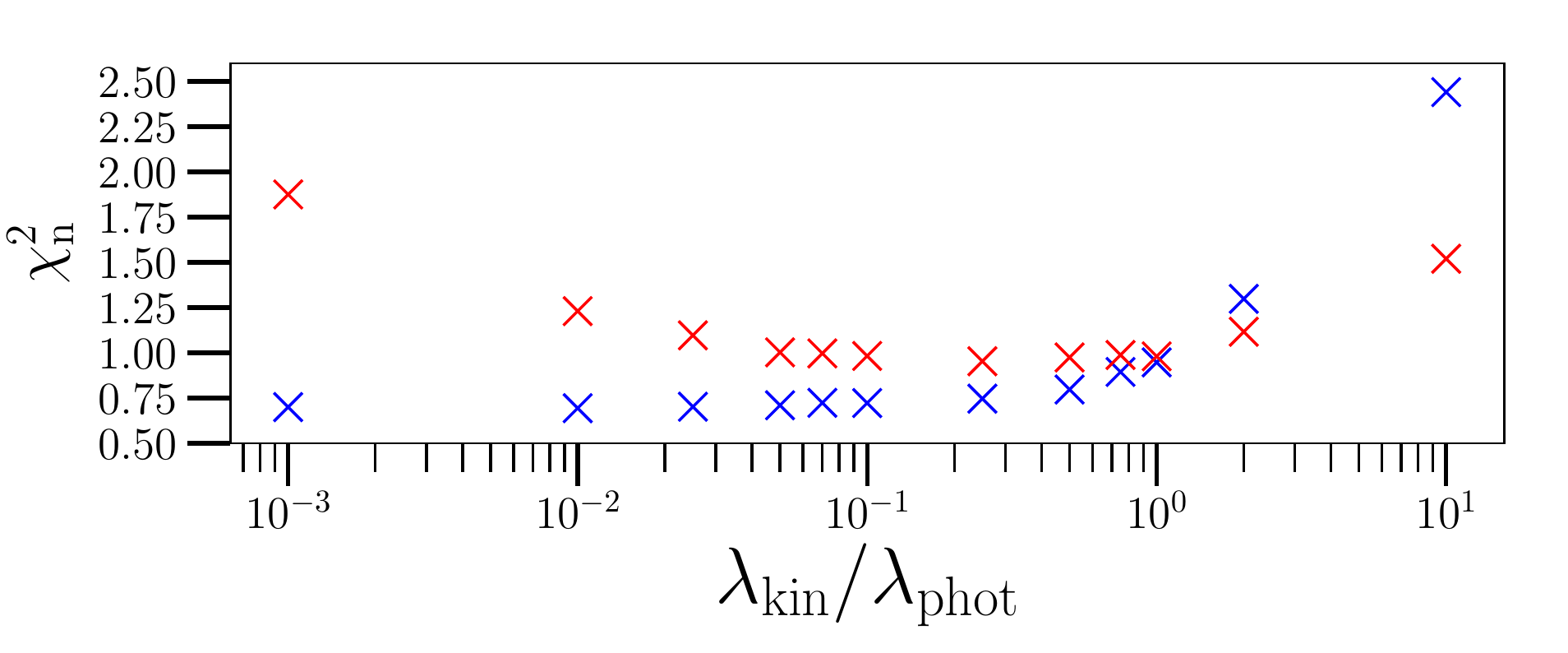}\\
\includegraphics[width=8cm,trim=0.1cm 0.35cm 0.3cm 0.4cm,clip]{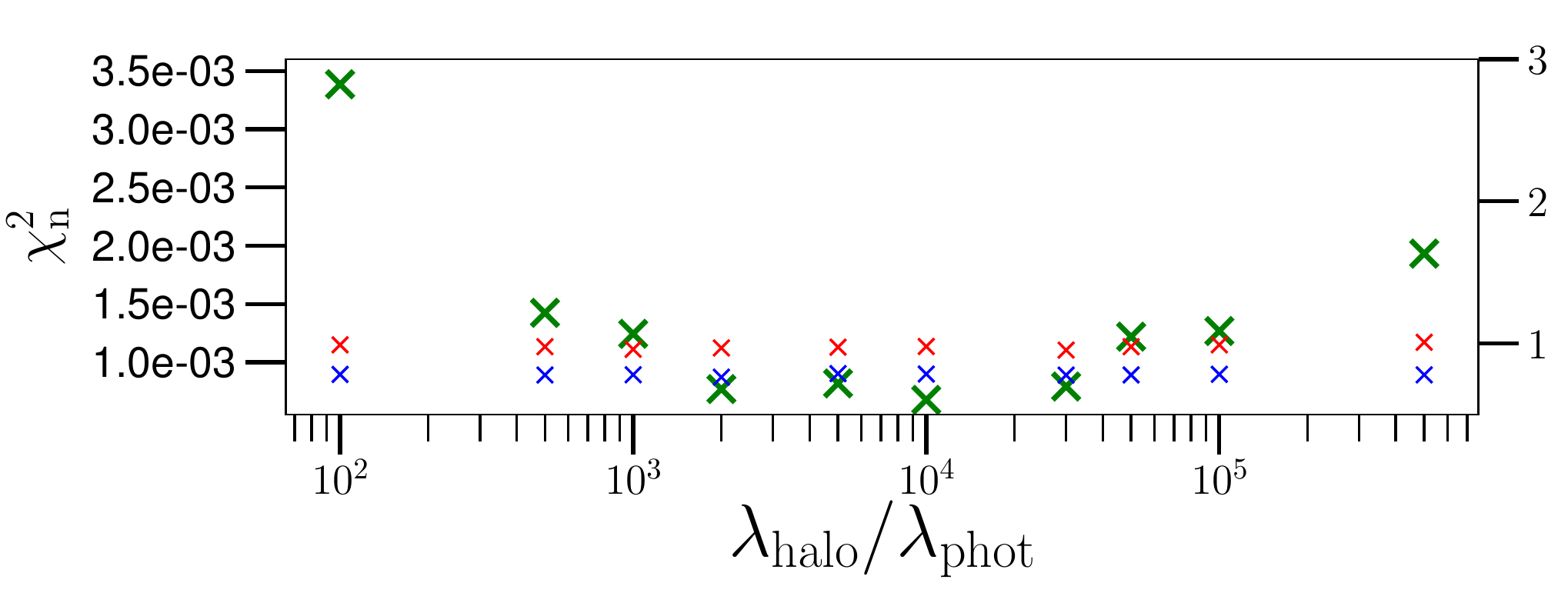}\\
\includegraphics[width=8cm,trim=0.1cm 0.35cm 0.4cm 0.4cm,clip]{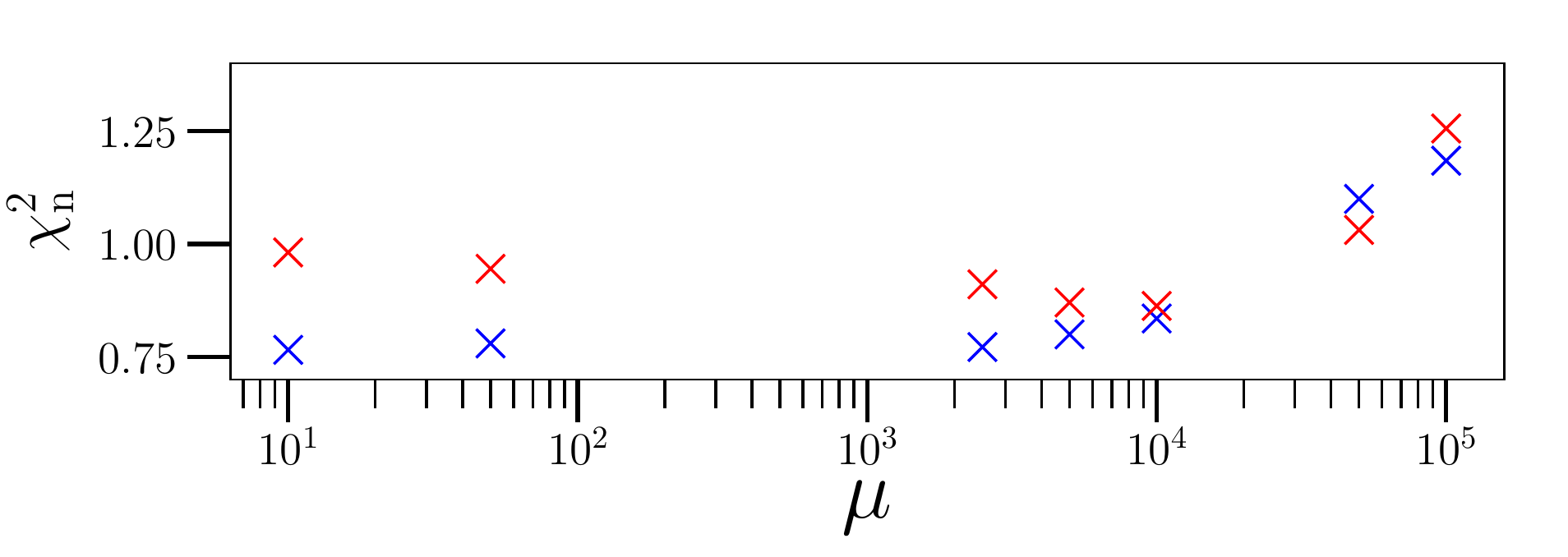}\\
\caption[FOC parameters]{Parameters for the FOC equation. Top panel: photometric $\chi^2_n$ values versus $\epsilon$ (blue crosses). 
Second panel: $\chi^2_n$ versus $\lambda_{\rm kin}/\lambda_{\rm phot}$ (for $\lambda_{\rm phot}\e1$) for the photometry (blue crosses) 
and the kinematics (red crosses).
Third panel: $\chi^2_n$ versus $\lambda_{\rm kin}/\lambda_{\rm phot}$ (also where $\lambda_{\rm phot}\e1$) for the dark matter density (green crosses).
We also show the $\chi^2_n$ for the photometry (blue crosses) and the kinematics (red crosses) where the $\chi^2_n$ values are in reference to the right Y-axis numbers. 
Bottom panel: $\chi^2_n$ versus $\mu$ for the photometry (blue crosses) and the kinematics (red crosses).}
\label{fig:paramFOC}
\end{center}
\end{figure}
The FOC parameters $\epsilon$, $\lambda_{k}$ and $\mu$ of equation \ref{eq:FOC2} are chosen sequentially. 
We first fit only the photometry, leaving the parameters $\lambda_{1\dots5}$ and $\mu$ fixed to zero and varying only $\epsilon$ 
(the parameter $\lambda_{k\e0}$, or $\lambda_{\rm phot}$, normalises $\epsilon$ and for simplicity is set to $\lambda_{\rm phot}\e1$).
We measure the reduced chi-square (terms $\chi^2\,_j^{k}$ in equation \ref{eq:X2}), for the photometry ($k\e0$) in the bulge region  
finding the relation between $\chi^2_n$ and  $\epsilon$ shown in Figure \ref{fig:paramFOC} in the top panel. 
For too small $\epsilon$ the photometry does not have the power to change the model and so the $\chi^2_n$ is large. 
For too large $\epsilon$ the photometry has too much power, changing the particle weights too quickly compared to the orbital timescale, so that  
only the local observable (or pixel) that the particle is crossing is fitted. An optimum $\epsilon$ value allows the weight to change an averaged amount once
it crosses all the observables that are along the particle's orbit, so that its weights converge to a constant value. 
We find this optimum value at the minimum $\chi^2_n$, when $\epsilon\e7.0\times10^{-9}$.

Second, we find the best $\lambda_k$ for the IFU kinematic observables (where $k\e1\dots4$) defined also as $\lambda_{\rm kin}$. 
We use the previous best $\epsilon$ 
and fit the photometry together with the IFU kinematics for several values of $\lambda_{\rm kin}$. 
We measure both the photometric and kinematic reduced chi-square in the bulge region,
obtaining the relations versus $\lambda_{\rm kin}$ shown in Figure \ref{fig:paramFOC} (second panel).
The photometric $\chi^2_n$ has smaller values for small $\lambda_{\rm kin}$, and  would be minimized for $\lambda_{\rm kin}\e0$, 
because then only the photometry would be fitted without the additional kinematic constraints. As $\lambda_{\rm kin}$ increases, the kinematic observations
have more power tailoring the model towards fitting the kinematics as well, as we see the kinematic $\chi^2_n$ decreasing for 
larger $\lambda_{\rm kin}$, which worsen the photometric $\chi^2_n$ if the kinematics get too much power. 
Similarly to $\epsilon$, if $\lambda_{\rm kin}$ increases too much, both the photometric and the kinematic $\chi^2_n$ get worse.
We find an optimal value of $\lambda_k\e2.5\times10^{-1}$ for the minimum kinematic $\chi^2_n$ while the photometric $\chi^2_n$ is still small.

To find the best parameter $\lambda_{\rm halo}$ for the dark matter halo fitting, 
we fix the previously found parameters $\epsilon$ and $\lambda_{\rm kin}$and test 
different values of $\lambda_{\rm halo}$ versus the reduced chi-square of the dark matter halo density (Figure 
\ref{fig:paramFOC} third panel). We find the minimum $\chi^2_n$ at $\lambda_{\rm 6}\e10^4$, where the photometric and the kinematic $\chi^2_n$ remain almost unchanged.

We determine the entropy magnitude term to be $\mu\e5\times10^3$ (Figure \ref{fig:paramFOC} bottom panel) in the same way, 
fixing the previous parameters and choosing the largest $\mu$ that still has small $\chi^2$ values for the photometry and the kinematics.

\begin{figure}
\begin{center}
\includegraphics[width=8.7cm]{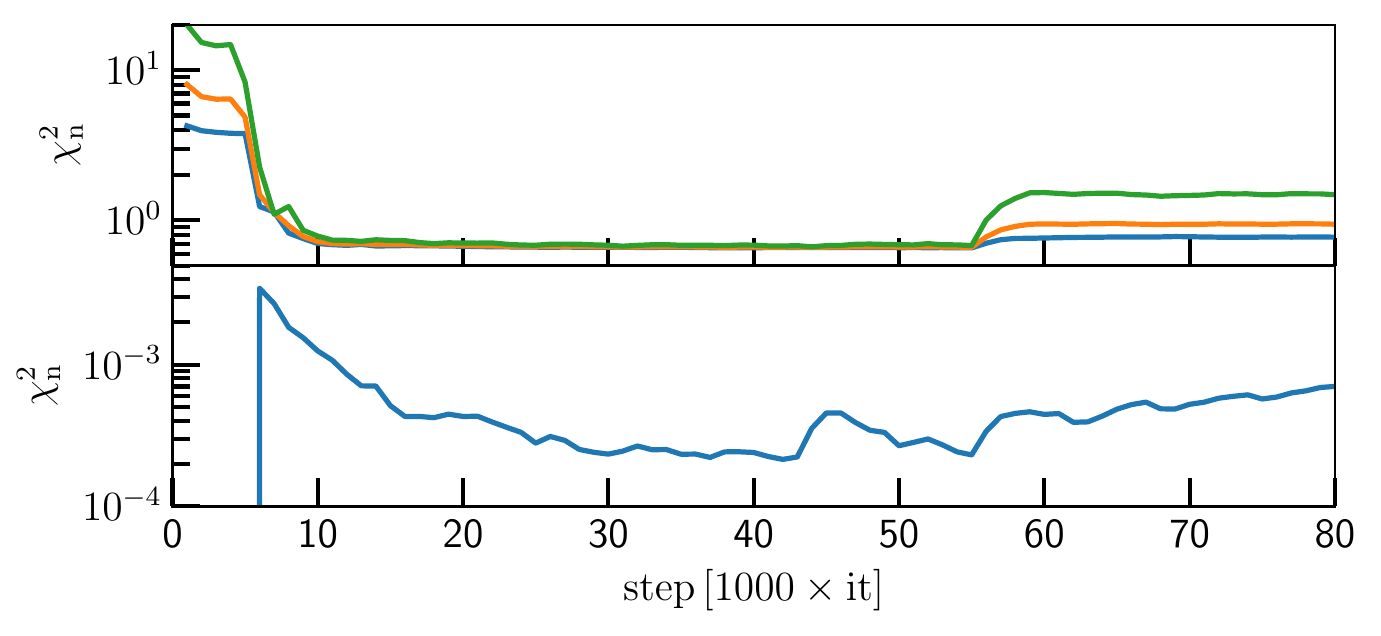}
\vspace{-0.7cm}
\caption[Chi-square versus time]{Reduced chi-square values as function of time for the M2M fitting of one model as defined in the main text. 
Top panel: photometry (blue), kinematics (green) and total (yellow) Bottom panel: dark matter halo density.}
\label{fig:chitime}
\end{center}
\end{figure}

After setting the fitting parameters, we run M2M fits, showing an example in Figure \ref{fig:chitime} 
where the reduced chi-squares of the model observables from equation \ref{eq:X2} are plotted versus time (iterations).
In the phase $T_{\rm obs}$ the model temporal smoothed observables are calculated decreasing $\chi^2_{\rm n}$ at first
and then staying constant. Then the fitting phase $T_{\rm M2M}$ starts where $\chi^2_{\rm n}$ of the photometry, kinematics and the dark matter halo
decrease in time. Finally in the stability check phase $T_{\rm stab}$ the values of $\chi^2_{\rm n}$ increase slightly.

\subsection{Exploring the effective potential parameters}
\label{sec:mod:tech}
\citetalias{Blana2017} find good constraints for the mass ratio between the \CB and the \BPB while
the dark matter distribution and the bar pattern speed are less constrained.  Here, we use stellar
kinematic and photometric observations as targets to better determine these properties.  While the
M2M method has the power to change the orbital distribution, thereby changing the model $\upsilon_{\rm
  los}$ $\sigma_{\rm los}$, $h3$, $h4$ and $L$ to match the observed kinematics, there are
macroscopic potential parameters that limit the orbital phase space, and therefore a particular
model will fit the data as well as these macroscopic parameters allow. Here we have three important
dynamical quantities that are inputs to the M2M modelling and that impact the effective potential:
the pattern speed of the bar $\pps$, the stellar mass-to-light ratio of the bulge in the 3.6\mum
band $\pml$ which, for a well fitted target observed luminosity, determines the total stellar mass
in the bulge $M^{\rm B}_{\star}$, and the amount of dark matter in the bulge region
$\pdm$. Therefore, we need to apply a method of meta-optimization where each M2M model is an
optimisation itself that finds the orbit distribution that best matches the observations for
fixed potential parameters. Then we vary $\pps$, $\pml$ and $\pdm$ 
{\bfun around reasonable values that we estimated from the literature,}
and then we find the range of values for which the M2M
models overall best reproduce all sets of observations. 
To explore these three global parameters we create
one cube (or grid) of model parameters for the Einasto dark matter profile, and a second cube for the NFW dark
matter halo profile, where each model $\vec{\rm M}$ has the coordinates:
\begin{align}
\label{eq:models}
&\vec{ \rm M}= \iz\pml,\pdm,\pps\de
\end{align}

For the Einasto cube we explore $\pml$ in the range of $0.5-0.85\ml$ in steps of $\Delta
\pml\e0.05\ml$ {\bfun to produce a low resolution grid that allows us to quickly find the best fitting region, and then we 
include} more values between $0.68-0.8\ml$ in steps of $\Delta \pml\e0.02\ml$.
For $\pdm$ we explore $0.6-2.4\times10^{10}\sm$ in steps of $\Delta M_{\rm
  DM}\e0.2\times10^{10}\sm$.  For $\pps$ we explore the range $20-55\psu$ in steps of
$\Delta\pps\e5\psu$, building then a cube of parameters with $13(\pml)\times 10(\pdm)\times
8(\pps)$, i.e., a total of 1040 M2M models with the Einasto dark matter profile.

For the NFW cube we explore $\pml$ in the range $0.62-0.8\ml$ in steps of $\Delta \pml\e0.02\ml$.
For $\pdm$ we explore $0.6-1.8\times10^{10}\sm$ in steps of $\Delta M_{\rm
  DM}\e0.2\times10^{10}\sm$.  For $\pps$ we explore $25-50\psu$ in steps of $\Delta\pps\e5\psu$,
giving a cube of parameters with $10(\pml)\times 7(\pdm)\times 6(\pps)$, i.e., a total of 420 M2M
models for the NFW cube.

Dark matter haloes are expected to be flattened in the central part of disc galaxies due to the
influence of the disc gravitational potential. \citet{Widrow2003, Widrow2005} explored different
flattening values for the dark halo of M31, finding reasonable fits between $q\simeq0.8$ and
1.0. Here we use a dark halo flattening of $q\e0.85$ as our fiducial value for both dark matter
density profiles, but we test the effects of different values on the final results.  We explore
$q\e0.7$ and $q\e1.0$, finding stellar mass distributions for the disc and the central region of 
the \CB similar to the fiducial model. This is discussed further in Section \ref{sec:res:param:mass}.

\subsection{Selection of best-matching models in effective potential parameter space}
\label{sec:model:select}

The selection of the best-matching models from the parameter grid just discussed cannot be done by
straightforward $\chi^2$-minimization, because with the extended, high-quality data available here,
systematic effects play a dominant role. These include uncertainties in the dust modelling,
intrinsic asymmetries in the observed surface brightness distribution (Figure~17 in Section~3.2.2
below), uncertainties in the parametrisation of the dark matter density distribution, and likely
gradients in $\pml$ especially between the BPB and adjacent disk. Because of these systematic
effects no model is found to give the best fit simultaneously in all regions of M31, and both
photometric and kinematic observables.

\begin{figure}
\begin{center}
\includegraphics[width=7.0cm]{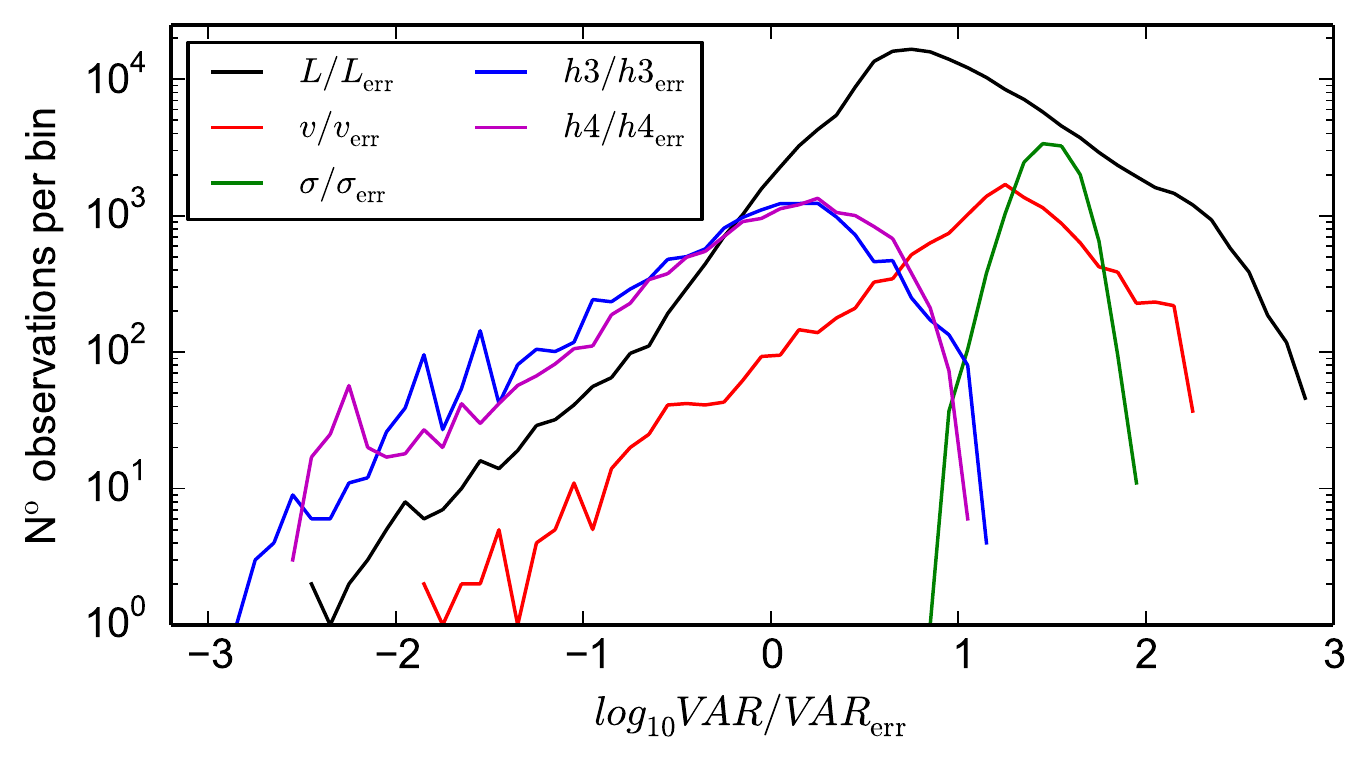}
\vspace{-0.4cm}
\caption[Photometric and kinematic relative errors]{Inverse relative errors of the photometric and the kinematic observations.
The minimum and maximum errors are: $10^{2.76}$ and $10^{8.05}\slu$ for the photometry, 
0.1 and 15.5\kms for the velocity, 1.1 and 15.6\kms for the dispersion, $0.6\times10^{-3}$ and $0.7\times10^{-1}$ for $h3$ 
and $0.7\times10^{-3}$ and $0.6\times10^{-1}$ for $h4$. 
The median errors are $10^{4.67}\slu$ for the photometry, 3.6\kms for the velocity, 3.8\kms for the dispersion, and 0.02 for $h3$ and $h4$.}
\label{fig:histerr}
\end{center}
\end{figure}

In addition, while the M2M models are fitted to an impressive number of 224251 photometric and
kinematic data values (pixels), the spatial distributions of photometric and kinematic pixels and
their residuals $\Delta_j^k$ are substantially different.  (i) Typical errors can differ
between different variables, e.g., between $L$ and $\sigma_{\rm los}$, or $\sigma_{\rm los}$ and
$h3$, leading to different ranges of $\Delta_j^k$; see Figure~\ref{fig:histerr}.  (ii) For
the same variable set, the errors depend on the spatial regions considered; e.g., relative
photometric errors are smaller in the central bulge than in its outer parts or in the disc region
(Figure \ref{fig:regions} below). Yet in all of these locations the data may contain signatures
important for specific physical properties of the system.

In consequence, combining all $(\Delta_j^k)^2$ values linearly in one total $\chi^2_{\rm tot}$ and
finding the M2M model with that minimum total chi-square will not adequately capture the entire
structure of M31; e.g., it will lead to a model providing a good fit of the \BPB region, but to an
unsatisfactory fit in the smaller central \CB region.  In the following we therefore describe an
alternative procedure which we believe leads to a more robust selection of the overall best-matching
models for M31 given the available data.

\subsubsection{Building a metric for the comparison with the observational data: five chi-square subsets}
\label{sec:mod:tech:regsub}
\begin{figure}
\begin{center}
\includegraphics[width=8.6cm]{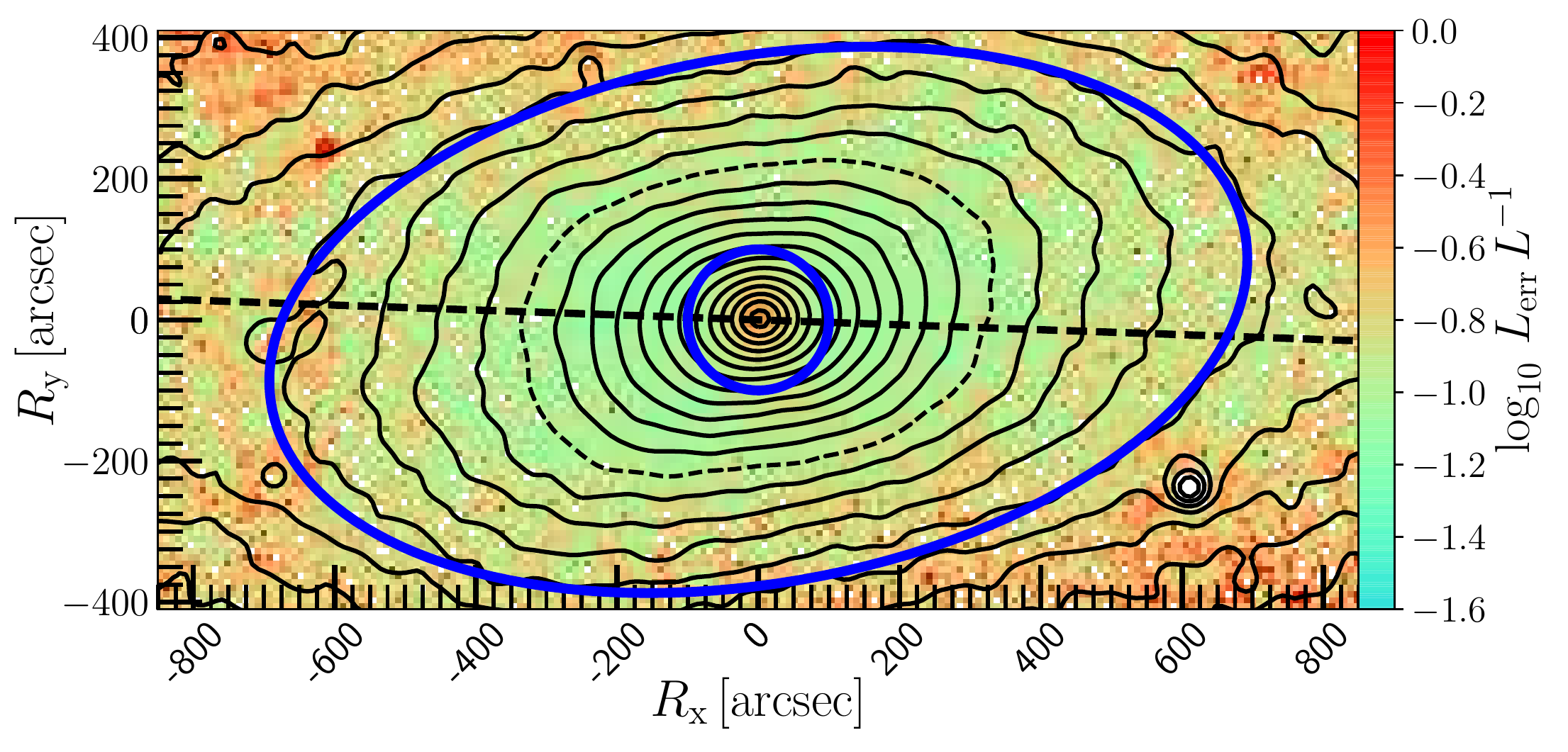} 
\vspace{-0.7cm}
\caption[Photometric relative error map]{The M31 bulge surface luminosity relative error map. We
  define the region of the \CB (CBR) within the blue circle, and the region of the
  \BPB (BPR) is between the circle and the ellipse, and everything within the ellipse
  comprehend then the bulge region (BR). The surface-brightness isophotes in the 3.6\mum band are
  shown spaced with $\Delta \mu_{3.6}\e0.25\,[{\rm mag}\,{\rm \as^{-2}}]$. The value
  $\mu_{3.6}\e16\,[{\rm mag}\,{\rm \as^{-2}}]$ is shown with a dashed isophote and the disc major
  axis is at $\pa\e38\degree$ (dash line). In white colour are the masked hot pixels and foreground
  stars.}
\label{fig:regions}
\end{center}
\end{figure}

Separating the observables $L$, $\upsilon_{\rm los}$, $\sigma_{\rm los}$, we first build five
subsets of normalized $\chi^2$. These are motivated by the properties of the system that we are
modelling, which is built from the three main substructures \CB, \BPB and disc that we want to fit
\textit{simultaneously} well.  The \CB dominates the light in M31 within
$R\lesssim100\as\,(380\pc)$, defined as region CBR.  Further out the \BPB dominates the light
within ellipses with semimajor axis $100\as\!<\!R_{\rm mj}\!<\!700\as$ (region BPR), and even
further out the disc dominates \citep[][B17]{Beaton2007}; see Figure \ref{fig:regions}. We define
five subsets of normalized $\chi^2$s:
\begin{itemize}
\item \CB central photometry (\setAo): we measure the normalized $\chi^2$ (per data point) of the
  photometry ($L_{3.6\mum}$) in the inner CBR within a diameter of 40\as (150\pc, $\sim R_{\rm e}/
  10$).  With this we search for models that match the cuspy light profile in the centre of M31's
  bulge.
\item \CB central dispersion (\setBo): the M31 dispersion profile shows two peaks of
  $\sigma_{\rm los}\si170\kms$ at $R\si50\as$, but drops to $\sigma_{\rm los}\si150\kms$ in the
  centre \citep{Saglia2010, Opitsch2017}.  Therefore, we also measure the normalized $\chi^2$ of
  $\sigma_{\rm los}$ within $R_{\rm e}/10$, to find models of the grid that reproduce this feature.
\item \BPB photometry (\setCo): we measure the normalized $\chi^2$ of the photometry in
  region BPR (Figure \ref{fig:regions}).
\item \BPB dispersion (\setDo): \citetalias{Blana2017} show that the \BPB and the
  \CB of M31 have different kinematic properties.  Hence, we calculate the normalized $\chi^2$ of
  the dispersion only in the BPR. This allows us also to find the dynamical mass within
  the bulge.
\item \BPB mean velocity (\setEo): \citet{Tremaine1984} showed that the bar pattern speed is
  related to the LOS velocity ($\upsilon_{\rm los}$) and the photometry. Therefore, we constrain the
  bar pattern speed with the normalized $\chi^2$ of the mean LOS velocity $\upsilon_{\rm los}$ in the
  bar region BPR.
\end{itemize}
In this way, each model $\vec{\rm M}$ is evaluated with five normalized $\chi^2$ parameters, 
$\vec\chi^2= \iz \setAo, \setBo, \setCo, \setDo,\setEo\de$.
While the Gauss-Hermite coefficients $h3$ and $h4$ and all observables in the disc region are also
fitted in each of the M2M models, we do not include $\chi^2$ subsets for them in the best model
selection; later we show that the best models selected by the five subsets defined above
satisfactory reproduce these observations as well.

\subsubsection{Model ranking}
\label{sec:mod:tech:sel}
\begin{figure}
\begin{center}
\includegraphics[width=7.0cm]{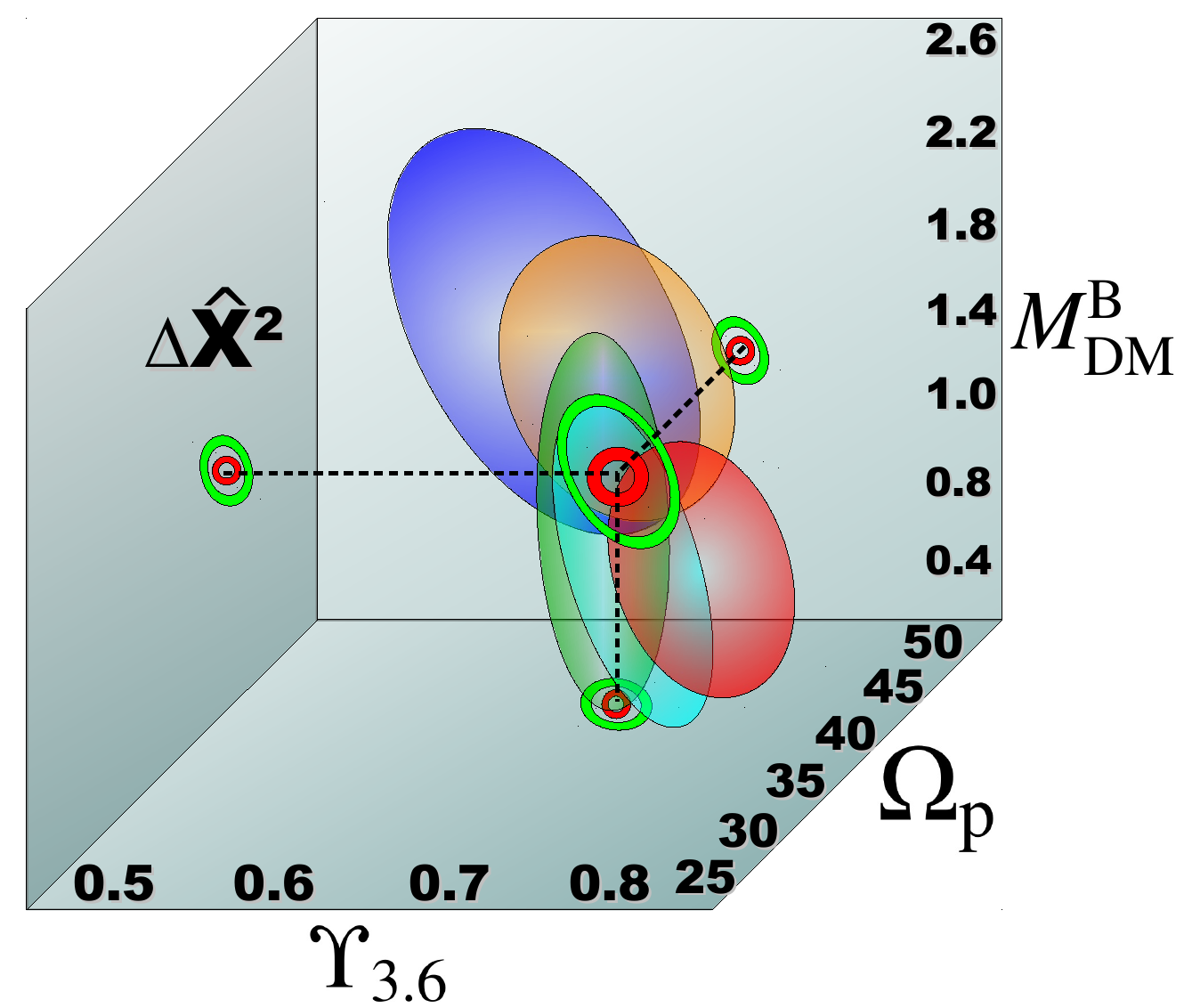}
\vspace{-0.3cm}
\caption[Chi-square subsets diagram]{Representation of the five chi-square subset volumes in the parameter space of $\pml$, $\pdm$ and $\pps$. 
While the models have chi-square values in the whole cube, the coloured ellipses represent volumes
where the chi-square values have the lowest values, showing \setA (green), \setB (cyan), \setC (blue), \setD (red) and \setE (orange).
The place where all ellipses intersect is where is located the overall best model $\vec{\rm M}_{\rm BM}$ (red circle).
We also show the range of the acceptable models $\vec{\rm M}_{\rm AM}$ (green ring). The projections of the best model 
on each of the planes of the effective potential parameters are indicated with the dashed lines.}
\label{fig:diagramX2}
\end{center}
\end{figure}

\begin{table*}
  \caption[Best Einasto models]{Main parameters for the range of acceptable models for the Einasto dark
    dark matter profile \bmein; see text. The overall best matching model is Model JR804, corresponding to
    the best parameter values (B.V.). Inferred error ranges $\Delta^{+}_{-}$ are based on all acceptable models.}
\vspace{-0.3cm}
\label{tab:MdelEIN}
\begin{tabular}{p{0.7cm}|p{0.4cm}|p{0.4cm}|p{0.4cm}|p{0.4cm}|p{0.4cm}|p{0.4cm}|p{0.4cm}|p{0.6cm}|p{0.6cm}|p{0.6cm}|p{0.6cm}|p{0.6cm}|p{0.6cm}|p{0.3cm}|p{0.6cm}}
\hline
Model &  $\pml$ & $\pdm$ & $\pps$& $M_{\star}^{\rm CB}$ & $M_{\star}^{\rm BPB}$ & $M_{\star}^{\rm B}$ & $M_{\rm DM}^{\rm B (p)}$ & $M_{\rm dyn}^{\rm B}$  & \setA & \setB & \setE & \setC &\setD & \setsum & $\Delta$\setsum\\ \hline
\textbf{JR804} & \textbf{0.72} & \textbf{1.2} & \textbf{40} & \textbf{1.18} & \textbf{1.91} & \textbf{3.09} & \textbf{1.16} & \textbf{4.25} & \textbf{0.57} & \textbf{1.27} & \textbf{0.43} & \textbf{1.04} & \textbf{0.61} & \textbf{3.92} & \textbf{0.00} \\
JR803  &  0.72  &  1.0  &  40  &  1.19  &  1.89  &  3.08  &  0.97  &  4.05  & 0.28  & 1.53  &  0.98  &  1.12  &  0.66  &  4.58  &  0.65 \\
JR813  &  0.74  &  1.0  &  40  &  1.22  &  1.97  &  3.19  &  0.99  &  4.18  & 1.67  & 0.72  &  0.88  &  1.31  &  0.18  &  4.77  &  0.84 \\
JR764  &  0.72  &  1.2  &  35  &  1.15  &  1.93  &  3.08  &  1.18  &  4.26  & 0.41  & 1.17  &  1.16  &  0.67  &  1.68  &  5.10  &  1.18 \\
JR763  &  0.72  &  1.0  &  35  &  1.16  &  1.91  &  3.07  &  0.98  &  4.05  & 0.21  & 0.98  &  2.24  &  0.89  &  1.07  &  5.39  &  1.46 \\
JR365  &  0.70  &  1.4  &  40  &  1.13  &  1.85  &  2.98  &  1.35  &  4.33  & 0.26  & 2.81  &  0.15  &  1.08  &  1.24  &  5.54  &  1.61 \\
JR285  &  0.70  &  1.4  &  35  &  1.11  &  1.86  &  2.97  &  1.38  &  4.35  & 0.14  & 2.32  &  0.52  &  0.40  &  2.31  &  5.68  &  1.75 \\
JR812  &  0.74  &  0.8  &  40  &  1.23  &  1.95  &  3.18  &  0.78  &  3.96  & 1.20  & 0.37  &  2.07  &  1.50  &  0.82  &  5.95  &  2.03 \\
JR853  &  0.74  &  1.0  &  45  &  1.24  &  1.95  &  3.19  &  0.99  &  4.18  & 1.58  & 0.44  &  0.94  &  2.64  &  0.51  &  6.12  &  2.19 \\
JR844  &  0.72  &  1.2  &  45  &  1.20  &  1.90  &  3.10  &  1.18  &  4.28  & 0.54  & 1.39  &  0.85  &  2.72  &  0.77  &  6.26  &  2.34 \\
JR284  &  0.70  &  1.2  &  35  &  1.12  &  1.85  &  2.97  &  1.18  &  4.15  & 0.47  & 2.59  &  1.23  &  0.35  &  1.68  &  6.32  &  2.40 \\
\hline
B.V.  &  0.72  &  1.2  &  40.0  &  1.18  &  1.91  &  3.09  &  1.16  &  4.25  &   &   &   &   &   &  & \\
$\Delta^{+}_{-}$  &  $^{+0.02}_{-0.02}$  &  $^{+0.2}_{-0.4}$  &  $^{+5.0}_{-5.0}$  &  $^{+0.06}_{-0.07}$  &  $^{+0.06}_{-0.06}$  &  $^{+0.10}_{-0.12}$  &  $^{+0.22}_{-0.38}$  &  $^{+0.10}_{-0.29}$  &   &   &   &   &   &  & \\  \hline
\end{tabular}\\
\textbf{Notes:} $M_{\star}^{\rm CB}$, $M_{\star}^{\rm BPB}$, $\pdm$, $M_{\rm DM}^{\rm B (p)}$ and $M_{\rm dyn}^{\rm B}$  in units of $10^{10}\sm$. 
Parameters $\pps$ and $\pml$ are in units of $\psu$ and \ml respectively. 
\end{table*}

\begin{table*}
\caption[Best NFW models]{Main parameters of the range of acceptable models for the NFW dark matter profile 
    \bmnfw.  The overall best matching model is Model KR241, corresponding to the best parameter values (B.V.). 
    Inferred error ranges $\Delta^{+}_{-}$ are based on all acceptable models with minimum given by the resolution
    of the model grid.}
\vspace{-0.3cm}
\label{tab:MdelNFW}
\begin{tabular}{p{0.7cm}|p{0.4cm}|p{0.4cm}|p{0.4cm}|p{0.4cm}|p{0.4cm}|p{0.4cm}|p{0.4cm}|p{0.6cm}|p{0.6cm}|p{0.6cm}|p{0.6cm}|p{0.6cm}|p{0.6cm}|p{0.3cm}|p{0.6cm}}
\hline
Model &  $\pml$ & $\pdm$ & $\pps$& $M_{\star}^{\rm CB}$ & $M_{\star}^{\rm BPB}$ & $M_{\star}^{\rm B}$ & $M_{\rm DM}^{\rm B (p)}$ & $M_{\rm dyn}^{\rm B}$  & \setA & \setB & \setE & \setC &\setD & \setsum & $\Delta$\setsum\\ \hline
\textbf{KR241} & \textbf{0.70} & \textbf{1.0} & \textbf{40} & \textbf{1.16} & \textbf{1.82} & \textbf{2.98} & \textbf{0.97} & \textbf{3.95} & \textbf{0.51} & \textbf{1.64} & \textbf{1.76} & \textbf{1.10} & \textbf{1.61} & \textbf{6.61} & \textbf{0.00} \\
KR248  &  0.72  &  1.0  &  40  &  1.18  &  1.90  &  3.08  &  0.98  &  4.06  & 0.80  & 3.27  &  1.66  &  1.16  &  0.75  &  7.64  &  1.03 \\
KR235  &  0.68  &  1.2  &  40  &  1.12  &  1.77  &  2.89  &  1.17  &  4.06  & 1.62  & 2.97  &  1.00  &  0.88  &  1.45  &  7.93  &  1.32 \\
KR171  &  0.70  &  1.0  &  35  &  1.13  &  1.85  &  2.98  &  0.98  &  3.96  & 0.31  & 1.26  &  3.87  &  1.13  &  1.45  &  8.03  &  1.41 \\
KR165  &  0.68  &  1.2  &  35  &  1.09  &  1.79  &  2.88  &  1.18  &  4.06  & 1.19  & 2.67  &  2.34  &  0.56  &  1.32  &  8.08  &  1.47 \\
KR247  &  0.72  &  0.8  &  40  &  1.20  &  1.88  &  3.08  &  0.78  &  3.86  & 0.27  & 1.35  &  3.43  &  1.85  &  1.99  &  8.89  &  2.28 \\
KR242  &  0.70  &  1.2  &  40  &  1.15  &  1.84  &  2.99  &  1.17  &  4.16  & 0.30  & 6.47  &  0.83  &  0.92  &  0.62  &  9.14  &  2.53 \\
KR159  &  0.66  &  1.4  &  35  &  1.06  &  1.74  &  2.80  &  1.37  &  4.17  & 3.34  & 2.84  &  1.31  &  0.20  &  1.52  &  9.21  &  2.60 \\
\hline
B.V.  &  0.70  &  1.0  &  40.0  &  1.16  &  1.82  &  2.98  &  0.97  &  3.95  &   &   &   &   &   &  & \\
$\Delta^{+}_{-}$  &  $^{+0.02}_{-0.04}$  &  $^{+0.4}_{-0.2}$  &  $^{+5.0}_{-5.0}$  &  $^{+0.04}_{-0.10}$  &  $^{+0.08}_{-0.08}$  &  $^{+0.10}_{-0.18}$  &  $^{+0.40}_{-0.19}$  &  $^{+0.22}_{-0.09}$  &   &   &   &   &   &  & \\  \hline
\end{tabular}\\
\textbf{Notes:} $M_{\star}^{\rm B}$, $M_{\star}^{\rm CB}$, $M_{\star}^{\rm BPB}$, $\pdm$, $M_{\rm DM}^{\rm B (p)}$ and $M_{\rm dyn}^{\rm B}$  in units of $10^{10}\sm$. 
Parameters $\pps$ and $\pml$ are in units of $\psu$ and \ml respectively.
\end{table*}

In the space of the parameters $\pml$, $\pdm$ and $\pps$, the normalized $\chi^2$ for each of the
five subsets defines a region of acceptable models and a minimum $\chi^2$ model. However, we find
that the subset chi-square values have stochastic local variations on top of the global trends,
similarly as \citet{Morganti2013} found for their M2M models. Thus there may be several models that
have $\chi^2$ values near the minimum. This stochasticity dominates the statistical uncertainty
measured by normal delta chi-square analysis, which is not unexpected given the large amount of high
quality data fitted and the remaining systematics.

Therefore, to better determine the global $\chi^2$ minimum in each subset, we follow
\citet[][]{Gebhardt2003} and obtain smoothed chi-square values for all models.  Specifically, we
average each model's normalized chi-square with those of its $3\times3\times3-1\e26$
neighbouring models (we also tested averaging with $3\times2\e6$ neighbouring models finding similar
chi-square volumes and the same range of selected models).  Then we find the minimum smoothed
chi-square value ($\chi_{\rm min}^2$) in each of the subsets (which do not necessarily correspond to
the same model $\vec{\rm M}$), obtaining for the Einasto halo grid
\begin{align}
\label{eq:subsetsmin}
{\vec \chi^2_{\rm min}} &= \iz \setAomin, \setBomin, \setComin , \setDomin, \setEomin\de \cr
                      &= \iz 0.195, 0.267, 0.774, 2.717, 3.544 \de.
\end{align}

We also quantify the scatter introduced by the
stochasticity described, calculating the standard deviation ($s$) of the {\sl original} chi-square values
of the models neighbouring the model with the minimum smoothed $\chi^2$ that is not on the border of
the grid. For the five subsets we obtain
$\setAos\e0.062$, $\setBos\e0.155$, $\setEos\e0.370$, $\setCos\e0.040$ and $\setDos\e0.097$.  
Then we compute normalized $\Delta {\hat \chi}^2$ values for each model $\vec{\rm M}\iz\pml, \pdm, \pps
\de$ in all data subsets, $\setA, \setB, \setC, \setD, \setE$, where
\begin{align}
\label{eq:delta}
&\Delta {\hat \chi}^2\iz {\rm subset}\de=\iz\chi_{\rm n}^2\iz{\rm subset}\de
    -\chi^2_{\rm n\, min}\iz{\rm subset}\de\de / s\iz{\rm subset}\de
\end{align}
based on the smoothed chi-squares and the standard deviation of the original chi-squares near
minimum.  In other words, we characterize the fit of a model to all the data in one of the five
subsets by a single goodness-of-fit $\Delta{\hat \chi}^2$ parameter. This is the difference between
the smoothed chi-square per data point relative to the minimum, normalized by the original scatter
between neighbouring models around minimum. In this way, all the $\Delta{\hat \chi}^2$ are of
similar magnitude, which allows us to compare models simultaneously with the signatures contained in
the different data subsets.

The range of good models in each independent subset is defined by a volume in the space of $\pml$,
$\pdm$ and $\pps$, with values $\Delta {\hat \chi}^2\iz {\rm subset} \de\lesssim1$, as illustrated
in Figure \ref{fig:diagramX2}. The volume where all subsets intersect with small chi-square values
is where the models \textit{simultaneously} have small deviations from the best model in all of the
subsets, and corresponds to the region of the best-matching parameters $\pml$, $\pdm$ and $\pps$.

We quantify the size of this region using a total delta chi-square $\Delta \setsum$ for each model,
obtained by summing the normalized delta chi-square values for the $N_{\rm sub}\e5$ subsets:
\begin{align}
\label{eq:ssetot}
&\Delta \setsum =\iz\setsum-\hat\chi^2_{\rm sum , min}\de; \qquad \setsum =\sum_i^{N_{\rm sub}} \Delta\hat{\chi}^2_i
\end{align}
where $\hat\chi^2_{\rm sum , min}$ is the minimum value of $\setsum$. $\Delta \setsum$ ranks the
models from the best fitting model with minimum $\hat\chi^2_{\rm sum , min}\e3.92$, up to the worst
fitting model on the grid with $\setsum\e954$.  Sorting the models by $\setsum$ results in Table
\ref{tab:MdelEIN} for the Einasto grid, where we show just the range of acceptable models. The first
model (JR804) is the overall best matching model $\vec{\rm M}_{\rm BM}$ and determines the best
values of the parameters \pml, \pdm and \pps. It does not necessarily has the minimum
chi-square in each data subset, but achieves the best compromise in matching
simultaneously all the observational datasets \citep[see also][]{Portail2017a}.

Errors for parameters \pml, \pdm and \pps are estimated from the maximum and minimum values 
in the acceptable models $\vec{\rm M}_{\rm AM}$ with
\begin{align}
\label{eq:del}
\vec{\rm M}_{\rm AM}=&\left\{\forall \vec {\rm M}\, |\,\, \Delta \setsum \leq \delta\right\}
\end{align}
where we choose $\delta\e2.7$, obtaining the range of models listed in Table \ref{tab:MdelEIN}.
While the exact value of this threshold is arbitrary, inspection of the models within this limit
shows that they reproduce all data satisfactorily including the most problematic outer bulge stellar
kinematics.  For $\delta\e2.7$, no individual subset of any model has $\Delta\hat{\chi}^2_i>3$,
while if we had chosen $\delta\e1.3$, all subsets would be fitted with
$\Delta\hat{\chi}^2_i<2$. These latter four models match the data even better, but we choose the
more conservative $\Delta\e2.7$ for the following reasons: (i) Compared to the number of models with
$\Delta\setsum<2.7$, there is only a small number of models with $2.7<\Delta\setsum<4$, and these
models allow only one new value of \pdm. (ii) At the same time, some individual subset $\Delta{\hat
  \chi}^2$ in this $2.7<\Delta\setsum<4$ range usually gets much worse, which is confirmed by
inspecting the model fits. Thus we consider $\delta\e2.7$ the most conservative choice consistent
with the data.  We note in passing that if the individual $\Delta\hat{\chi}^2_i$ were the square residual of single,
Gaussian distributed measurements (which they are not), then $\Delta\setsum\e2.7$ would correspond
to 90 per cent of the $\chi^2$-distribution.

We also tested a different selection criterium to find the range of acceptable models. There we
selected models for which each data subset has a maximum allowed deviation from the minimum in each
subset, finding a similar range of models $\vec{\rm M}_{\rm AM}$ and consequently, a similar
uncertainty range for the parameters \pml, \pdm and \pps. 

We finally applied the same procedure to the grid of NFW models (Table~\ref{tab:MdelNFW}). The
chi-square comparisons of the subset values and \setsum between the Einasto and the NFW models
indicate that the Einasto dark matter profile provides a better fit to the observations (the best
NFW model KR241 has $\setsum=6.62$, already outside the range of acceptable models in the Einasto
grid).  Nonetheless, the range of parameters \pml, \pdm and \pps obtained within the NFW models on
their own is very similar to that found previously.

\section{Results}
\label{sec:res}
Here we first describe the results of our parameter study for M31, and discuss the values we obtain
for the mass-to-light ratio, dark matter mass in the bulge, and pattern speed, as well as the
implied dark matter density and rotation curve decomposition (Section \ref{sec:res:param}).  In the
second part (Section \ref{sec:res:bm}), we compare the photometric and kinematic maps and profiles
of M31 with our best matching model.

\begin{figure*}
\begin{center}
\includegraphics[height=5.cm]{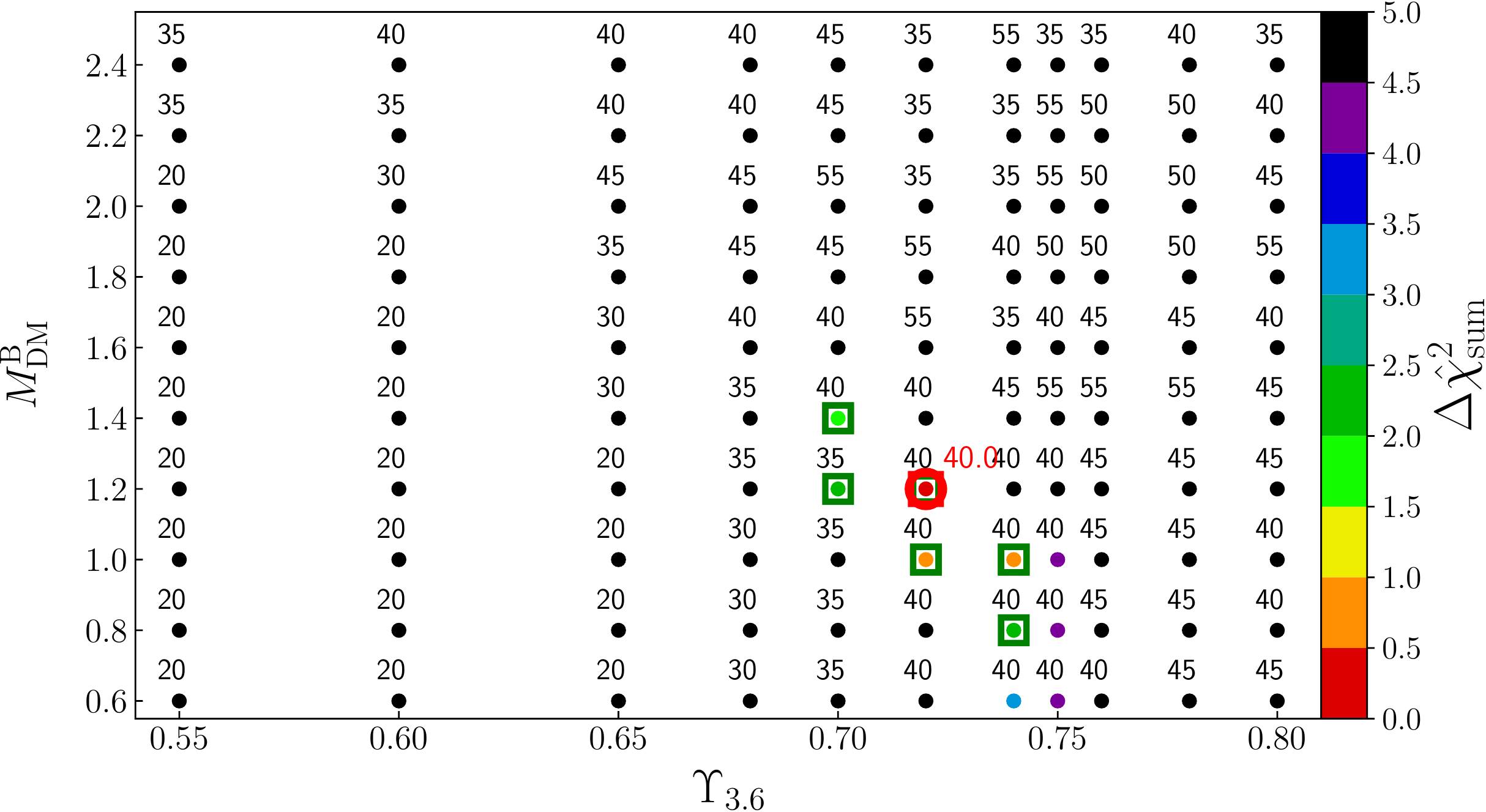}
\qquad 
\includegraphics[height=5.cm]{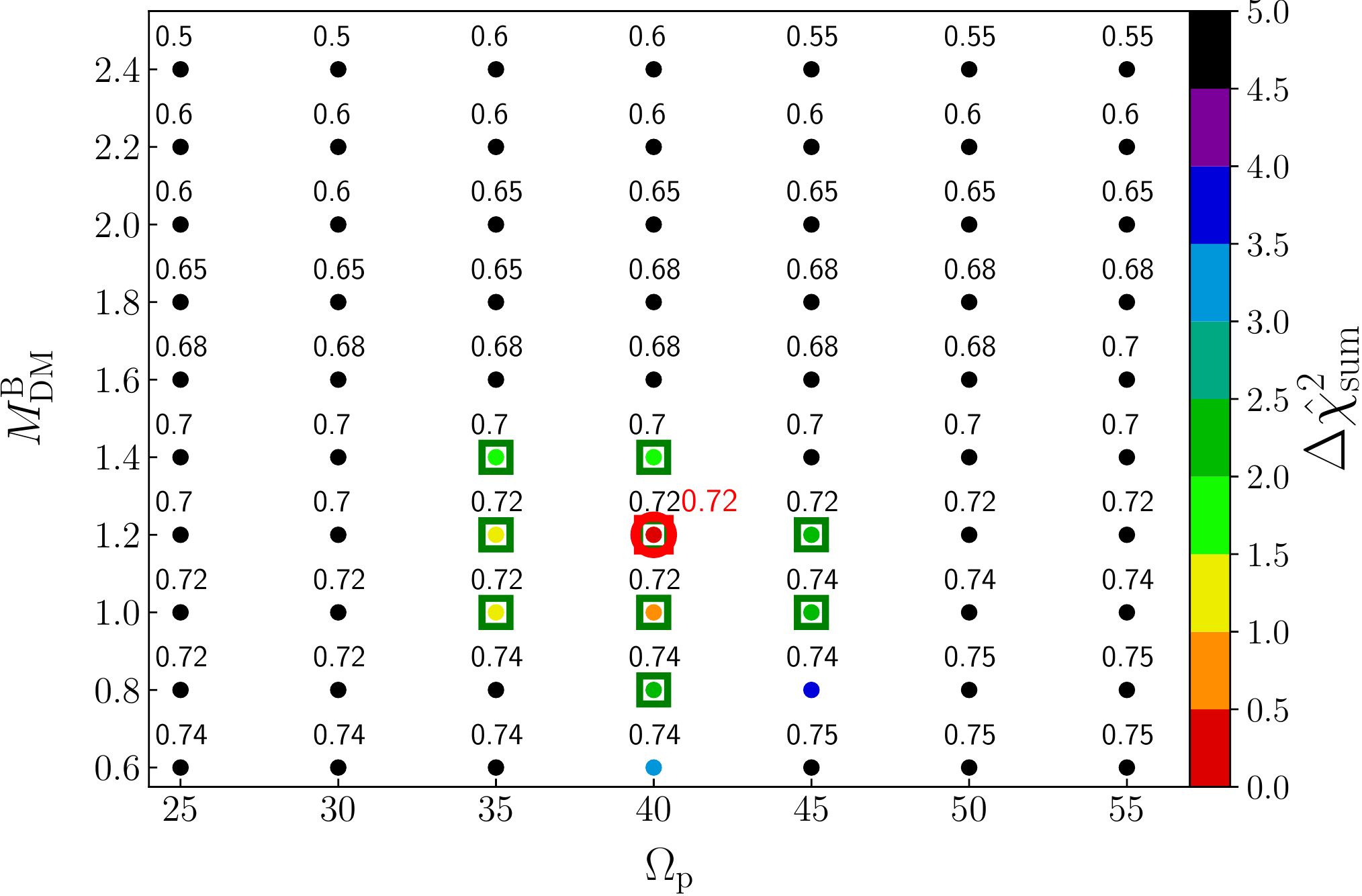} 
\vspace{-0.2cm}
\caption[Parameter \setdsum: best models]{Range of acceptable models defined by the total
  goodness-of-fit \setdsum, for an Einasto halo.  The green squares mark the range of acceptable
  models, with the red circles marking the overall best matching model JR804.  Left panel: \setdsum
  in the \pml, \pdm plane, always selecting the minimum value along the parameter \pps axis.  Right
  panel: \setdsum as function of \pps and \pdm, selecting the minimum value along the parameter \pml
  axis.  In black are shown models with $\setdsum\ge5$), with the largest value in the grid being
  $\setsum\e954.7$.}
\label{fig:Xsum}
\end{center}
\end{figure*}

\subsection{M31 potential parameters from the best M2M models}
\label{sec:res:param}

From the model grid with Einasto dark matter halo profiles and the selection procedure explained in
Section \ref{sec:mod:tech:sel}, we find the allowed range of values for the  3.6\mum mass-to-light ratio,
the dark matter mass in the bulge, and pattern speed:
\begin{align}
\label{bestvalues}
& \bml, \\
& \bdm, \\
& \bps.
\end{align}
Models with an NFW halo fit the data significantly worse (Section \ref{sec:mod:tech:sel}), but
result in similar parameter values: \bnfwml, \bnfwdm, and \bnfwps. In both cases the central value is the best model and the errors are based on the range of acceptable models; see Table
\ref{tab:MdelEIN} (Einasto) and Table \ref{tab:MdelNFW} (NFW). In the subsequent discussion we will
therefore use the Einasto models.

Figure \ref{fig:Xsum} shows the total goodness-of-fit \setdsum as function of the parameters \pml,
\pdm and \pps for the Einasto models. A small degeneracy between \pml and \pdm remains within the
range of allowed values. This is discussed further below. Figure \ref{fig:XsumNFW} in the Appendix
shows similar information for the NFW models, where the degeneracy is slightly increased because the
more concentrated NFW profile has more mass within the bulge than the Einasto profile.

In the next subsections we explain how the physical parameters \pml, \pdm and \pps are constrained
by different subsets of the data. The corresponding signatures in the chi-square values between M31
data and models allow us to determine these parameters and, for example, break the degeneracy
between the stellar mass and the dark matter mass in the bulge.

\subsubsection{Constraining  \pml and \pdm}
\label{sec:res:param:MLDM}

\begin{figure*}
\begin{center}
\includegraphics[width=8.82cm]{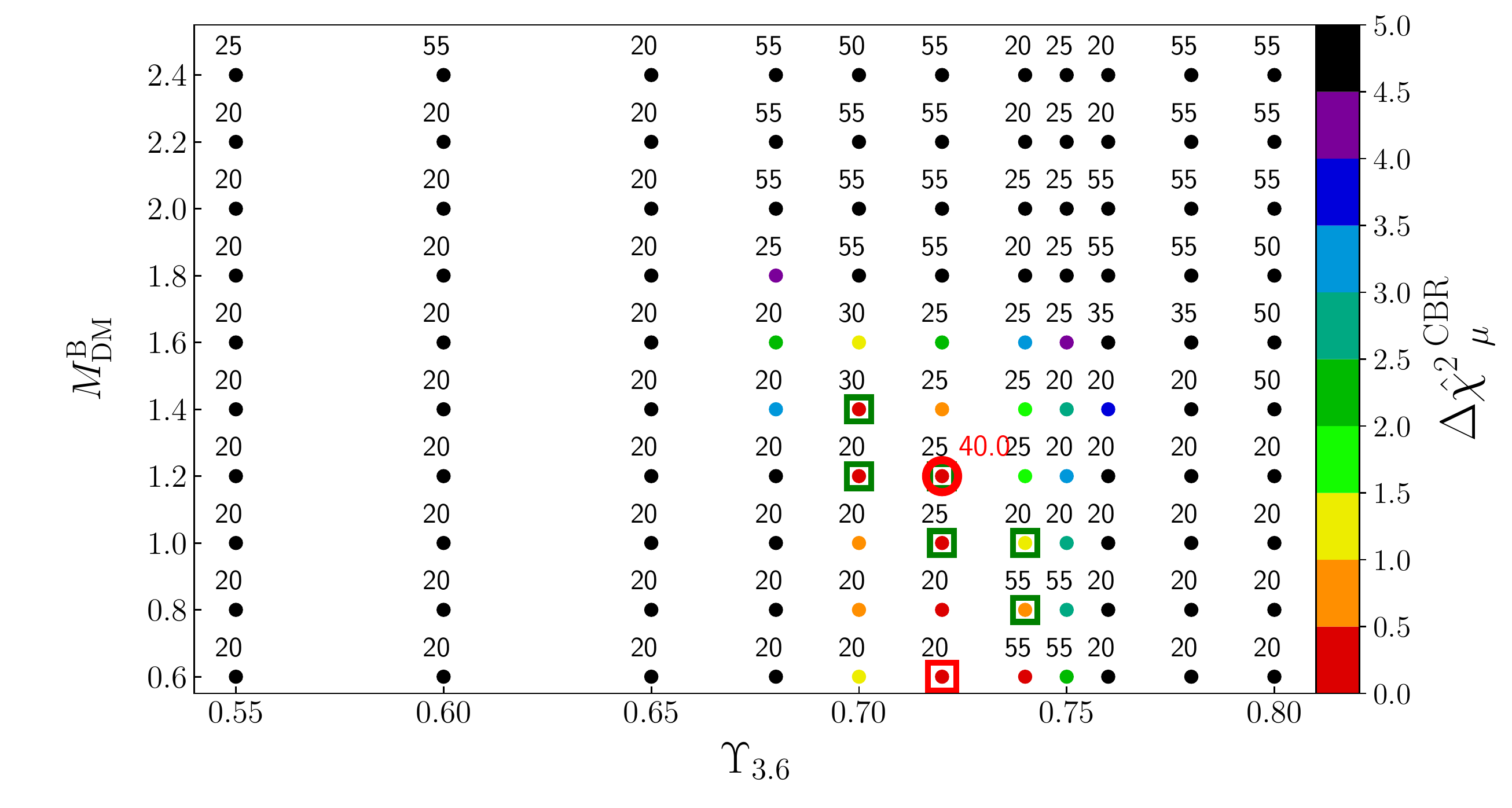} 
\hspace{-0.2cm}
\includegraphics[width=8.82cm]{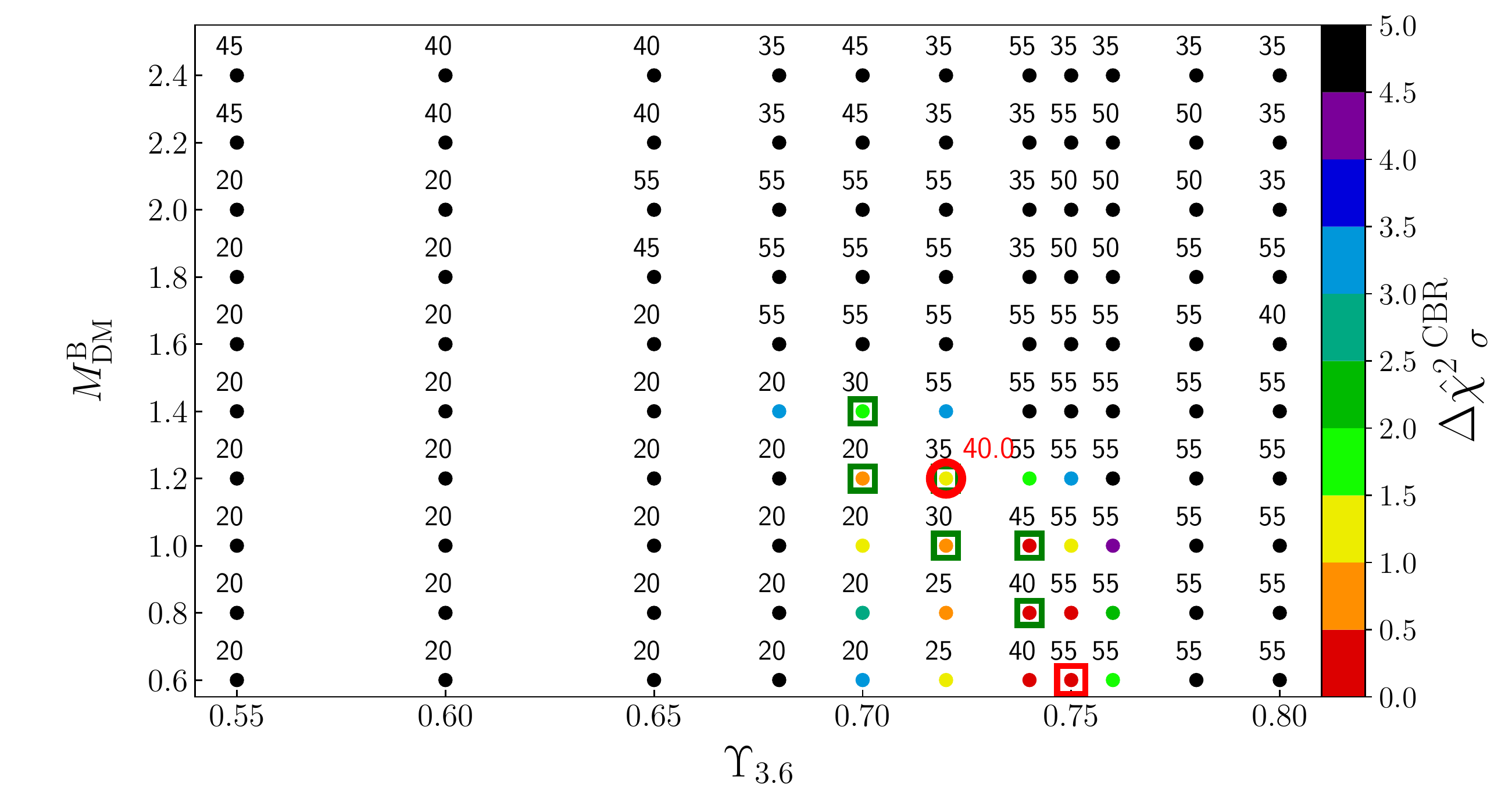}\\
\includegraphics[width=8.88cm]{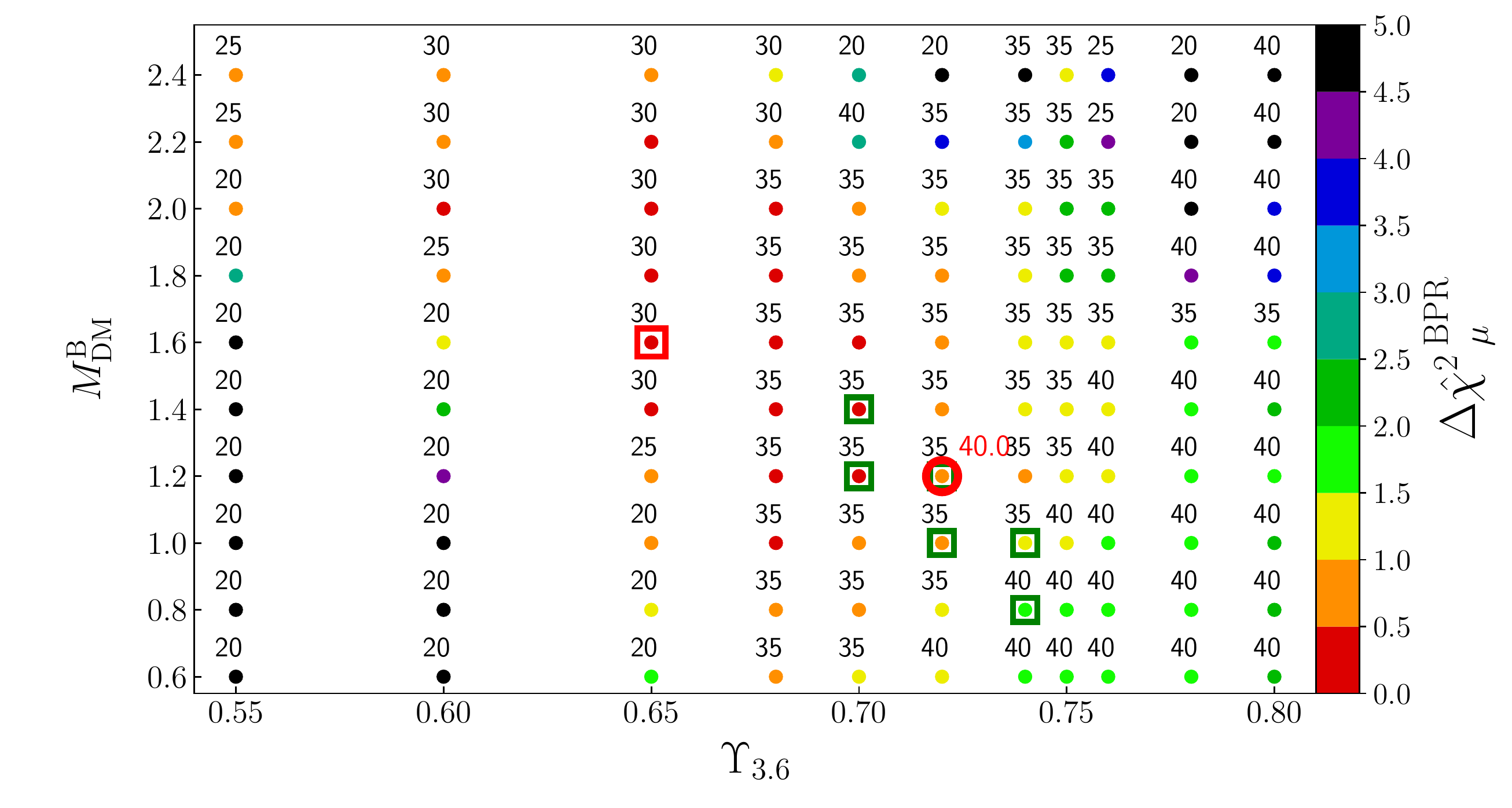} 
\hspace{-0.2cm}
\includegraphics[width=8.88cm]{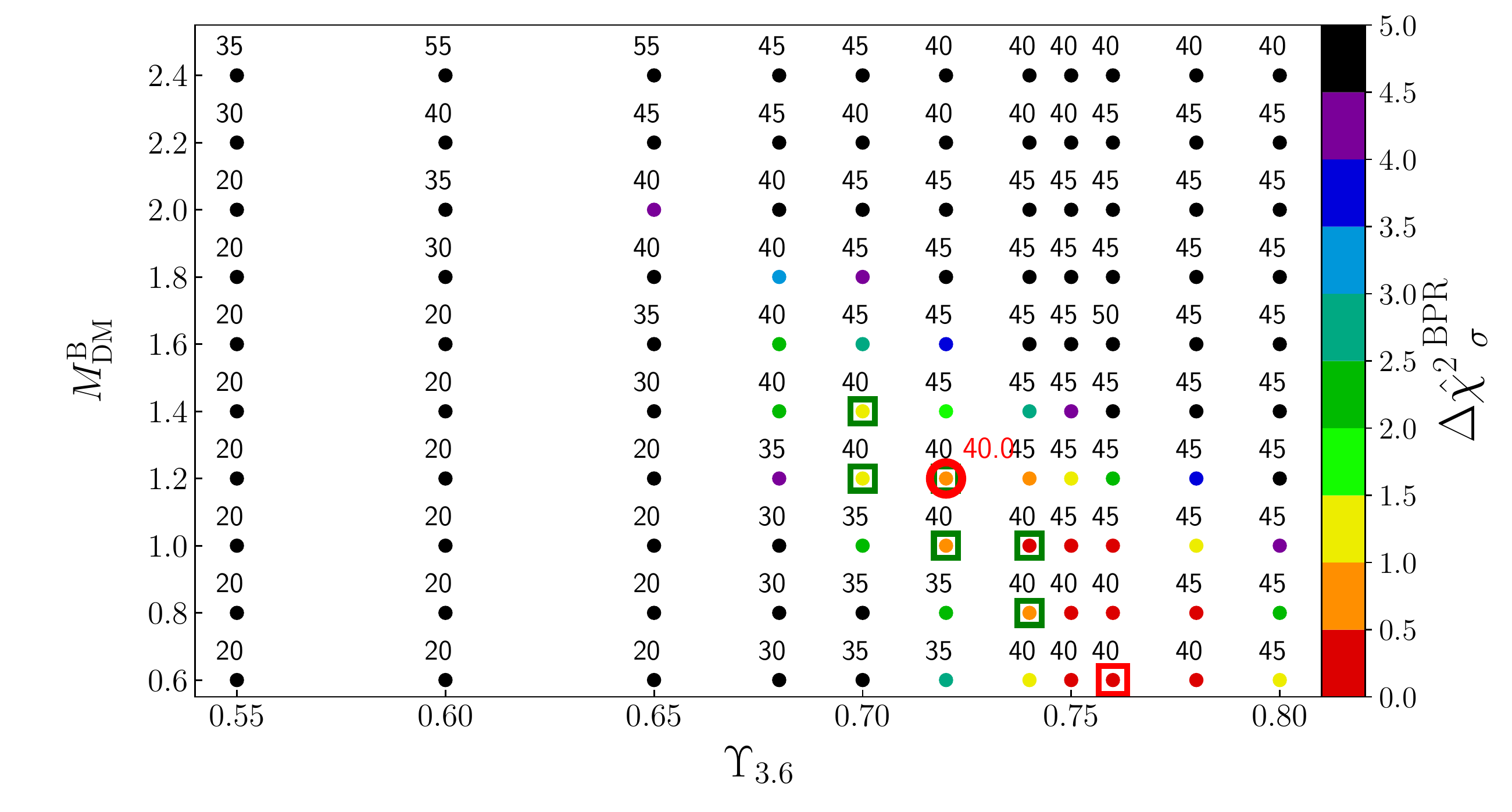}
\vspace{-0.2cm}
\caption[Subsets part 1]{Breaking the degeneracy of \pml and \pdm with different data sets.  Plotted
  are the subset \setA (top left), \setB (top right), \setC (bottom left) and \setD (bottom right)
  for the Einasto models as function of the parameters $\pml$ and $\pdm$, selecting for each pair of
  values the model with the minimum $\Delta\hat{\chi}^2_i$ along the $\pps$ axis.  Numerical values
  for the points are given by the colour scale. The number next to each point corresponds to the
  $\pps$ value of the model with the lowest chi-square.  We mark the best model JR804 (red circle), the models
  with the minimum values in each subset (red squares), and the range of acceptable models \bmein
  (green squares). The green squares do not necessarily have the pattern speed shown.}
\label{fig:sspart1EIN}
\end{center}
\end{figure*}

\begin{figure*}
\begin{center}
\includegraphics[width=8.6cm,trim=0cm 2cm 0cm 0cm,clip]{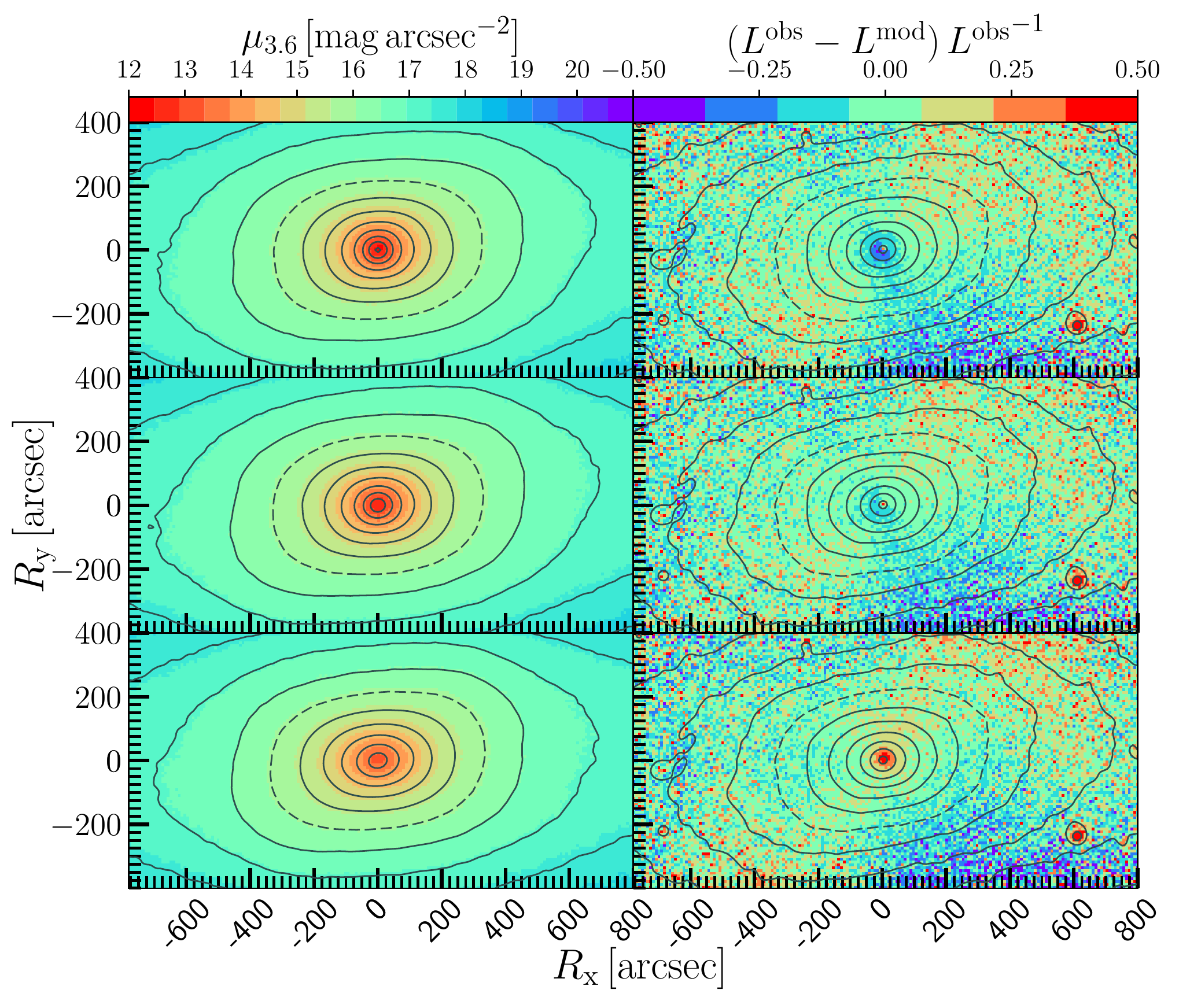}
\includegraphics[width=7.7cm,trim=2.05cm 2cm 0cm 0cm,clip]{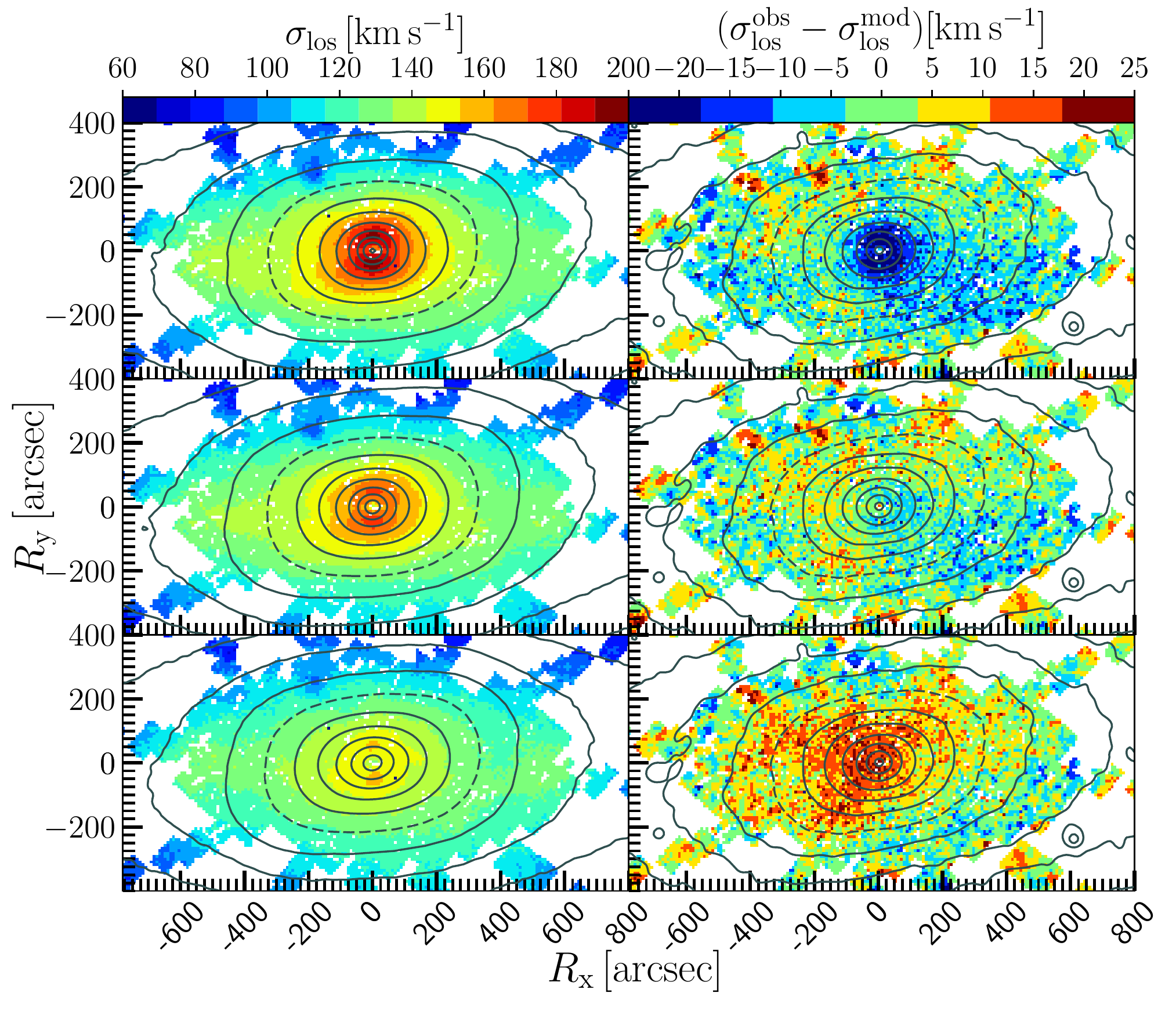}
\hspace{-0.3cm}
\includegraphics[height=5.6cm,trim=0cm 0cm 0cm 0cm,clip]{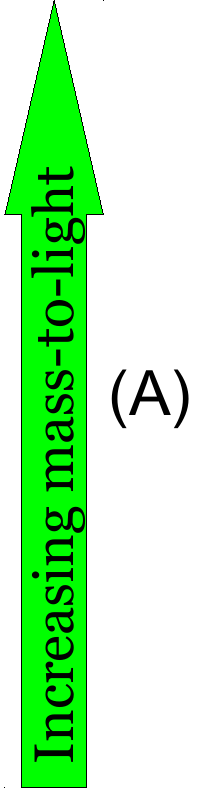}\\
\vspace{0.2cm}
\includegraphics[width=8.6cm,trim=0cm 2cm 0cm 2.15cm,clip]{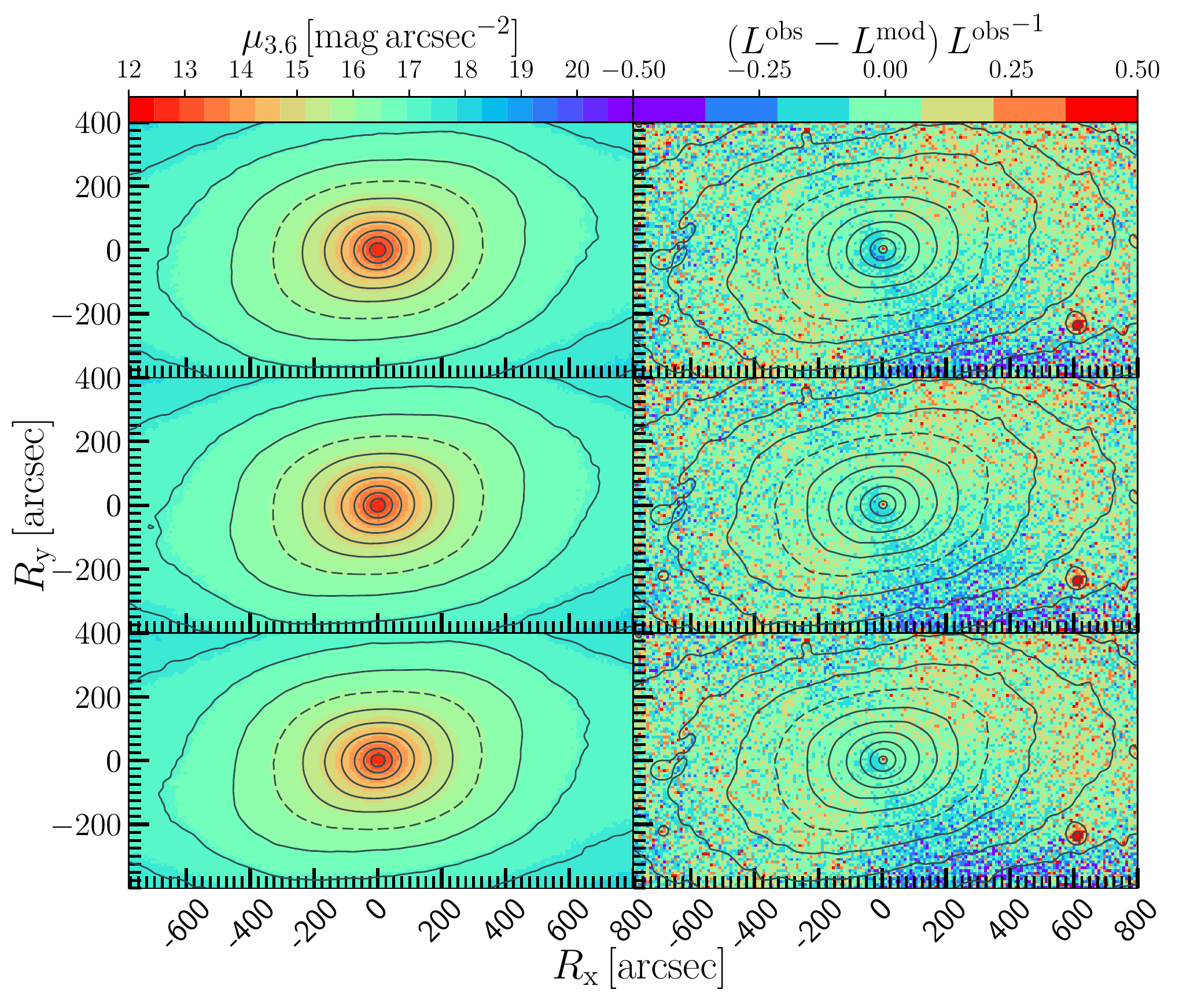}
\includegraphics[width=7.7cm,trim=2.05cm 2cm 0cm 2.15cm,clip]{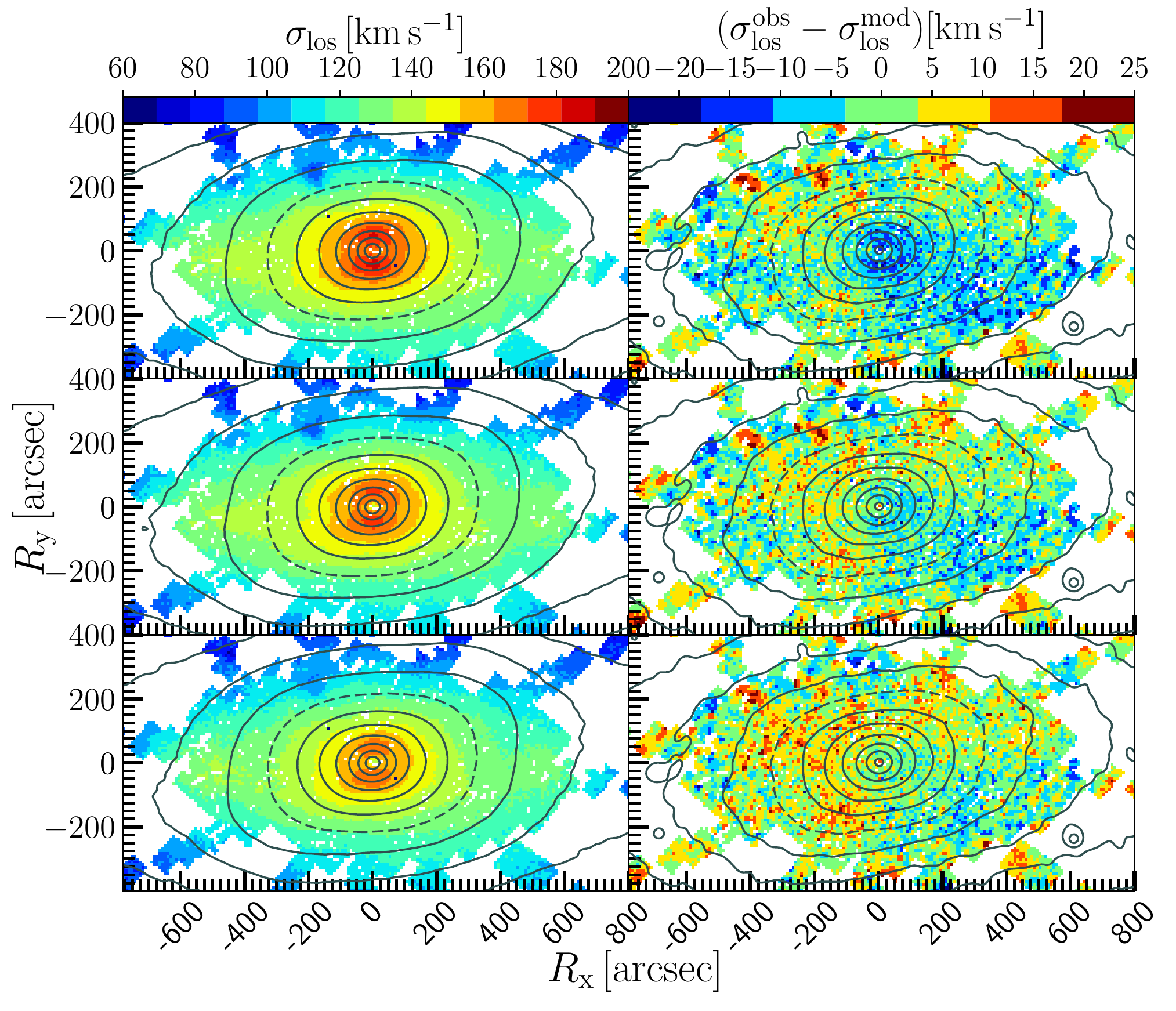}
\hspace{-0.3cm}
\includegraphics[height=5.6cm,trim=0cm 0cm 0cm 0cm,clip]{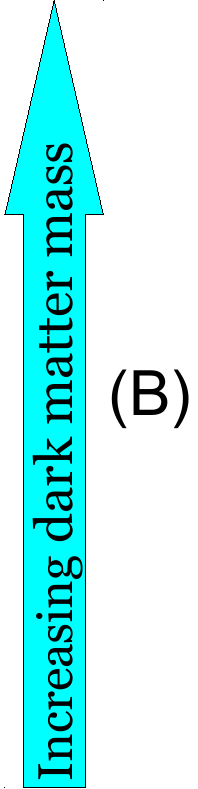}
\hspace{5cm}\\
\vspace{0.2cm}
\includegraphics[width=8.6cm,trim=0cm 0cm 0cm 2.15cm,clip]{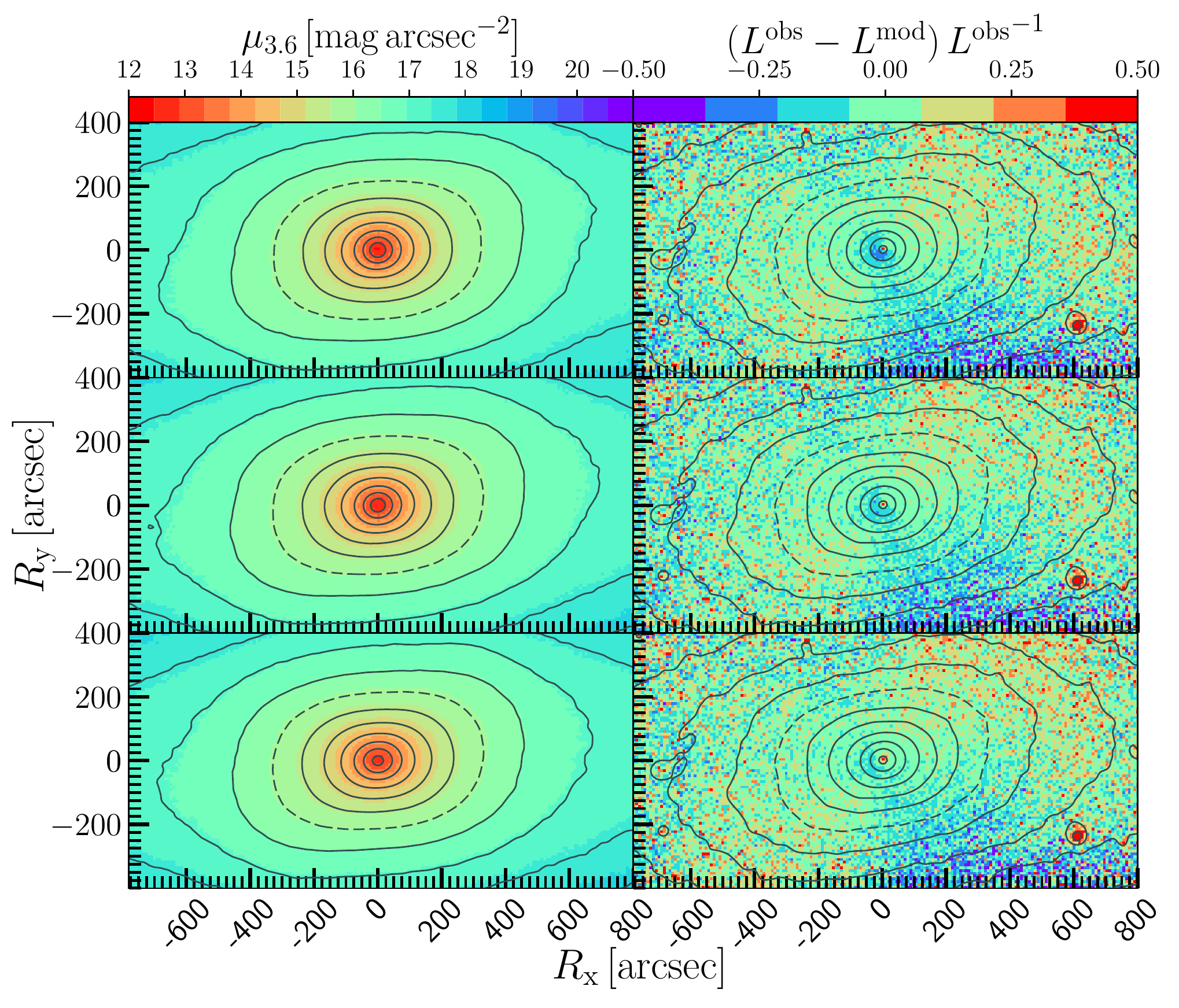}
\includegraphics[width=7.7cm,trim=2.05cm 0cm 0cm 2.15cm,clip]{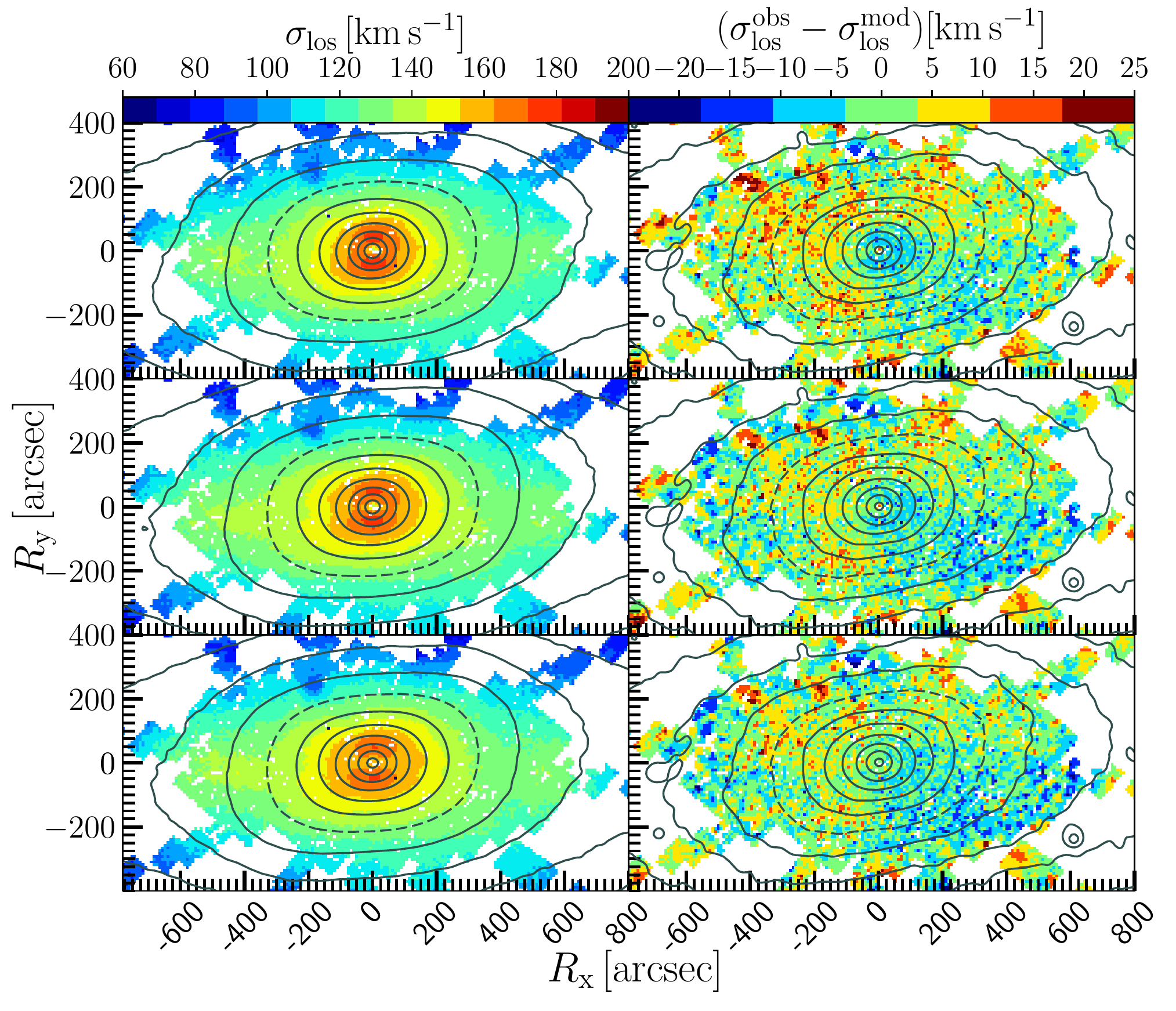}
\hspace{-0.3cm}
\includegraphics[height=6.5cm,trim=0cm -1.2cm 0cm 0cm,clip]{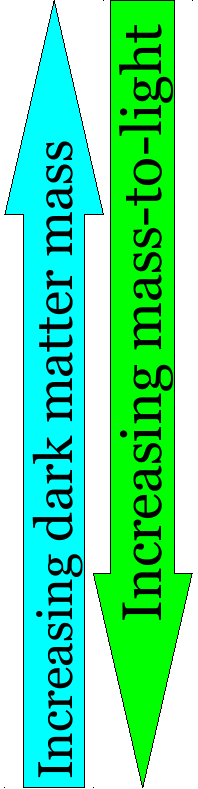}
\vspace{-0.0cm}
\caption[Photometry and dispersion maps with residuals]{Model maps and their residuals with the
  observations for the surface-brightness (1st, 2nd columns) and the dispersion and residual (3rd
  and 4th columns). Case A with variation only in \pml showing a model with $\pml\e0.80\ml$ (1st
  row), the best model with \bmml (2nd row) and a model with $\pml\e0.65\ml$ (3rd row). Case B with
  variation only in \pdm showing a model with $\pdm\e1.6\!\times\!10^{10}\sm$ (4th row), the best
  model with \bmdm (5th row) and a model with $\pdm\e0.8\!\times\!10^{10}\sm$ (5th row). Case C (last three rows) with
  variation of \pml and \pdm showing a model with $\pml\e0.75\ml$ and
  $\pdm\e0.8\!\times\!10^{10}\sm$ (6th row), the best model (7th row), and a model with
  $\pml\e0.68\ml$ and $\pdm\e1.6\!\times\!10^{10}\sm$ (8th row). We show the isophotes of the models
  (1st and 3rd column) and M31 (2nd and 4th column) spaced with $\Delta \mu_{3.6}\e0.5\,[{\rm
    mag}\,{\rm \as^{-2}}]$, with the $\mu_{3.6}\e16\,[{\rm mag}\,{\rm \as^{-2}}]$ ($I_{3.6}\e3.4\!\times\!10^{3}\,\slu\,\pc^{-2}$) isophote shown
  with a dashed contour.}
\label{fig:LDmaps}
\end{center}
\end{figure*}

Figure \ref{fig:sspart1EIN} shows the separate goodness-of-fit values for the \CB photometry and
dispersion, \setA, \setB, and for the \BPB photometry and dispersion, \setC and \setD, as function
of the stellar mass-to-light ratio and bulge dark matter mass in the Einasto models.  For each \pml
and \pdm, we show the lowest $\Delta{\hat \chi}^2$ value along the $\pps$ axis.  Equivalent results
for the NFW model grid are shown in the appendix (Figure \ref{fig:sspart1NFW}).

The \CB region (CBR): we see from the top panels of Figure \ref{fig:sspart1EIN} that the parameter
\pml is strongly constrained by the dynamical properties of the \CB in M31, where \setA and \setB
have very confined regions of low chi-square in \pml.  This is expected because in the very centre
of the bulge the dynamics is governed mainly by potential of the stellar mass, 
{\bfun which is set by \pml}, while the dark matter matters more
in the outer part of the bulge, in the \BPB region. The models that best match the photometry in
the centre of the CBR (lowest \setA) are in the range $\pml\e0.70 - 0.74\ml$, while the models that
best match the central velocity dispersion in the CBR (lowest \setB) are within $\pml\e0.70 -
0.75\ml$.  \setA and \setB constrain the dark matter mass to be within $\pdm\!\leq\!1.4\times
10^{10}\sm$, while the pattern speed has only a small effect in the CBR, which translates into
having low values of \setA, \setB for a wide range of values of \pps.

The \BPB region (BPR): {\bfun the photometry in this region is less constraining with low values of \setC over a wider range of \pml
and \pdm. This is because the stellar and dark matter can be exchanged to some degree and the M2M fitting can 
adjust rather well the stellar luminosity density within some range of values. Therefore, the acceptable models for \setC are 
limited to $\pml\!\lesssim\!0.74\ml$ and $\pdm\!\gtrsim\!0.8\times10^{10}\sm$.}  The \BPB velocity dispersion parameter \setD 
has a constrained region of low chi-square values in the range $\pml\e0.70 - 0.78\ml$ and $\pdm\!\leq\!1.4\times10^{10}\sm$, so both
\BPB data sets together constrain \pdm. We show later in Figure \ref{fig:sspart2EIN} that the
pattern speed is also constrained by \setD.  We note that, while the lowest chi-square values for
each subset have slightly different locations in the space of \pml and \pdm, the region of
acceptable models overlap defining the range of best models, like our didactic Figure \ref{fig:diagramX2} illustrates.

The most important result shown by Figure \ref{fig:sspart1EIN} is that the degeneracy between the
stellar mass-to-light ratio and the dark matter is broken by combining the different data subsets,
particularly the \CB photometry (\setA) and dispersion (\setB) which are sensitive to \pml and
imply a tight range of values, which then narrow the bounds for the dark matter \pdm, strengthening
the combined results from the \BPB data (\setD and \setC). 

Figure \ref{fig:LDmaps} illustrates how \pml and \pdm influence the velocity dispersion maps, and
how the degeneracy between them is limited by the different data subsets. At lowest order, the mass
in stars and dark matter can compensate. However, for given luminosity distribution and pattern
speed, the gradient of the dispersion is changed with the steepness of the gravitational potential
that depends on the stellar mass in the central bulge region and the dynamical mass in the outskirts
of the bulge. Thus, for example, models that have too much dark matter mass within the bulge and low
mass-to-light ratios result in a too flat dispersion profile.
 
Figure \ref{fig:LDmaps} shows photometric and kinematic maps of the best model (\bein) and of models
with modified values of \pml and \pdm. Residual maps are also shown that illustrate how these
physical parameters are connected with goodness-of-fit parameters \setA, \setB, \setC and \setD.  We
consider three main cases: (A) variations of only the mass-to-light ratio ($\Delta\pml$), (B)
variations of only the dark matter mass in the bulge ($\Delta\pdm$), and (C) varying both
simultaneously ($\Delta\pml$, $\Delta\pdm$) showing how the degeneracy between these parameters is
constrained:

(A) The top panels in Figure \ref{fig:LDmaps} show the best model compared to two models with the
same dark matter mass and pattern speed, but with different mass-to-light ratios.  The model with a
larger $\pml$ has a slightly worse fit to the photometry in the \BPB region (BPR) (larger \setC),
and a worse fit to the inner dispersion, which is higher in the model than in the data (larger
\setD).  The high \pml results in too much mass in the centre of the bulge, hence a too deep central
potential, which has the consequence of a velocity dispersion that is higher than the observations.
For the model with lower $\pml$ (3rd row) the effects are the opposite.  The most important result
here is that the mass-to-light ratio has the strongest effect in the central region where the \CB
is, showing the important signature of the chi-square variables \setA and \setB.

(B) If we change only the dark matter mass within the bulge, we obtain similar effects on the
velocity dispersion but on larger scales.  The middle panels of Figure \ref{fig:LDmaps} show the
best model and two models that have the same \pml and \pps, but different \pdm.  These two models
overpredict (underpredict) the observed dispersion map outside the central bulge for too high (low)
\pdm.  In the \BPB region the mass of the dark matter is comparable to the stellar bulge mass (typically 25
per cent of the stellar mass depending on the model), contributing significantly to the total
dynamical mass, which is connected to the dispersion and is constrained by the data through the \setC
and \setD variables.  Because the stellar mass is determined by $\pml$ which is fixed by the
central regions of the bulge, \setC and \setD thus constrain the dark matter mass \pdm.
 
(C) Finally, considering the case of \pml-\pdm jointly: what happens if we decrease (increase) the
mass-to-light ratio, but also increase (decrease) the dark matter mass content? Using our selection
criteria in Section \ref{sec:mod:tech:sel} we found a range of acceptable models around the best
model parameters \bml and \bdm, in the elongated region of Figure \ref{fig:Xsum} (left panel).  The
stellar \pml is constrained mostly by the data in the CBR, while the influence of the \pdm is
strongest in the \BPB. Here we show two models just outside the range of acceptable models
along this elongation. Therefore the differences between these models and the data are subtle, but
they are still visible directly in the maps.

\subsubsection{\pml for the two bulge components}
\label{sec:res:param:ML}
\begin{table}
\caption[Cases for different $\pml$ values for the bulge components and the outer disc.]{Cases for different $\pml$ values for the bulge components and the outer disc.}
\vspace{-0.3cm}
\label{tab:testsml}
\begin{tabular}{lcccccc}
\hline
$\pml\,\ml$			&	i) 	&	 ii)   & 	iii) 	& 	iv) & v) & vi) \\\hline
$\Upsilon^{\rm CB}$		&	0.72	&	0.72	&	0.72	&	0.72		& 0.72	& 0.72 \\
$\Upsilon^{\rm BPB}$	&	0.70	&	0.68	&	0.72	&	0.72	 	& 0.72 & 0.72	\\
$\Upsilon^{\rm d}$		&	0.70	&	0.68	&	0.55	&	0.65	 	&	0.80  & 0.85\\
\hline
\end{tabular}
\end{table}

\begin{figure}
  \begin{center}
  \includegraphics[width=8.8cm]{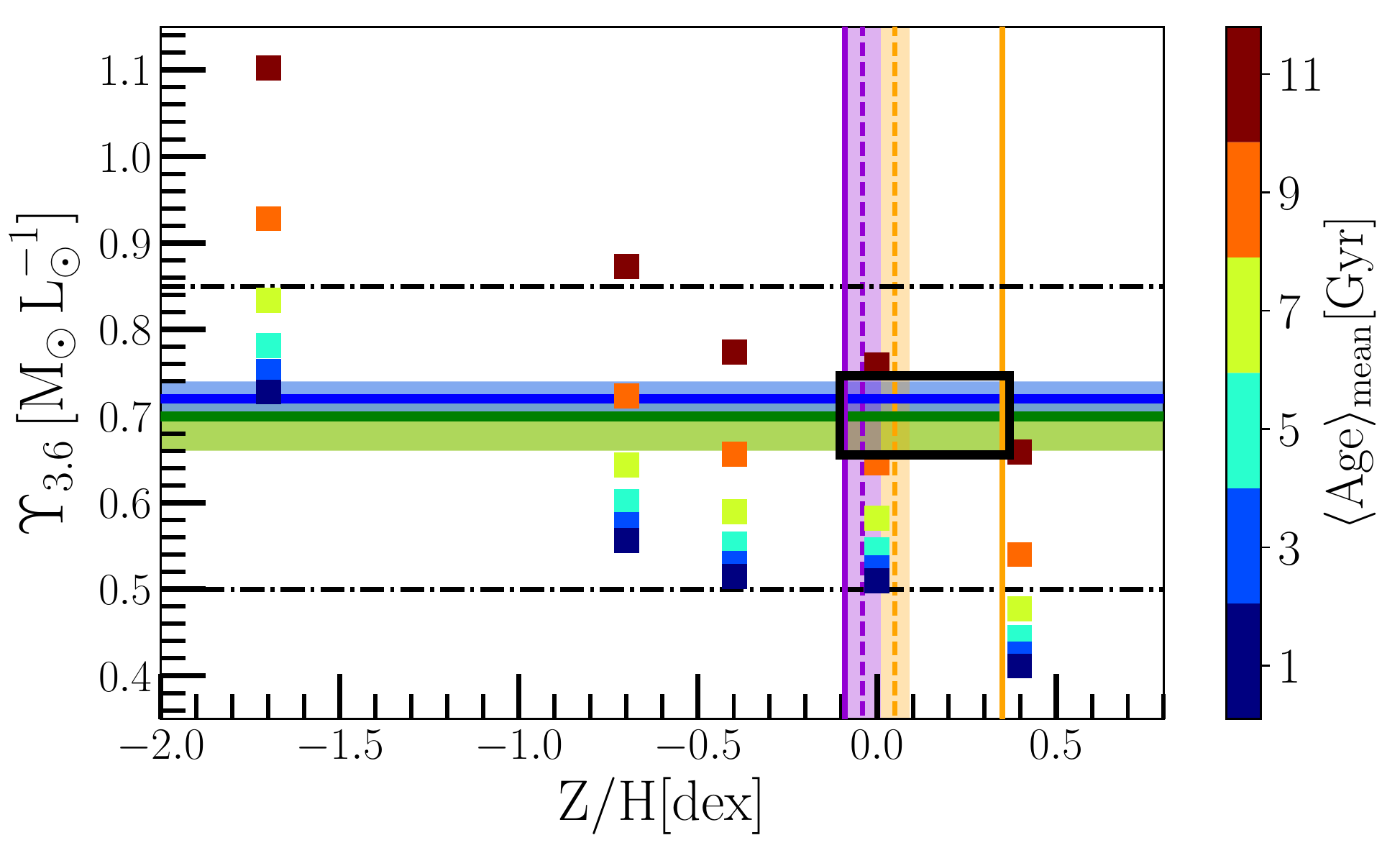}
    \vspace{-0.6cm}
    \caption[Mass-to-light versus metallicity and age]{
    Stellar mass-to-light ratio in the 3.6\mum band as function of the metallicity. 
    The best values are \bml (blue line and shaded region) for the Einasto grid of models, and \bnfwml for the NFW grid (green line and shaded region),
    and the range of explored values of \pml is shown with the dotted dashed horizontal lines.
    From \citet{Saglia2018} we show the \zh of their model estimations (solid vertical lines), the average (vertical dashed line) and the root mean square (vertical shaded region) of the \BPB (purple) and \CB (orange). 
    \citet{Meidt2014} estimate from stellar population evolution analysis predictions, relations for \pml, the metallicity and the mean stellar age (squares).
    The black rectangle indicates the region of \pml values that we expect to intersect with the values from the stellar populations analysis, given the 
    metallicities estimated within M31's bulge.}
    \label{fig:MLZH}
  \end{center}
\end{figure}

We find for the Einasto grid of models that the best range of values for the stellar mass-to-light ratio in the 3.6\mum band is \bml.
Given that the bulge of M31 has two components: a \CB that likely formed very early from a hierarchical process, 
and a \BPB formed by the redistribution of a disc component, we might expect different values of \pml for each component.
However, we now show that due to their measured metallicities and ages, their expected mass-to-light ratios in the 3.6\mum band are rather similar 
and that the best value represents well both bulge components.

In Figure \ref{fig:MLZH} we show the stellar mass-to-light in the 3.6\mum band as function of
metallicity and age computed by \citet{Meidt2014}\footnote{values taken directly from their Figure 2} using a stellar population
analysis. These values assume a Chabrier initial mass function (IMF). 
Analysis of IMF sensitive absorption features in high signal-to-noise spectra by \citet{Zieleniewski2015}
indicate that the IMF is consistent with Chabrier across the M31 bulge. 
We also over-plot in Figure \ref{fig:MLZH} the ranges of metallicities within the bulge components of M31 measured by \citet{Saglia2018} \citep[see also][]{Opitsch2016} who {\bfun find a \CB with a metal rich and very old centre with an average of $\langle\zh\rangle^{\CB}\e0.06\pm0.05\dex$ (and as high as $\zh^{\CB}_{\rm model}\e0.35\dex$) and ($12.9\pm0.3\Gyr$); and a comparably old \BPB ($12.8\pm0.3\Gyr$) with a slightly sub-solar averaged metallicity of $\langle\zh\rangle^{\BPB}\e-0.04\pm0.01\dex$.}
Our range of best values for \pml are in agreement with what is expected for stellar populations with these metallicities and average ages {\bfun for a Chabrier IMF}.

Note from Figure \ref{fig:MLZH} that, in the 3.6\mum band, an old and slightly more metal-rich population could have a mass-to-light similar to that of a slightly younger and less metal-rich population, {\bfun which is relevant given the negative metallicity gradient measured by \citet{Saglia2018} of $\nabla{\zh}^{\CB}\e-0.5\pm0.1\dex\kpc^{-1}$.
This is not uncommon,} as other classical bulges and elliptical galaxies show metallicity gradients with the most metal rich part in their centres \citep{Koleva2011}.
The \BPB is indeed slightly younger and less metal rich.
Consequently, our assumption of a common value of $\pml$ for both bulge components is not unexpected and is sufficient to 
reproduce the most important dynamical properties of the M31 bulge,
while the narrow range of valid values suggests that any difference in mass-to-light between the two bulge components must 
be small. \citet{Saglia2018} also compute from stellar population analysis the expected V-band $\Upsilon$ for both bulge components, 
finding differences in mass-to-light by less than 10 per cent, reinforcing that our common mass-to-light is not unexpected.

However, in the outer disc region, beyond the bar, younger stars can decrease the mass-to-light ratio. 
Colour gradients also suggest a metallicity gradient between the more metal rich bulge and the outer disc \citep{Courteau2011}.
To test these assumptions we also performed M2M fits with different \pml values for the bulge components 
($\Upsilon^{\CB},\,\Upsilon^{\BPB}$) and the disc ($\Upsilon^{\rm d}$), considering six cases shown in Table \ref{tab:testsml}. 
We only find small changes in the dynamical properties of the model within the bulge region.
As we show in the next section, even in the outer part of the disc ($R>10\kpc$) for lower \pml in the outer disc we require small 
variations of \si10 per cent of dark matter mass at that radius in order to match the \HI rotation curve.
{\bfun These $\Upsilon^{\rm d}$ variations also encompass the changes which would be caused by the mass of the gas in the disc, 
which would increase the mass in the outer disc by less than 10 per cent.}

\subsubsection{Stellar and dark matter mass distribution}
\label{sec:res:param:mass}
\begin{figure*}
\begin{center}
\includegraphics[width=8.81cm]{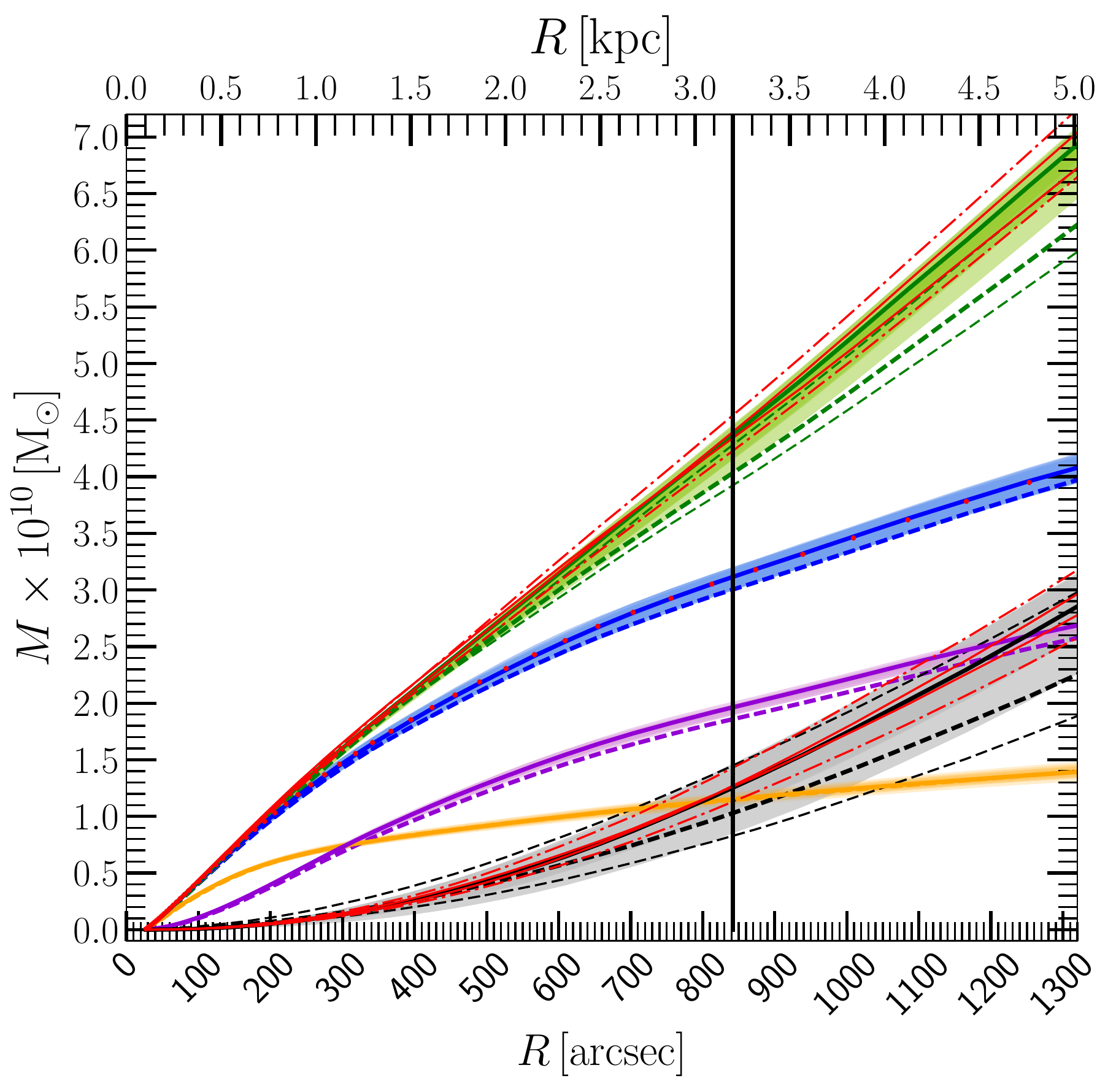}
\includegraphics[width=8.81cm]{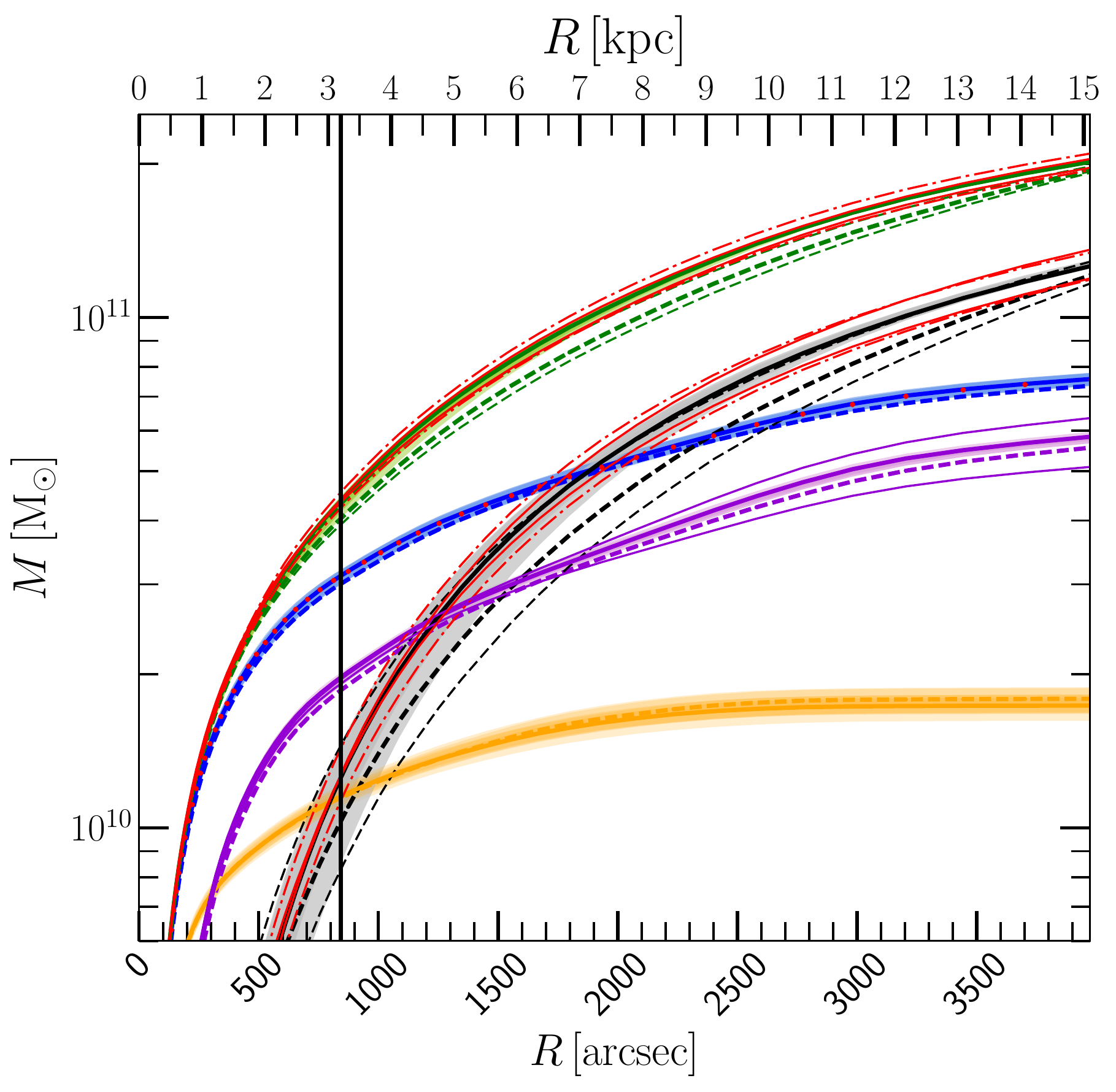}
\vspace{-0.7cm}
\caption[Mass profiles]{Cumulative mass profiles within 5\kpc and (left panel) 15\kpc (right panel) for the best model (JR804) of the 
grid with the Einasto dark matter profile (solid curves) and the best model (KR241) of the NFW grid (thick dashed 
curves), for the different components: \CB (orange), \BPB and disc (purple), total stellar (blue), dark 
matter (black) and dynamical mass (green). The range of acceptable models of the Einasto grid \bmein is 
shown in shaded regions. The most extreme values of \pdm profiles from the range of the models \bmnfw are shown with the 
thin dashed curves. The end of the de-projected \BPB is at 3.2\kpc (vertical black solid). 
We also show the \BPB and disc cumulative mass profiles of the tests with 
$\Upsilon_{\rm d}\e0.55\ml$ and $0.85\ml$ (lower and upper purple thin solid lines in right panel) (see cases iii and vi in Table \ref{tab:testsml}) 
and the respective dark matter and dynamical mass profiles (upper and lower red solid curves). 
The tests for different flattening show masses within the bulge that lay within the range of models for $q\e1.0$ and 0.7 (upper and lower 
red dot dash lines), and the stellar component in red dots. The profiles are function of the cylindrical radius 
$R$ summing the mass within a ellipsoidal volume with our fiducial flattening of $q\e0.85$.}
\label{fig:massprof}
\end{center}
\end{figure*}

\begin{figure}
\begin{center}
\includegraphics[width=8.5cm]{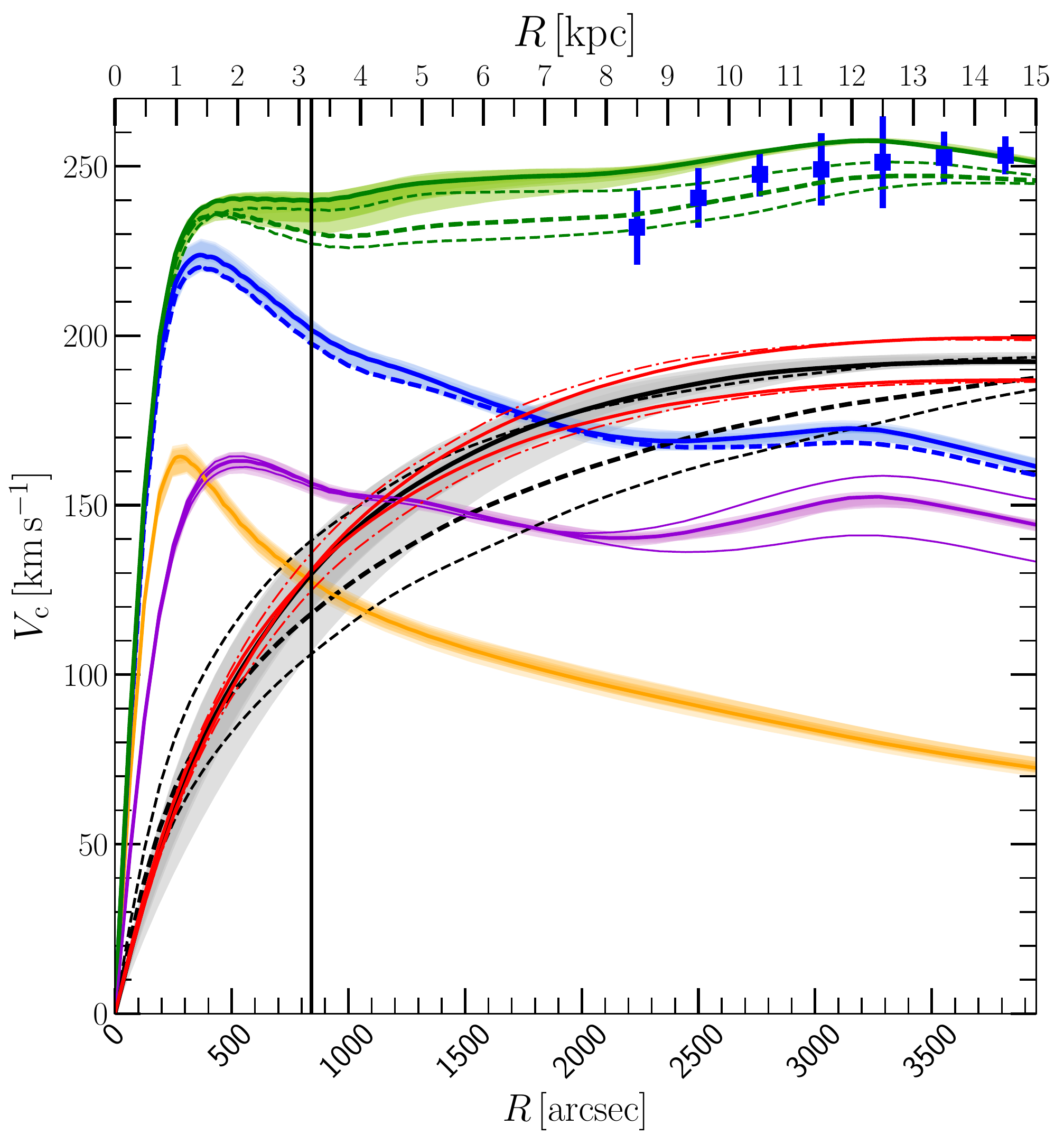}
\vspace{-0.5cm}
\caption[Circular velocity profiles]{
Azimuthally averaged circular velocity in the plane of the disc of the model JR804 (solid curves) and model KR241 (dashed curves) 
for the CB (orange), the \BPB and the disc (purple), total stellar mass (blue curve), the dark matter (black curve), and the total 
circular velocity (green curve). The \HI data of \citet{Corbelli2010} is shown out 15\kpc (blue squares). 
The shaded regions correspond to the models with the Einasto profile \bmein.
The dashed thin curves indicate the profiles of the models with the maximum and the minimum \pdm of the models \bmnfw.
We show the circular velocity of the \BPB and disc components of the tests with 
$\Upsilon_{\rm d}\e0.55\ml$ and $0.85\ml$ (lower and upper thin purple solid curves) and corresponding dark matter halo circular velocities 
(upper and lower solid red curves), corresponding to cases iii and vi of Table \ref{tab:testsml}. 
The test of the different flattening for $q\e1.0$ (lower red dot dash line) and 0.7 (upper red dot dash line).
The vertical black line marks the end of the \BPB.}
\label{fig:profvc}
\end{center}
\end{figure}

In the previous section we found the range of 3.6\mum mass-to-light ratios and dark matter masses within the bulge that best reproduce the observations, thereby
obtaining the range of stellar masses for each bulge component.
Table \ref{tab:MdelEIN} contains the resulting masses within 3.2\kpc for the range of acceptable models with the 
Einasto dark matter haloes \bmein, with the best values being: 
$M_{\star}^{\rm CB}\e1.18^{+0.06}_{-0.07}\times10^{10}\sm$ for the classical bulge,
$M_{\star}^{\rm BPB}\e1.91\pm0.06\times10^{10}\sm$ for the \BPB, making a total bulge stellar mass of
$M_{\star}^{\rm B}\e3.09^{+0.10}_{-0.12}\times10^{10}\sm$. For the bulge dark matter mass we find \bdm 
finding then a total dynamical mass within the bulge of $M_{\rm dyn}^{\rm B}\e4.25^{+0.10}_{-0.29}\times10^{10}\sm$.
Integrating the mass of the \CB out to 10\kpc we obtain $M_{\star}^{\rm CB,10\kpc}\e1.71^{+0.10}_{-0.09}\times10^{10}\sm$.
Other bulge mass estimations in the literature neglect the composite nature of M31's bulge, and therefore they recover
similar values to our bulge total stellar mass \citep[$M_{\star}^{\rm B}\e4\times10^{10}\sm$;][]{Kent1989},
\citep[$M_{\star}^{\rm B}\e2.5\times10^{10}\sm$;][]{Widrow2003}. 
Our \CB mass estimation is the lowest value in the literature for M31, which can be used to constrain the early formation history of M31.

The models with NFW haloes result in a similar range of values (Table \ref{tab:MdelNFW}), with $M_{\star}^{\rm CB}\e1.16^{+0.04}_{-0.10}\times10^{10}\sm$ and  
$M_{\star}^{\rm BPB}\e1.82\pm0.08\times10^{10}\sm$, and a total stellar mass of $M_{\star}^{\rm B}\e2.98^{+0.10}_{-0.18}\times10^{10}\sm$. 
The dark matter is \bnfwdm with the total mass within the bulge being $M_{\rm dyn}^{\rm B}\e3.95^{+0.22}_{-0.09}\times10^{10}\sm$. 

\begin{figure}
\begin{center}
\includegraphics[width=8.8cm]{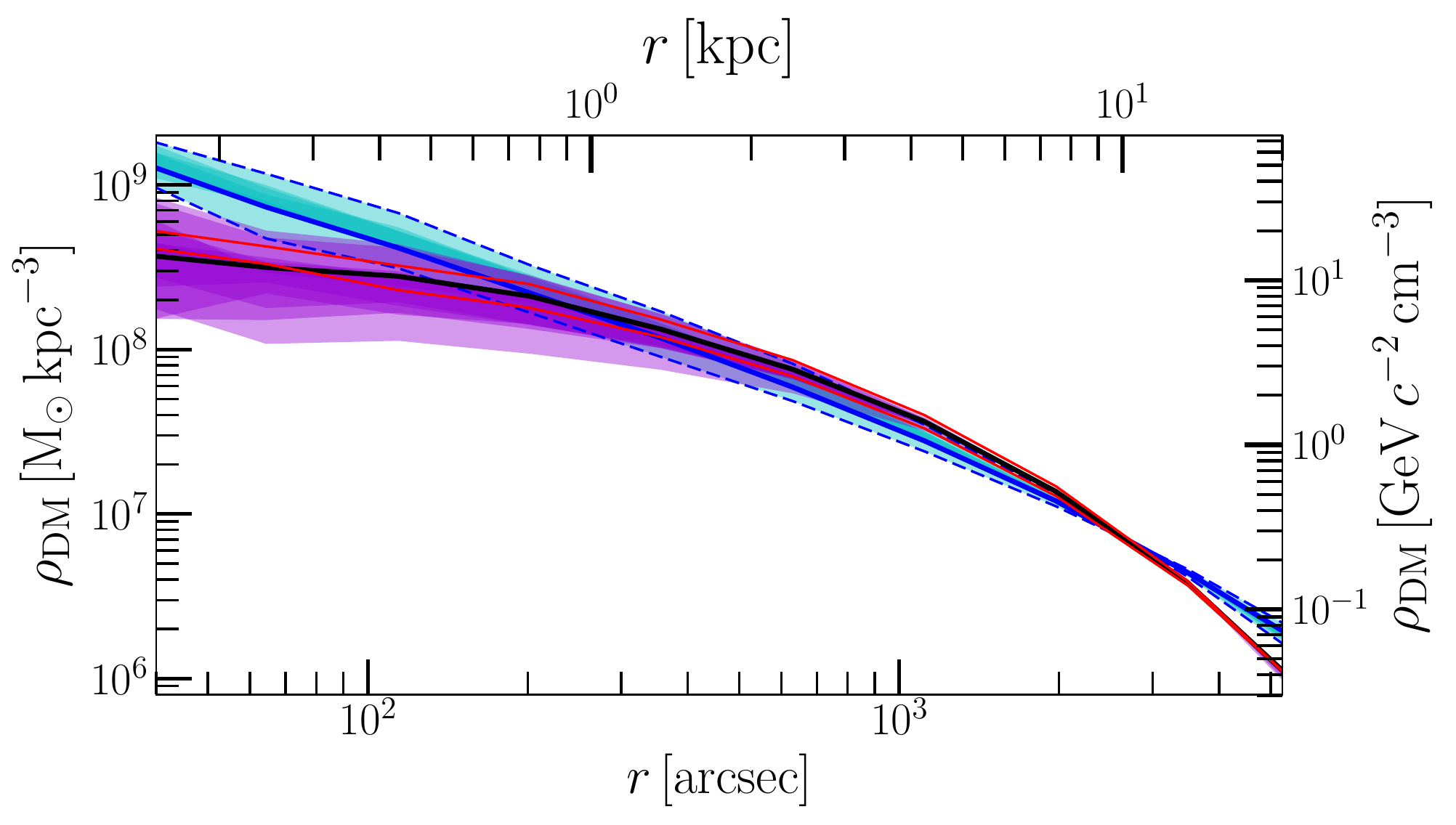}
\vspace{-0.5cm}
\caption[Dark matter density profiles]{Dark matter density profiles of the 
best matching Einasto model (black curve) and the best NFW model (blue curve).
The shaded regions correspond to the models with the Einasto profile \bmein (violet shade) and the NFW profile \bmnfw (cyan shade).
The dotted blue dashed curve indicate the profile of the models with the maximum and the minimum \pdm of the models \bmnfw.}
\label{fig:profileden}
\end{center}
\end{figure}

In Figure \ref{fig:massprof} we present the cumulative mass profiles of the best models and the acceptable range models of the Einasto grid (\bmein) and the NFW grid (\bmnfw).
The resulting range of models have very similar stellar mass profiles, and most of the total mass variation is due to the dark matter.
The \CB dominates the centre reaching the same mass of the \BPB at 1.2\kpc (300\as).
Further out the \BPB dominates the stellar mass, and is almost double the mass of the \CB at the end of the \BPB.
Interestingly, the profiles show that the dark matter masses reach a similar value to the \CB at end of the \BPB
at 3.2\kpc (850\as). 
The best values of the Einasto grid of models are similar within the errors to the best NFW models, with the best matching NFW models 
requiring slightly lower masses within 3.2\kpc. This is explained by the more cuspy density profile of the NFW profile:
for the same mass at the end of the bulge (3.2\kpc) the NFW models have more dark matter distributed in the very centre than the Einasto models, 
as is shown by the density profiles in Figure \ref{fig:profileden}.

We show in Figure \ref{fig:profvc} the circular velocity profiles of the models \bein and \bnfw within
15\kpc $i.e.$ the radius where we fit the photometry. 
While the total dark matter within the bulge is fixed to a value \pdm during each M2M fit, where we select the values that 
best reproduce the photometry and the stellar kinematic observations, the dark matter in the disc region 
is determined during each run by fitting the \HI rotation curve. 
We find that for the Einasto profile the range of dark matter masses and the resulting circular velocity values are more constrained
than the range of values of the NFW profile.

We include in the mass profile and in the circular velocity figures variations of model JR804 with a flattening $q\e0.7$ and 1.0, having dark matter mass and circular velocity values within the range of the acceptable models.
As expected the dark matter mass profile deviates for different flattening values; however, the stellar mass profile remains within the range of the acceptable models.
We also include in these figures tests with different \pml values for the disc from Table \ref{tab:testsml}, 
showing that even the extreme values $\Upsilon^{\rm d}\e0.55\ml$ and
$\Upsilon^{\rm d}\e0.85\ml$ remain within the range of the acceptable models. 
The variation of the circular velocity in the disc region at \si10\kpc is small because most of the stellar mass is contained within this radius and 
the dark matter dominates at this distances, making the local variation of the stellar mass at \si10\kpc only a small contribution to the total circular velocity.
{\bfun We note that the tests of $\Upsilon^{\rm d}$ generate variations of stellar mass and surface mass density in the disc region that are larger than the mass 
contribution of the gas at this radius. Therefore, we do not need to include the gas contribution in the modelling.}

In Figure \ref{fig:profileden} we present the particle dark matter density profiles of the best models of the Einasto and the NFW grids, and the 
range of acceptable models.
Fitting equation \ref{eq:haloein} to the density of the best Einasto model we recover the parameters $\rho_{\rm E}\e1.29^{+0.12}_{-0.28}\times10^{7}\,\sm\,\kpc^{-3}$, 
$m_{\rm E}\e7.8^{+1.1}_{-0.5}\kpc$ and $\alpha\e0.51^{+0.22}_{-0.12}$ (or $n_{\rm Ein}\e\alpha^{-1}\e1.96\pm0.6$), with errors from the range of best models.
Similarly, a fit from equation \ref{eq:halonfw} to the best NFW model, we recover the values $\rho_{\rm N}\e1.54^{+1.9}_{-0.7}\times10^{7}\,\sm\,\kpc^{-3}$, 
and $m_{\rm N}\e10.4^{+4.0}_{-3.4}\kpc$.

We find a dark matter mass of \bdm within 3.2\kpc for the Einasto grid of models and \bnfwdm for the NFW models,
where the bulge stellar kinematics favours the cored Einasto profile. 
We find that the central dark matter masses are in agreement with cosmologically motivated haloes.
Haloes with the virial mass M31 of $M_{{\rm DM}\,200}\e1.04\times10^{12}\sm$ \citep{Tamm2012} in cosmological simulations
are expected to have an average concentration of $c_{200}\e8.8$ and virial radius of $R_{200}\e277\kpc$ 
\citep[][with Planck cosmology; \citealt{PlanckCollaboration2013}]{Correa2015a,Correa2015c}.
For such halo, the expected mass within 3.2\kpc for a pure NFW halo is $M_{{\rm DM}\,200}^{3.2\kpc}\e0.34\times10^{10}\sm$, lower than our measurement. 
However, the baryonic mass accretion can cause an adiabatic contraction of the halo that increases the central dark matter 
mass up to $M_{{\rm DM}\,200}^{3.2\kpc}\e1.88\times10^{10}\sm$ in the most extreme case \citep{Blumenthal1986}, 
or a lower value of $M_{{\rm DM}\,200}^{3.2\kpc}\e0.97\times10^{10}\sm$, as more recent hydrodynamical cosmological simulations 
show less contraction \citep[][implemented with $\nu\e0.4$ prescription from \citealt{Dutton2011}]{Abadi2010}. 
Our results then agree with a moderate adiabatic contraction in the centre 
of the halo, but also favour a cored nature of the halo's central distribution. 

\subsubsection{The box/peanut bulge and thin bar pattern speed (\pps).}
\label{sec:res:param:ps}

\begin{figure}
\begin{center}
\includegraphics[width=8.8cm]{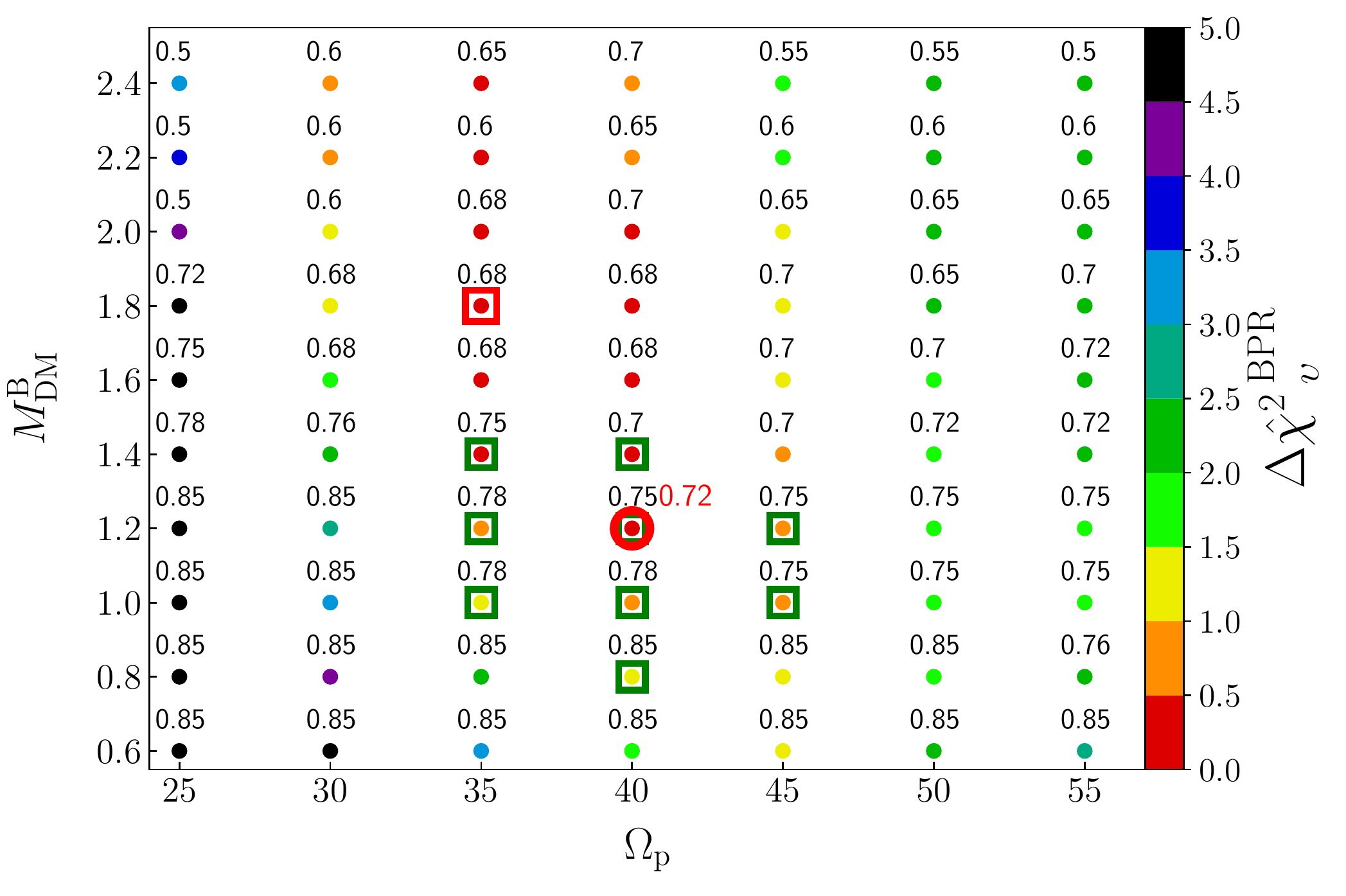}\\
\vspace{-0.1cm}
\includegraphics[width=8.8cm]{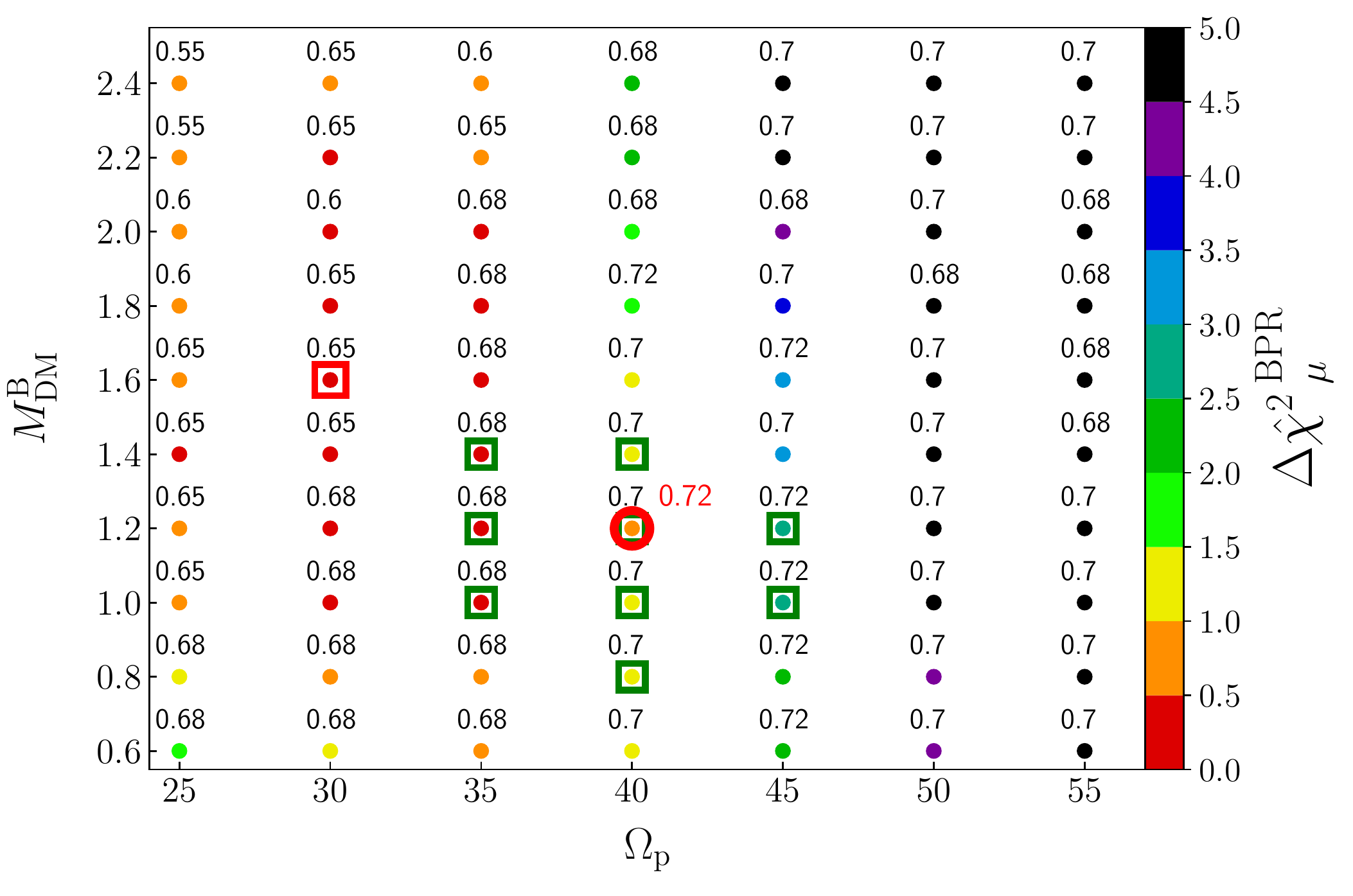}\\
\vspace{-0.1cm}
\includegraphics[width=8.7cm]{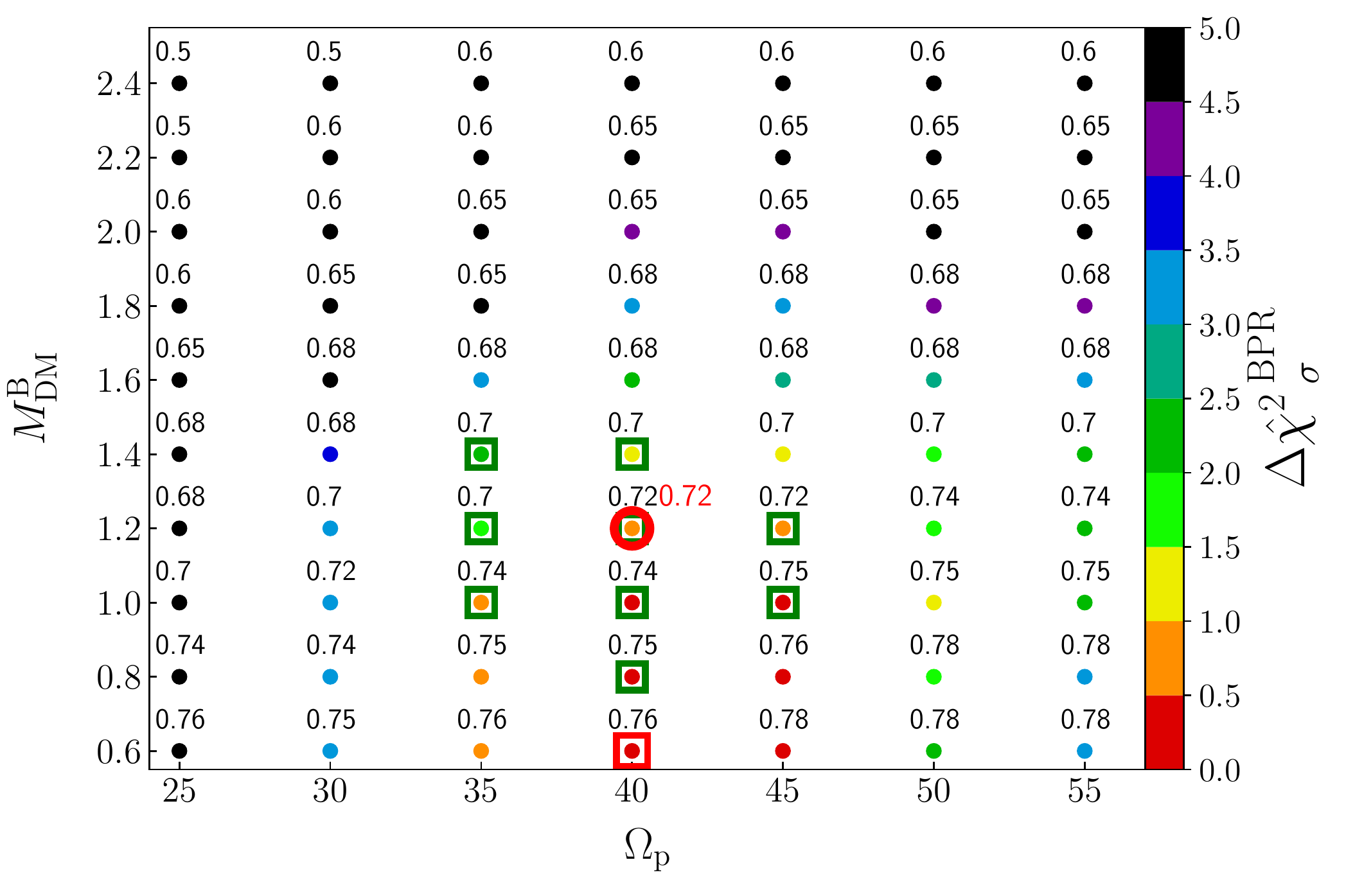}
\vspace{-0.7cm}
\caption[Subsets part 2]{Results of the grid of models for the Einasto dark matter halo:
\setE (top), \setC (middle) and \setD (bottom) as function of the parameters 
$\pps$ and $\pdm$, selecting the lowest value along the axis of the parameter $\pml$.
The values of each subset are the points that are coded in the coloured bar, and the number corresponds to the selected $\pml$. 
We mark the best model JR804 (red circle), the models with the minimum values in each subset (red squares), 
and the range of acceptable models \bmein (green squares). 
The green squares do not necessarily agree with the shown \pml.}
\label{fig:sspart2EIN}
\end{center}
\end{figure}

The bar of M31 consists of a vertically thick structure that is the box/peanut bulge (\BPB) component, and the thin bar component that is 
mostly concentrated in the disc's plane, where both structures are aligned and rotate at the same pattern speed. 
Most estimations of the M31 bar pattern speed are based on comparisons with gas kinematics, finding typically 
$\pps\!\!\!\!\approx\!\!50-60\psu$ \citep{Stark1994,Berman2001, Berman2002}.
\iftoggle{paper}{
\citet{Tremaine1984} derived a relation from the continuity equation to determine the pattern speed of a two dimensional bar in disc galaxies 
directly from the observations using the information of the line-of-sight velocity ($\upsilon_{\rm los}$) and the photometry ($L_{3.6}$). 
Here we have the unique possibility to use new IFU stellar kinematics of the M31 bulge from \citetalias{Opitsch2017} to determine the
bar pattern speed. However, the disc inclination is too high to robustly determine it directly from the data using the Tremaine-Weinberg method.
Therefore, we use this relation indirectly by comparing with models that have been fitted to the photometric and IFU observations, 
which have different pattern speed values. Then, we select the models with a good match of the velocity field in the bar region (\setE), 
and the surface luminosity density (\setC).
Furthermore, the velocity dispersion ($\sigma_{\rm los}$) can also change the velocity through the total kinetic energy
($\sigma_{\rm los}^2+\upsilon_{\rm los}^2$), and therefore it also constrains the bar pattern speed.
And so, combining these two variables with the variables \setA, \setB and \setD we
are able to find the range of best matching models that also reproduce the velocity field in M31's bulge.
From the explored range of $\pps\e20-55\psu$, we find $\pps\e40\pm5\psu$ for both grids of Einasto and NFW models (tables \ref{tab:MdelEIN} and \ref{tab:MdelNFW}).
}{%
\citet{Tremaine1984} derived a relation from the continuity equation to determine the pattern speed of a two dimensional bar in disc galaxies 
directly from the observations with the relation (re-written for our case):
\begin{align}
\label{eq:ps}
\pps&= \frac{\iint\! dR_{\rm x} dR_{\rm y}\,\,\, L\iz R_{\rm x},\,R_{\rm y}\de \,\upsilon_{\rm los}\iz R_{\rm x},\,R_{\rm y}\de}{\sin i \iint\! dR_{\rm x} dR_{\rm y}\,\,\,L\iz R_{\rm x},\,R_{\rm y}\de\,R_{\rm x}}
\end{align}
However, the disc inclination is too high to robustly determine it directly from the data.
Therefore, we proceed to use this relation indirectly by comparing with models with 
a good matching of the velocity field in the bar region (\setE), and the surface luminosity density (\setC).
Furthermore, the velocity dispersion ($\sigma_{\rm los}$) is connected to the velocity through the total kinetic energy
($\sigma_{\rm los}^2+\upsilon_{\rm los}^2$), and therefore it also constrains the bar pattern speed.
And so, combining these two variables with the variables \setA, \setB and \setD we
are able to finding the range of best matching models that also reproduce the velocity field in M31's bulge.
}

In Figure \ref{fig:sspart2EIN} we show the results for \setE, \setD, and \setC as function of \pps and \pdm for the Einasto 
grid of models, with the best model located at \bmdm and \bmps (NFW grid results in Figure \ref{fig:sspart2NFW}).
The variable \setE has low values in the range of $\pps\e30 - 45\psu$ and for $\pdm\geq1.0\times10^{10}\sm$.
\setC has low values within $\pps\e25 - 40\psu$ and within $\pml\e0.55 - 0.75\ml$.
The variable \setD has low values within $\pps\e35 - 50\psu$ and $\pdm\leq 1.2\times10^{10}\sm$.
Taking into account the restrictions given by the variables \setA, \setB and \setD that 
constrain the best values for the mass-to-light ratio and the dark matter mass to be \bmml and \bmdm,
we find that the best value for the bar pattern speed is \bmps.

\begin{figure}
\begin{center}
\includegraphics[width=7.7cm]{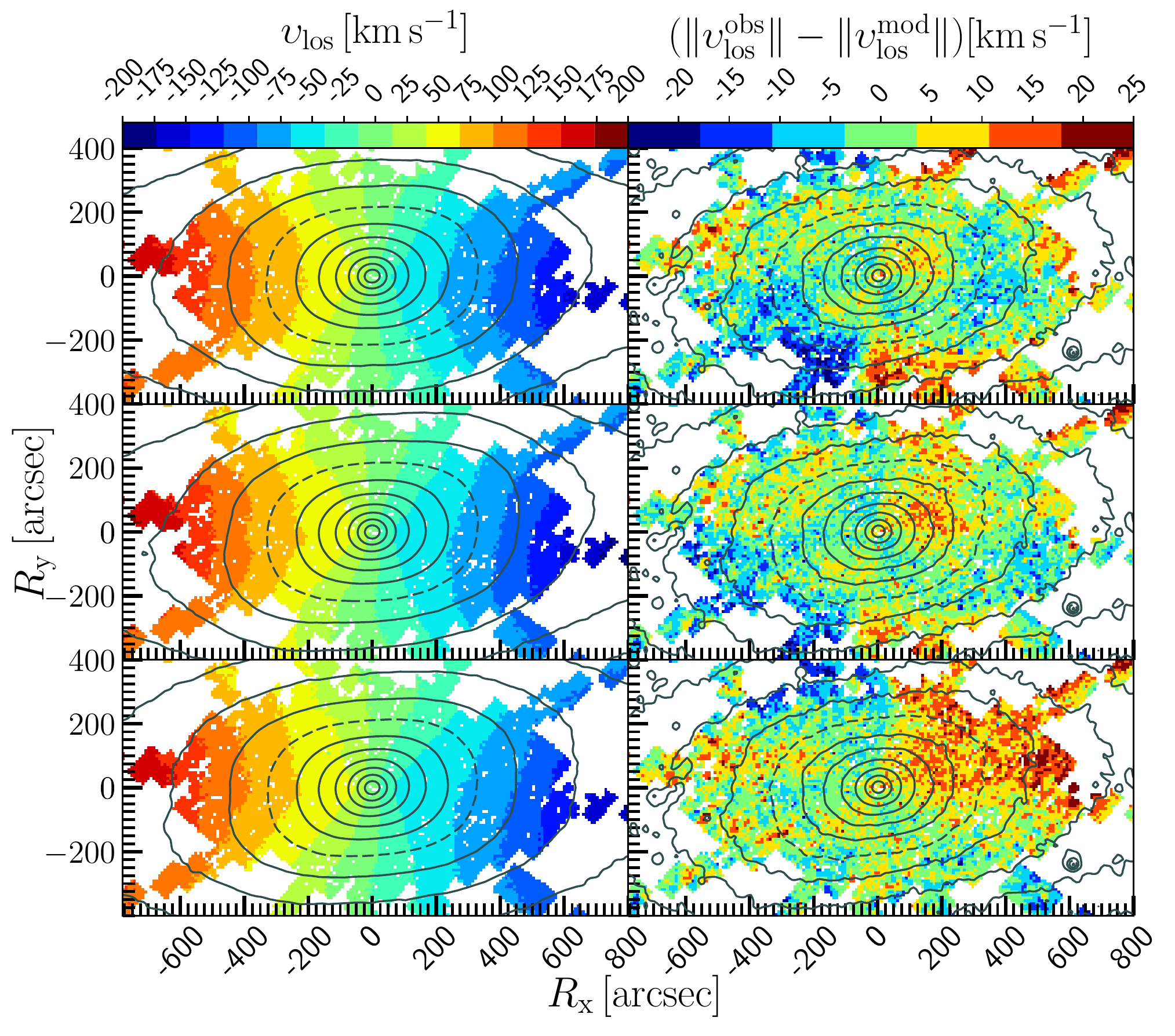}\hspace{-0.1cm}\includegraphics[height=6.2cm,trim=0cm -1.2cm 0cm 0cm,clip]{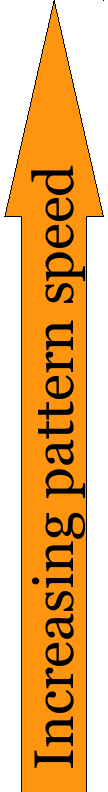}\\
\vspace{-0.5cm}
\caption[Velocity maps and residuals]{Model velocity maps (left column) and velocity residual with the observations (right column) 
for models with $\pml\e0.72\ml$ and $\pdm\e1.2\times10^{10}\sm$ with different 
pattern speeds, with 55 (top), 40 (middle) and 25\psu (bottom).
We show the isophotes of the models (first column) and M31 (second column) spaced with $\Delta \mu_{3.6}\e0.5\,[{\rm mag}\,{\rm \as^{-2}}]$ 
and the value $\mu_{3.6}\e16\,[{\rm mag}\,{\rm \as^{-2}}]$ is shown with a dashed isophote.}
\label{fig:PSmaps}
\end{center}
\end{figure}

In order to show the effects of changing the bar pattern speed we present in Figure \ref{fig:PSmaps} the isophotes, the velocity maps and velocity 
residual maps of the best model (\bein) and compare them with maps of two models with the same \pml and \pdm, but with 
$\pps\e25\psu$ and $\pps\e55\psu$. The best model shows smaller residuals than the other two models.
The isophotes slightly change in the outer parts of the \BPB in response to the change of \pps, where 
the model with $\pps\e25\psu$ shows slightly more boxy isophotes than the model with $\pps\e55\psu$.\\

\textit{Could the M31 bulge be a triaxial elliptical galaxy?} 
Classical bulges are often considered to be akin to elliptical galaxies sitting in the centres of disc galaxies \citep{Kormendy2013}. 
Triaxial elliptical galaxies can also show rotation, but contrary to box/peanut bulges they show very little or no configuration rotation
or no pattern speed. The historic consideration of the M31 bulge as a classical bulge implies that the bulge has no pattern speed.
Many studies estimate the pattern speed of M31's bulge \citep{Stark1994, Berman2001, Berman2002}.
The recent kinematic analysis of \citetalias{Opitsch2017} (see their Section 5.3.) identify several signatures directly from the data, 
such as the bulge cylindrical rotation, which favours the barred nature of the M31 bulge over the triaxial elliptical galaxy bulge scenario.
We compared our best matching model with the extreme cases of a model with a slowly rotating bar with $\pps\e15\psu$ and 
another with $\pps\e0\psu$, which is fundamentally a triaxial ``elliptical'' galaxy.
{\bfun In Figure \ref{fig:kinmap_PS0} in the appendix we show the kinematic maps and residuals of the model with $\pps\e0\psu$.}
The resulting models do indeed have a central triaxial bulge substructure; however, the fits are much worse in all the five subsets:
{\bfun the central stellar dispersion is higher than the observations, the dispersion plateaus reproduced by the best model are much weaker
(see Section \ref{sec:res:bm:kin}) and the stellar velocities are much lower.
In addition, the fits to $h3$ and $h4$ are also worse, where the $h3-v_{\rm los}$ correlation observed in the bar region cannot be well reproduced. 
This test therefore demonstrates the barred nature of M31's composite bulge.}

\subsubsection{The bar angle ($\theta_{\rm bar}$)}
\label{sec:res:param:ba}
\begin{figure}
\begin{center}
\includegraphics[width=8.8cm]{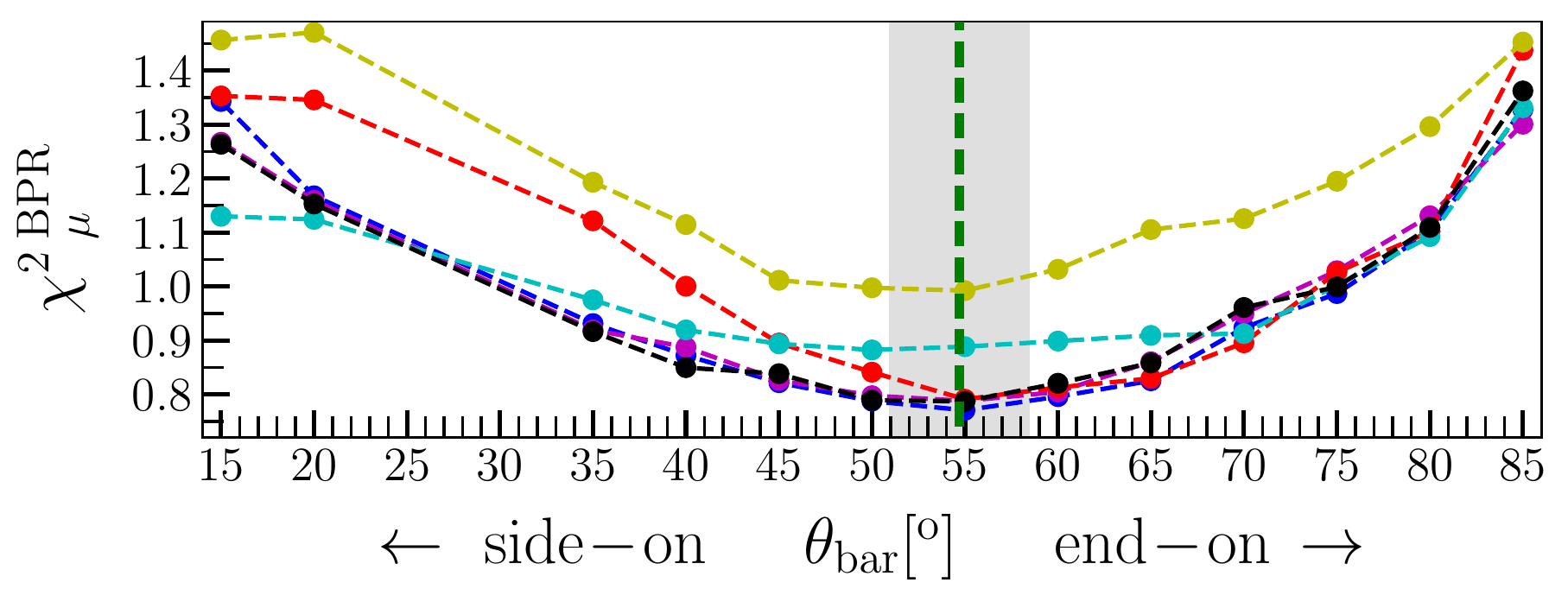}
\vspace{-0.4cm}
\caption[Bar angle]{Variable \setCo (chi-square of the photometry in the \BPB region) for six different M2M models fitted with different bar angle values: 
the best model JR804 (black), model JR355 (red), model JR813 (magenta), model JR364 (blue), model JR683 (cyan) and model JR923 (yellow). 
Their properties are given in the main text. The fiducial value for our runs is $\theta_{\rm bar}\e54\degree\!.7\pm3\degree\!.8$ (vertical green line) from 
\citetalias{Blana2017}, which matches within errors with the minimum in all the tested models.}
\label{fig:ba}
\end{center}
\end{figure}

Here we show that the fiducial bar angle value chosen for the Einasto and NFW grid of models of $\theta_{\rm bar}\e55\degree$
gives the best photometric fits in the \BPB region compared to other values of $\theta_{\rm bar}$.
In Figure \ref{fig:ba} we show different values of the bar angle versus \setCo for the best matching model JR804, 
confirming that our fiducial value $\theta_{\rm bar}\e54\degree\!.7\pm3\degree\!.8$
(Section \ref{sec:mod:input}) from \citetalias{Blana2017} best matches the observations within the errors. 
The minimum value \setCo depends on the bar angle to reproduce the observed twist of the bulge isophotes
with respect to the projected major axis of the isophotes in the disk region, while the allowed range of angles is given by 
the flexibility of the made-to-measure technique to adapt the orbital distribution to match the twist.
Furthermore, we have also considered models with very different dynamical properties such as model JR355 with 
$\iz \pml, \pdm, \pps\de \e \iz 0.65,1.4,40\de$\footnote{$\pml$, $\pdm$ and $\pps$ in units of \ml, $10^{10}\sm$ and \psu},
and models neighbouring the best model in variations of the mass-to-light ratio, such as JR813 with  $\iz \pml, \pdm, \pps\de \e \iz 0.70,1.0,40\de$,
JR364 with  $\iz \pml, \pdm, \pps\de \e \iz 0.74,1.0,40\de$, 
and variations of the bar pattern speed, like JR683 with $\iz \pml, \pdm, \pps\de \e \iz 0.72,1.0,25\de$ 
and JR923 with $\iz \pml, \pdm, \pps\de \e \iz 0.72,1.0,55\de$, 
finding that these models also have a minimum values of \setCo at $\theta_{\rm bar}\approx55\degree$.
This confirms that the fiducial bar angle value found by \citetalias{Blana2017} is located in a global chi-square minimum, 
making unnecessary to vary the bar angle during our parameter search exploration.

\subsection{Properties of the best M2M model}
\label{sec:res:bm}
In the following section we compare the photometric and kinematic properties of M31 with the 
best model from the Einasto grid of models (JR804), showing the contribution of the \CB and the \BPB components separately as well.
{\bfun We find similar properties for the photometric and kinematic substructures in the best model of the grid with NFW 
haloes (KR241).}

\subsubsection{Surface-brightness maps}
\label{sec:res:bm:SB}
\begin{figure*}
\begin{center}
\includegraphics[width=18.4cm]{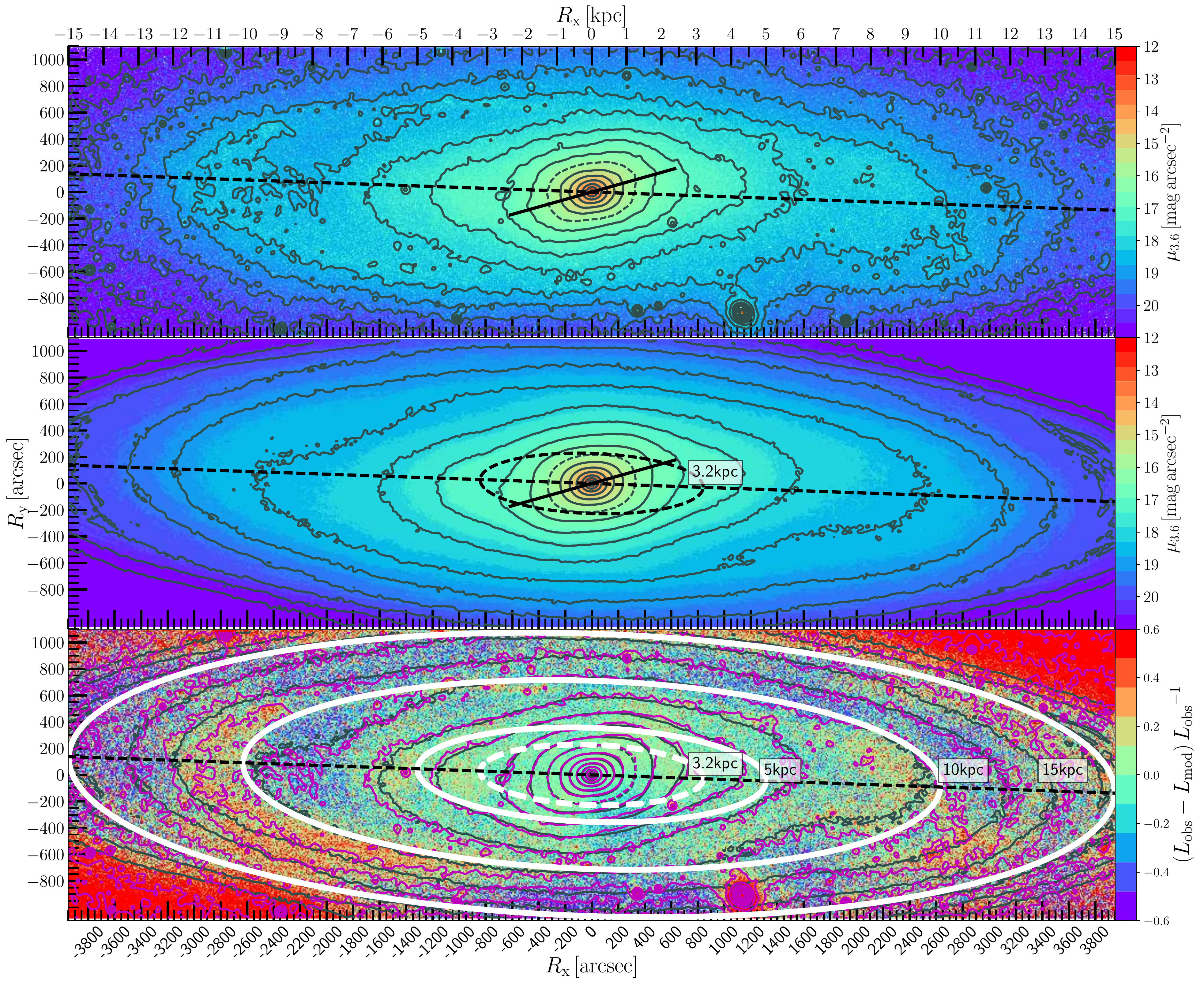}
\vspace{-0.5cm}
\caption[Surface-brightness maps and residuals]{3.6\mum band surface-brightness maps and isophotes spaced with 
$\Delta \mu_{3.6}\e0.5\,[{\rm mag}\,{\rm \as^{-2}}]$. The isophote with $\mu_{3.6}\e16\,[{\rm mag}\,{\rm \as^{-2}}]$ ($I_{3.6}\e3.4\!\times\!10^{3}\,\slu\,\pc^{-2}$) 
is shown with a dashed contour.
Top panel: M31 with the disc projected major axis at $\pa_{\rm disc}\e38\degree$ (dash line) and the projected bar major axis at
$\pa_{\rm bar}\e55\degree\!.7$ (solid line), 
where the de-projected thin bar semimajor axis $r_{\rm bar}^{\rm thin}\e4.0\kpc\,\iz1000\as\de$ is in projection $R_{\rm bar}^{\rm thin}\e2.3\kpc\,\iz600\as\de$ \citepalias{Blana2017}.
The north-east and the near side of the disc are in the top part of the panel (positive $R_{\rm y}$).
Some foreground stars are visible as well as M32 in the bottom at $R_{\rm x}\si1100\as$.
Middle panel: Model JR804 with the disc major axis (dash line) and bar major axis (solid line). 
We indicate the end of the \BPB with a circle projected in the plane of the disk with $i\e77\degree$ at the radii 3.2\kpc (840\as) (black ellipse).
Bottom panel: fractional difference of the luminosity per pixel normalised by the observations. 
We also show the isophotes of M31 (magenta) and model JR804 (black). We show with circles projected 
into the disc the different substructures observed in M31 where the largest deviations occur, where the spiral arms are located
 5\kpc (1300\as), the ring-like structures at 10\kpc (2600\as) and at 15\kpc (3950\as) (white ellipses). Note: model surface-brightness calculated 
 from the temporal smoothed model observable $L$ with a pixel size of $8.63\as$, as in the observations.}
\label{fig:SBmapD}
\end{center}
\end{figure*}

\begin{figure}
\begin{center}
\includegraphics[width=9.0cm]{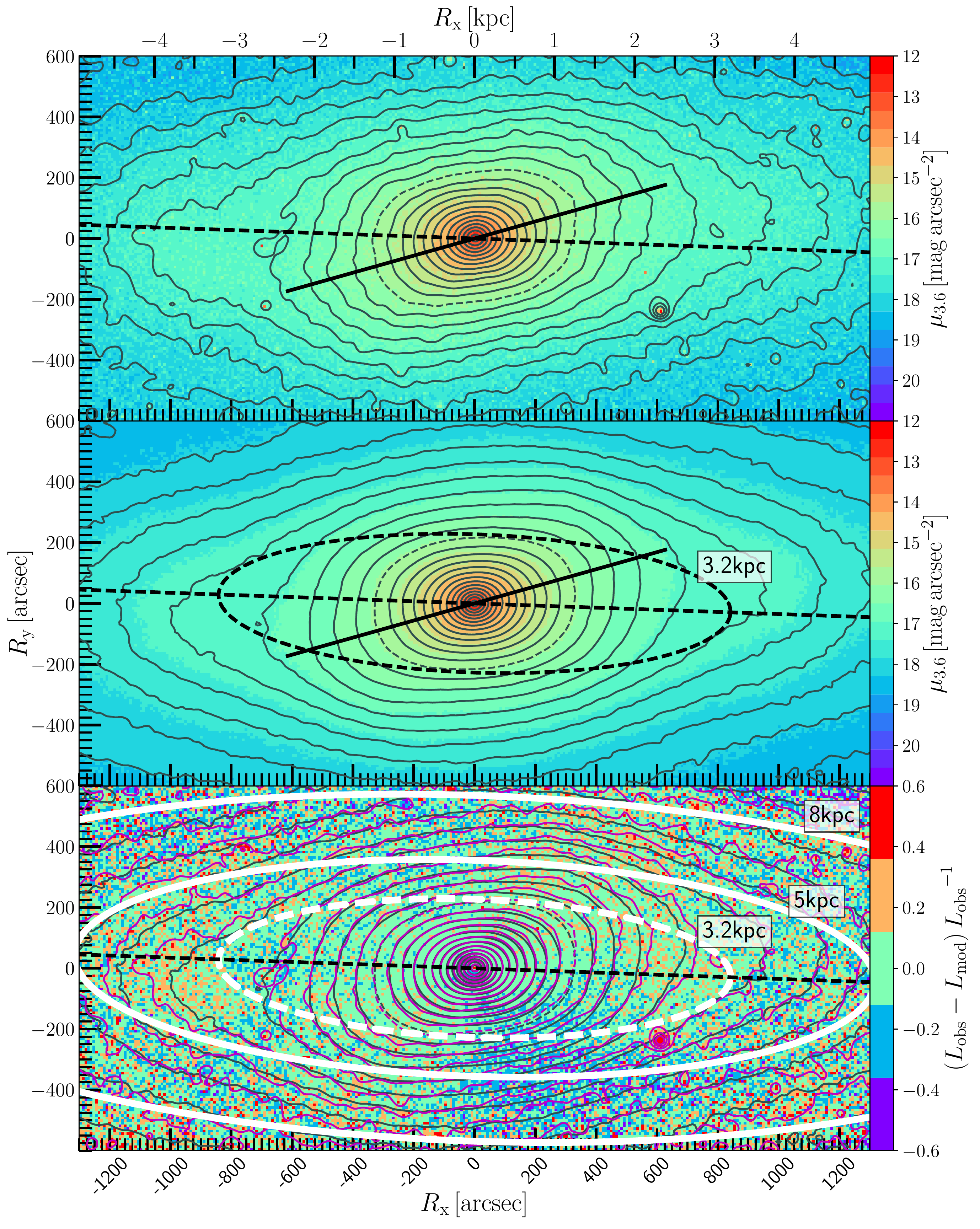}
\vspace{-0.5cm}
\caption[Bulge surface-brightness maps and residuals]{Bulge 3.6\mum band surface-brightness maps and isophotes spaced with $\Delta \mu_{3.6}\e0.25\,[{\rm mag}\,{\rm \as^{-2}}]$.
The value $\mu_{3.6}\e16\,[{\rm mag}\,{\rm \as^{-2}}]$ ($I_{3.6}\e3.4\!\times\!10^{3}\,\slu\,\pc^{-2}$) is shown with a dashed isophote.
Top panel: M31 with the disc projected major axis at $\pa\e38\degree$ (dash line) and the projected bar major axis $\pa\e55\degree\!.7$ (solid line), 
where the de-projected thin bar semimajor axis $r_{\rm bar}^{\rm thin}\e4.0\kpc\,\iz1000\as\de$ is in projection $R_{\rm bar}^{\rm thin}\e2.3\kpc\,\iz600\as\de$ \citepalias{Blana2017}.
The north-east and the near side of the disc are in the top part of the panel (positive $R_{\rm y}$). 
Middle panel: Model JR804 with the disc major axis (dash line) and the projected bar major axis (solid line).
We indicate the end of the \BPB with a circle projected in the plane of the disk with $i\e77\degree$ at the radii 3.2\kpc (840\as) (black ellipse).
Bottom panel: fractional difference of the luminosity per pixel normalised by the observations. We also show the isophotes of M31 (magenta) and the model (black). 
We show circles in the plane of the disk projected for $i\e77\degree$ at radii 3.2kpc(840arcsec), 5kpc(1300arcsec) and 8kpc(2100arcsec)
(white ellipses). Note: model surface-brightness calculated from the temporal smoothed model observable $L$ with a pixel size of $8.63\as$, as in the observations.}
\label{fig:SBmapB}
\end{center}
\end{figure}

We present our photometric M2M fitted map of the best model in Figure \ref{fig:SBmapD}, 
compared to M31 in the 3.6\mum band, and a close-up of the bulge in Figure \ref{fig:SBmapB}.
The model fits in general well, particularly in the bulge. Note that  because the model is a system in dynamical equilibrium and 
it has a symmetric structure (to 180\degree rotations) where the larger differences arise where substructures such as the spiral arms at 
$\si5\kpc\,\iz1300\as\de$ and the ring at $\si10\kpc\,\iz2600\as\de$ are found.
Even the bulge region of M31 is not entirely symmetric, showing asymmetries between the near side (upper) of the bulge and the far side (bottom), 
where the near side has slightly higher luminosity than the far side, more noticeable for the isophotes with $\mu_{3.6}\geq16\magasq$.
The dust extinction is too weak in the 3.6\mum band to cause this asymmetry,
with typical V band extinction in the bulge of $A_{\rm V}\approx1\mags$ \citep{Draine2014} which
corresponds to a 3.6\mum band extinction of $A_{3.6}\si0.07\mags$ \citep{Schlafly2011}.
Moreover, the expected dust extinction effect is the opposite of what is observed, where the luminosity on the far side 
should be systematically higher than in the near side, unlike the asymmetry observed in the map of Figure \ref{fig:SBmapB}.
The 3.6\mum photometric asymmetry also does not show a spatial correlation with high dust density regions 
(Figure \ref{fig:AV}) where the dust could have more emission. Another possibility is that the outer parts of the 
\BPB are not in complete dynamical equilibrium, perhaps related to transient material in the disc, or even a possible passage of a 
satellite galaxy near its centre \citep{Block2006, Dierickx2014, DSouza2018}.

\begin{figure}
\begin{center}
\includegraphics[width=9.0cm]{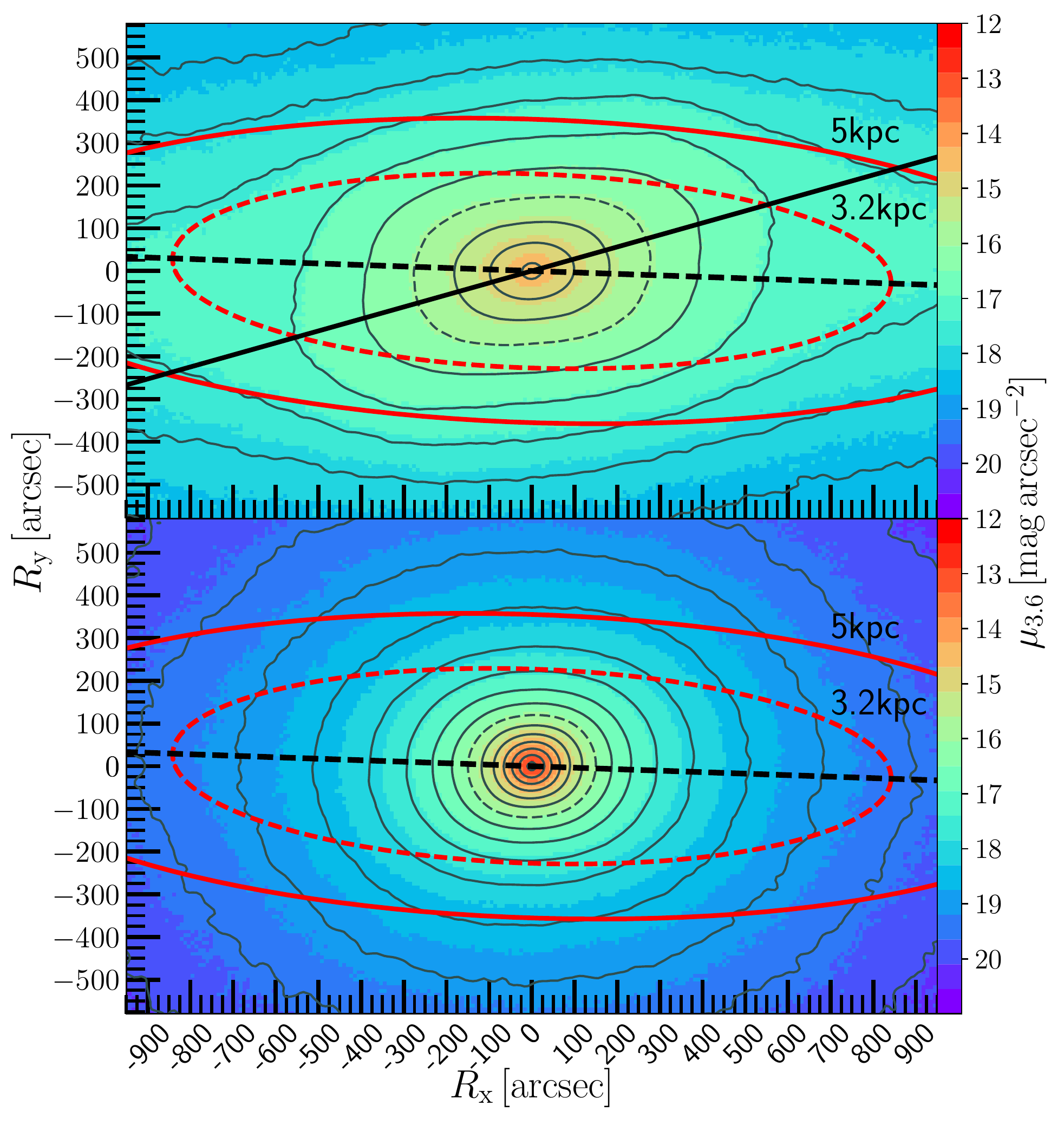}
\vspace{-0.7cm}
\caption[Bulge components surface-brightness maps]{Surface-brightness maps of the \BPB component (top panel) and the \CB component (bottom panel). 
showing also their isophotes, spaced with $\Delta \mu_{3.6}\e0.5\,[{\rm mag}\,{\rm \as^{-2}}]$ 
and the value $\mu_{3.6}\e16\,[{\rm mag}\,{\rm \as^{-2}}]$ ($I_{3.6}\e3.4\!\times\!10^{3}\,\slu\,\pc^{-2}$) is shown with a dashed isophote.
We show circles at radii 3.2 and 5\kpc in the plane of the disk projected for an inclination of $i\e77\degree$ (red ellipses).
The projected bar major axis is shown at $\pa\e55\degree\!.7$ (solid black line).}
\label{fig:SBmapB_comp}
\end{center}
\end{figure}

\begin{figure*}
\begin{center}
\includegraphics[width=18cm]{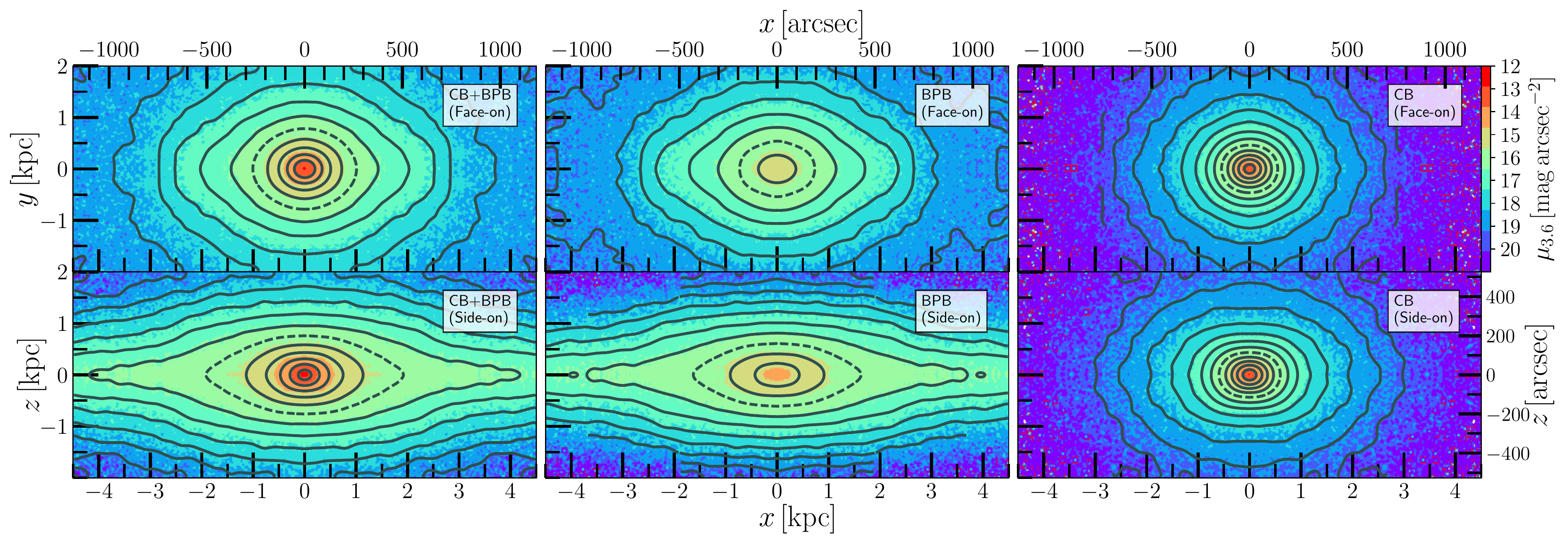}
\vspace{-0.5cm}
\caption[Surface-brightness maps views]{Different views of the best model bulge components. 3.6\mum band surface-brightness maps and 
isophotes spaced with $\Delta \mu_{3.6}\e0.5\,[{\rm mag}\,{\rm \as^{-2}}]$ and the value $\mu_{3.6}\e16\,[{\rm mag}\,{\rm \as^{-2}}]$ 
($I_{3.6}\e3.4\!\times\!10^{3}\,\slu\,\pc^{-2}$) 
is shown with a dashed isophote. The orientations and bulge components are shown in the corners of each panel.
Note: figures generated from the eight-folded model particles to decrease the noise.}
\label{fig:SBmap_views}
\end{center}
\end{figure*}

In Figure \ref{fig:SBmapB_comp} we show separately the \CB component and the \BPB component of model JR804.
As we show with the surface-brightness profile in Section \ref{sec:res:bm:SBpro}, the \CB dominates in light and mass
in the centre. Within $R\lesssim 100\as$ it has roundish ellipses isophotes with their major axis roughly aligned with the disc major axis. 
The \BPB is more extended and it has boxy isophotes that give to the combined bulge a twist of the isophotes 
as observed in M31, shifted away from the disc major axis by $\Delta\pa\si13\degree$ \citepalias{Blana2017}.
The \CB has a more oblate shape and therefore it cannot reproduce the triaxial structure and the twist.
This is better revealed in Figure \ref{fig:SBmap_views}, where we show surface-brightness maps of the best model and its 
bulge components from different orientations.

\subsubsection{Surface-brightness profiles}
\label{sec:res:bm:SBpro}
\begin{figure}
\begin{center}
\includegraphics[width=8.8cm]{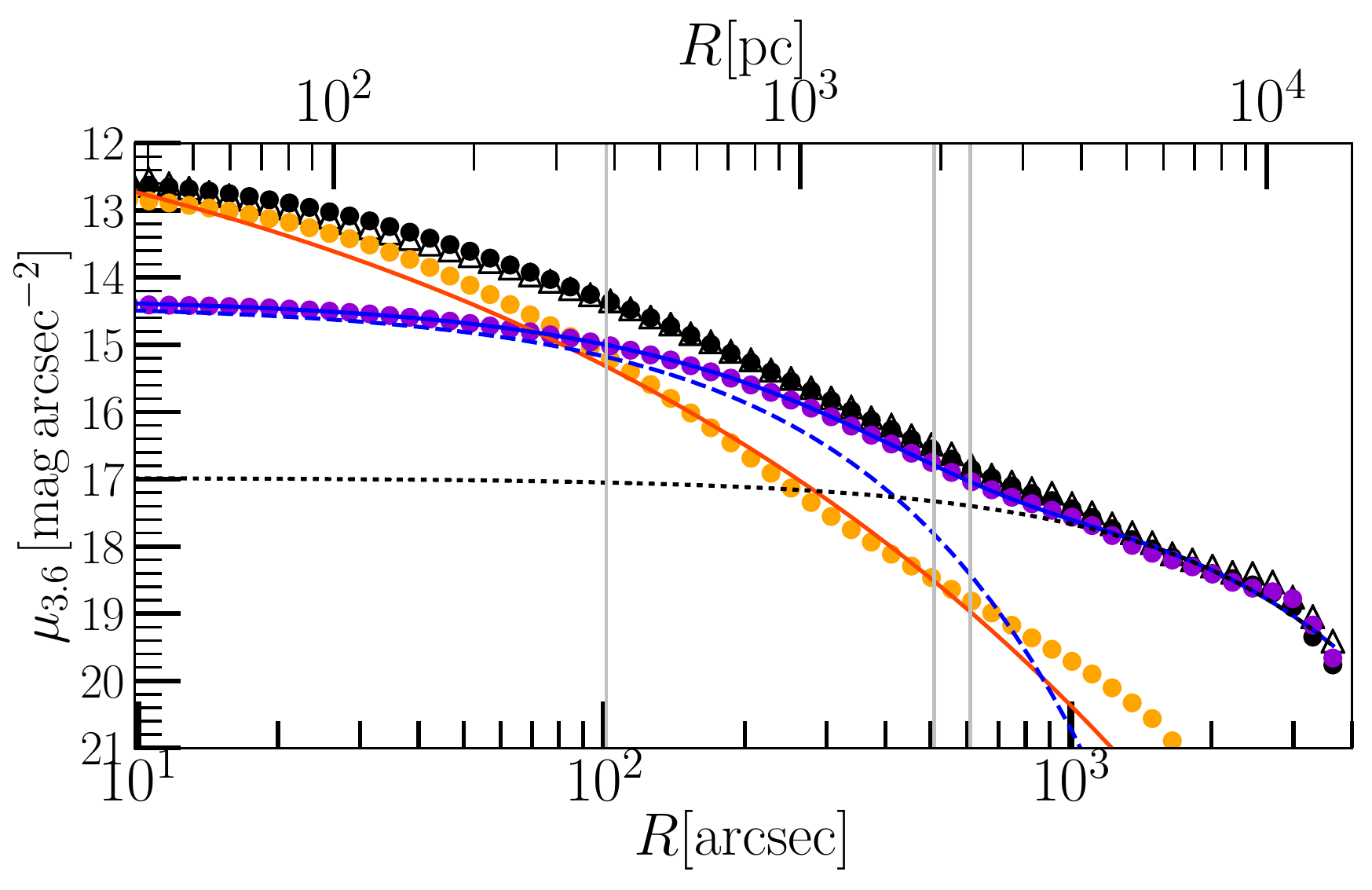}
\vspace{-0.4cm}
\caption[Surface-brightness profiles]{Azimuthally averaged surface-brightness profiles in the 3.6\mum band from ellipses fitted to the images of Figure \ref{fig:SBmapD} 
as function of the ellipse major axis for M31 (white triangles) and model JR804 (black circles) plotted out to 15\kpc. 
We also plot separately the \CB component (orange circles) with its S\'ersic fit (red curve) and the \BPB and disc component
(purple circles), with its S\'ersic fit (dash blue curve), an exponential fit (black dotted curve), and the combined 
(solid blue curve). The vertical line at 100\as marks the end of region CBR. The vertical lines at 510\as and 600\as 
mark the end of the projected semimajor axis of the \BPB and the thin bar \citepalias{Blana2017}. Region BPR ends at 700\as.}
\label{fig:SBprofile}
\end{center}
\end{figure}

\begin{table}\small
\caption[Photometric parameters]{Photometric profile fit parameters for the azimuthally averaged surface-brightness in the 3.6\mum band for M31 and the best model.}
\vspace{-0.3cm}
\label{tab:phot}
\begin{tabular}{lrrrr}
\hline
Parameter			& M31 & CB+BP+disc & BP+disc & CB \\\hline
$n$ 				&$2.58\!\pm\!0.04$		&	$2.24\!\pm\!0.04$ & $1.10\!\pm\!0.01$	& $4.3\!\pm\!0.2$ \\
$\mu_{\rm e}$ [a] 	&$16.50\!\pm\!0.04$		&	$15.96\!\pm\!0.04$ & $16.42\!\pm\!0.01$	& $17.4\!\pm\!0.1$\\
$R_{\rm e}$ [\kpc]	&$1.38\!\pm\!0.04$		&	$0.98\!\pm\!0.03$	& $1.09\!\pm\!0.02$ & $1.22\!\pm\!0.06$ \\
$\epsilon_{R_{\rm e}}$& $0.37\!\pm\!0.01$		& $0.33\!\pm\!0.01$	& $0.40   \!\pm\!0.01$	& $0.25\!\pm\!0.02$\\
$\mu_{\rm o}$ [a]	& $16.94\!\pm\!0.03$	&	$16.80\!\pm\!0.03$ & 	$16.98\!\pm\!0.02$ & -\\
$R_{\rm d}$ [\kpc]		& $5.71\!\pm\!0.08$		& $5.31\!\pm\!0.07$	& $6.02\!\pm\!0.08$	& -\\
\hline
\end{tabular}\\
\textbf{Notes:} parameters from top to bottom are the S\'eric profile parameters: index $n$, surface-brightness $\mu_{\rm e}$ in units of $\magasq$, effective radius $R_{\rm e}$ and ellipticity $\epsilon_{R_{\rm e}}$; and the exponential profile parameters: the surface-brightness $\mu_{\rm o}$ in units of $\magasq$ and the disc scale length $R_{\rm d}$. 
Each parameter error is calculated from the range of solutions taking 90 per cent of the chi-square distribution.
\end{table}

In Figure \ref{fig:SBprofile} we show the azimuthally averaged (AZAV) surface-brightness profiles of the best model and M31 in the 3.6\mum band calculated with \texttt{ellipse-IRAF} \citep{Jedrzejewski1987} directly from the images shown in Figure \ref{fig:SBmapD}.
We also plot separately the \BPB component and the \CB component. 
We fit the total AZAV surface-brightness profiles of the best M2M model JR804 and M31 with a S\'ersic profile \citep{Sersic1968, Capaccioli1989} and an exponential profile out to 15\kpc using a
non-linear least squares (NLLS) minimization method, obtaining the parameters in Table \ref{tab:phot}. 
We also fit the model bulge components serparately, fitting the \BPB and the disc with a S\'ersic profile and an exponential profile; and the \CB component alone with another S\'ersic profile (Table \ref{tab:phot}).

We also use \texttt{imfit} \citep{Erwin2015b} to perform a 2D fit to the image of the \CB component 
(Figure \ref{fig:SBmapB_comp} bottom panel) with a S\'ersic profile, finding values similar to the 1D fit, with $R^{\CB}_{\rm e}\e273.3\as$, 
$\mu^{\CB}_{\rm e}\e 17.1\magasq$ and a S\'ersic index of $n^{\CB}\e3.4$.  If we do not parameterise the contribution of the \BPB in the fitting with an additional Sersic profile, the 
resulting S\'ersic index from the usual photometric decomposition of one S\'ersic profile and one exponential profile component is 
$n\approx 2$ as shown by \citet{Courteau2011}  and also \citetalias{Blana2017}. \citet{Fisher2008} show that the S\'ersic 
index value of $n\si2$ is a threshold that can distinguish galaxies with pseudobulges or classical bulges, the latter typically 
showing values larger than 2. However, in our scenario we have a composite bulge with a \CB with a high S\'ersic 
index $n^{\rm CB}\si4$ and a \BPB with a lower value $n^{\rm BPB}\si1$, that when fitted with a single S\'ersic 
and an exponential for the disc results in an intermediate value of 2.

The most important properties revealed in Figure \ref{fig:SBprofile} are: 
\renewcommand{\labelenumii}{\roman{enumii})}
\begin{enumerate}[leftmargin=4pt ,itemsep=0pt,labelsep=0pt]

\item~The \CB dominates in the central region $R\lesssim100\as$, and it is required in order to reproduce the central light concentration in M31, 
and, as we show later in more detail in Section \ref{sec:res:bm:kin}, this component also reproduces the central dispersion profile observed in M31.

\item~The \BPB dominates in projection between $\si100\as$ and $R^{\rm BPB}\e510\as$; and the thin bar extends out to $R^{\rm thin}_{\rm bar}\e600\as$ \citepalias{Blana2017}.

\item~The surface-brightness bump at $R\si1000$ $-$ $1300\as$ $(4$ $-$ $5\kpc)$ is caused by spiral arms and material trailing the bar (see Section 4.6.2 in \citetalias{Blana2017}), 
and it is reproduced in the M2M model by a slightly increase of the disc surface density.

\item~The surface-brightness profile also reveals a second ``bump" at 10\kpc \citep{Barmby2006, Courteau2011} from which point the surface brightness decreases at a faster rate. This is generally attributed to an additional contribution of the 10\kpc-ring structure, 
however it is also possible to attribute this to a change in the SB profile. 
M31 is a barred galaxy, which are systems that often develop such a break due to the 
secular evolution of the disc due to the angular momentum transfer with the bar through the Lindblad resonances, and also due to the redistribution of the disc 
material by the bar formation \citep{Debattista2006b}. 
As we show later, in Section \ref{sec:res:bm:kin:vcps}, we find that the outer Lindblad resonance is indeed located at $11\pm1\kpc$, supporting this scenario.
This indicates that the disc of M31 could be a mild Type II.o-OLR disc, with of a SB break at $\si10\kpc$ related 
to a ring-like structure near the OLR resonance \citep{Erwin2008,Kim2014}, like the galaxy NGC3504, but more difficult to detect due to 
the high disc inclination. This would imply that a broken profile would be better suited for the photometric parametrisation of the outer 
M31 stellar disc, rather than the standard single exponential profile.
\end{enumerate}

\subsubsection{Dust extinction effects on the observed kinematics}
\label{sec:res:bm:dust}

\begin{figure}
\begin{center}
\includegraphics[width=8.4cm]{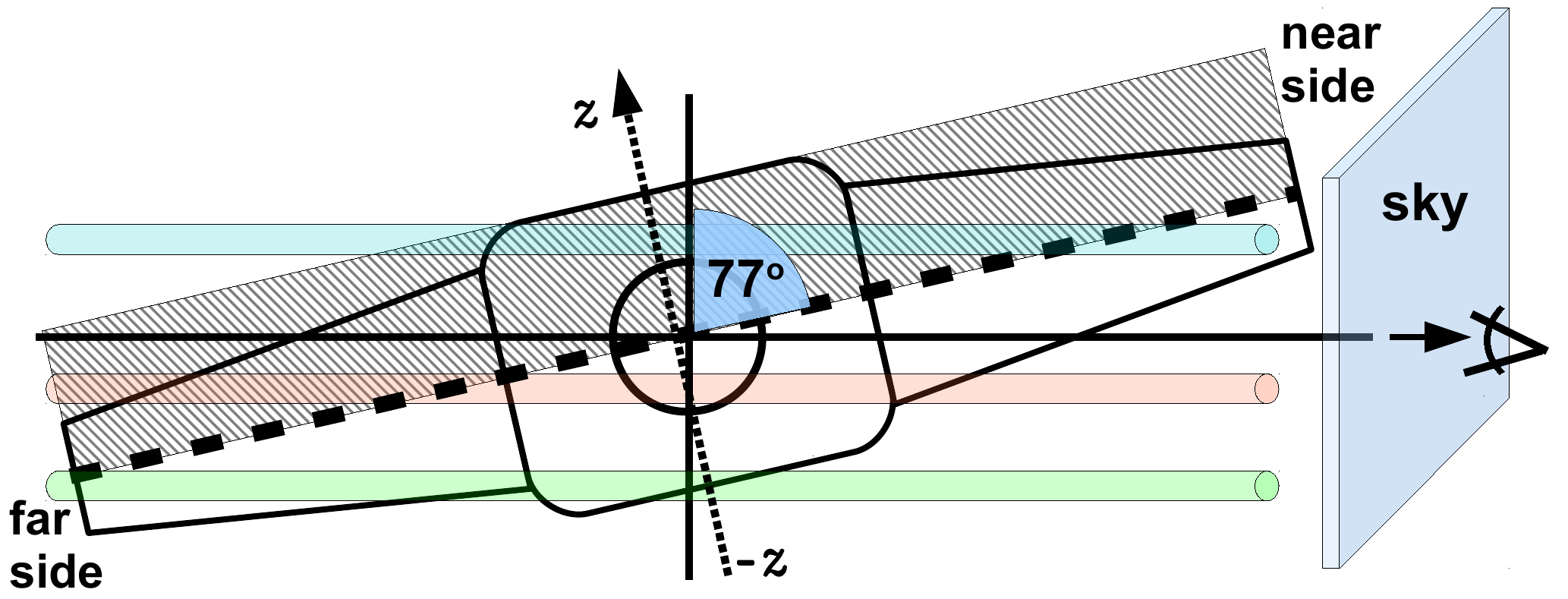}
\vspace{-0.5cm}
\caption[M31 orientation diagram]{Diagram of M31 asymmetric projection effects due to the dust and the geometrical orientation with a disc inclination $i\e77\degree$. 
Without extinction M31 projects into an image where the near and the far side are symmetric to the observer (right). However, if the light has a strong extinction by the dust located in the plane of the disc (dashed line) the observer detects an asymmetry.
With strong extinction the light integrated along the near side of the disc (blue upper tube) will be dominated by the outer and younger part of the disk, 
while the material within the dashed area will be obscured by the dust in the plane of the disc. 
The opposite occurs in the far side of the disk, where the inner part of the disc dominates (green  bottom tube). The bulge also projects asymmetrically and as a consequence the region where most of the light of the bulge is detected
is slightly shifted to the far side (bottom) from the bulge centre (red middle tube), being then the deepest part of the bulge.}
\label{fig:dust}
\end{center}
\end{figure}

\begin{figure}
\begin{center}
\includegraphics[width=9.0cm]{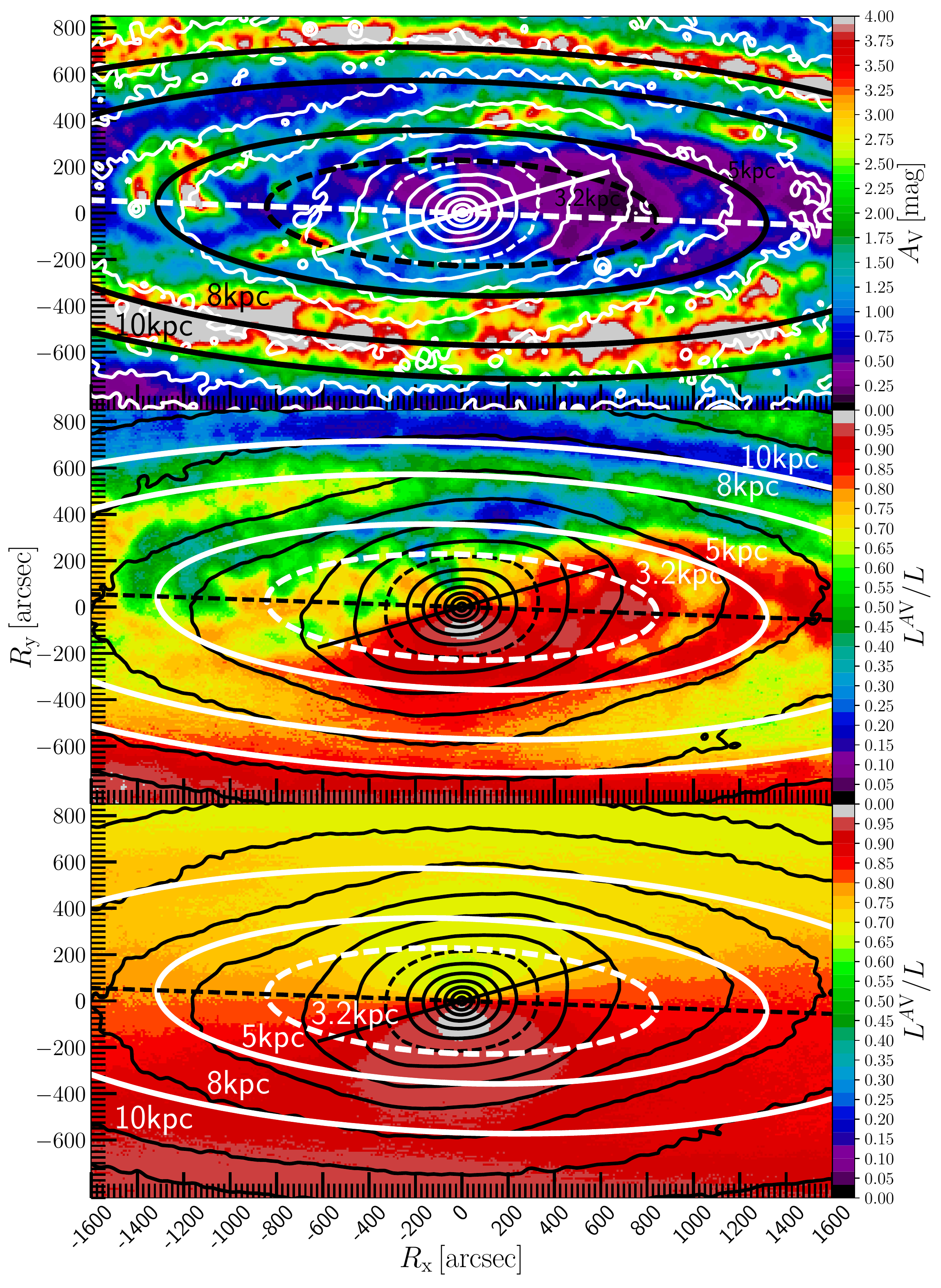}
\vspace{-0.7cm}
\caption[M31 absorption map]{Top panel: M31 absorption map in the V band calculated from equation \ref{eq:Av} and the dust surface mass 
map \citep{Draine2014}, with overplotted white isophotes of the M31's 3.6\mum band image.
Middle panel: $L^{A{\rm V}}\,L^{-1}$ map of model JR804 with overplotted model black isophotes in the 3.6\mum band.
Note the light grey region near the bulge centre where the effects of the extinction in the V band are minimal.
Bottom pane: $L^{A{\rm V}}\,L^{-1}$ map and 3.6\mum band isophotes for a model with the parameters of JR804, but observed and fitted 
through a dust map with a constant absorption of $A_{\rm V}\e0.5\mags$.
All panels: the disc projected major axis is shown with a dash line, with the near side of the disc in the upper part of the figures (positive $R_{\rm y}$),
and the projected bar major axis is shown with a solid line.
We show circles at the radii 3.2, 5, 8 and 10\kpc in the plane of the disk projected for $i\e77\degree$  (white and black ellipses).
The isophotes are spaced with $\Delta \mu_{3.6}\e0.5\,[{\rm mag}\,{\rm \as^{-2}}]$ and $\mu_{3.6}\e16\,[{\rm mag}\,{\rm \as^{-2}}]$ is shown with a dashed isophote.}
\label{fig:AV}
\end{center}
\end{figure}

Given that the IFU M31 bulge stellar kinematic observations \citepalias{Opitsch2017} are in the V band, we have included the effects of the dust extinction 
in our modelling implemented according to Section \ref{sec:mod:obs:kin2}. 
The diagram in Figure \ref{fig:dust} qualitatively shows that, when some of the light of the galaxy is absorbed by the dust located in the plane of the disc, 
the projected image can have asymmetries between the near side of the disc and the far side.
These asymmetries are strongly reflected in the stellar kinematics, as we show in the following sections.
The line-of-sight to the far side of the disc penetrates more deeply into the galaxy than the near side.
Thus, for example, the deepest region in the bulge is located slightly towards the far side from the bulge centre. 
A similar effect has also been detected in the reddening of RGB stars in M31's disk \citep{Dalcanton2015}.

As the dust effects in the 3.6\mum band are very weak, we observe the model without dust extinction to fit the light in this band, which corresponds to the 
model light observable $L$. We also observe the model through the V band extinction map shown in Figure \ref{fig:AV} (top panel), which results in the 
model observable $L^{A{\rm V}}$. In Figure \ref{fig:AV} (middle panel) we show a map of the fraction of the absorbed and non absorbed light of the 
model ($L^{A{\rm V}}\,L^{-1}$).
Without dust the near and the far side of the disc are symmetric, as in Figure \ref{fig:SBmapB} (middle panel);
however with dust extinction the model produces asymmetries between both sides, as shown in Figure \ref{fig:AV} (middle panel). 
The regions of the map with a ratio of $L^{A{\rm V}}\,L^{-1}\e1$ are where all the light is detected, while for a ratio of zero the light is completely 
absorbed. Note that the ratio $L^{A{\rm V}}\,L^{-1}$ is proportional to the ratio between the light in the V band and the 3.6\mum band.

The map in Figure \ref{fig:AV} (middle panel) reveals interesting features that are caused not only by the dust absorption itself, but also by the 
geometrical orientation of M31 with its disc inclination $i\e77\degree$ and its bar angle $\theta_{\rm bar}\e54\degree\!.7$.
The least absorbed (or deepest) region in the M31 bulge is shifted from the centre to the far side of the disc (light grey region at $R_y\si-100\as$), 
as expected from the diagram in Figure \ref{fig:dust}, and the most extreme effect of extinction near the M31 bulge is in the near side of the disc, 
between $R_{\rm y}\si200\as$ and 400\as (green and blue regions), which are produced by the dust accumulated in the spiral arms. 
In addition, regions with large amounts of dust, but at the far side of the disc, can have weak effects on the light extinction like, for example, the 
far side at $R_{\rm y}\si-500\as$ where the outer ring is with a large dust lane. 

In the bottom panel of Figure \ref{fig:AV} we show a fit where we used a constant light absorption of $A_{\rm V}\e0.5\mags$ to estimate how 
the heterogeneity of the M31 dust map \citep{Draine2014} affects the ratio $L^{A{\rm V}}\,L^{-1}$, finding that the general features and the 
asymmetry are also reproduced. \citet{Dalcanton2015} finds lower absorption values in M31 than \citet{Draine2014}; and so to investigate this 
we reduced the absorption values of the dust map by 50 per cent. We found a model with properties similar to the overall best model that again 
produced the observed asymmetries, but weaker than the fiducial model.

We conclude that the most important consequence of the dust extinction for the kinematics in the V band is that kinematic asymmetries
are generated between the near side of the disc and the far side, because the light integrated along the line-of-sight can be dominated by different 
structures with different intrinsic kinematic properties. 
An example of this is shown by \citet{Baes2000} for elliptical galaxies.
Furthermore, it is important to consider that neither the near side of the disc, nor the far side, have the complete signature along the line of sight,
although the far side is much less affected by light extinction.

\subsubsection{Stellar kinematics}
\label{sec:res:bm:kin}

\begin{figure*}
\begin{center}
\includegraphics[width=18.2cm]{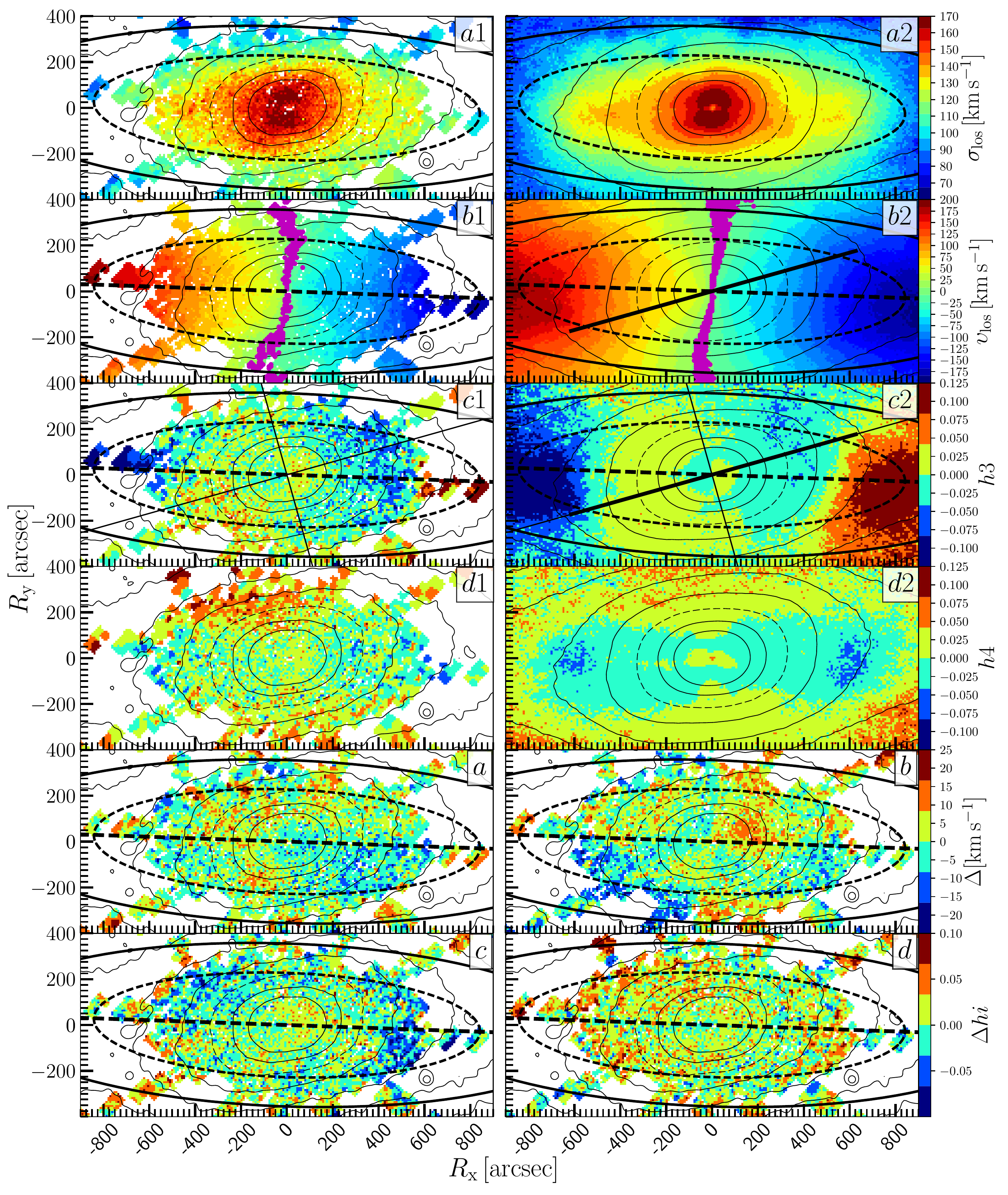}
\vspace{-0.55cm}
\caption[Kinematic maps and isophotes of M31 and the best model]{Isophotes and kinematic maps of $\sigma_{\rm los}$, $\upsilon_{\rm los}$, $h3$, $h4$
of M31 \citepalias{Opitsch2017} ($a1$, $b1$, $c1$, $d1$) 
and model JR804 (panels $a2$, $b2$, $c2$, $d2$), showing isophotes spaced every $\Delta \mu_{3.6}\e0.5\,[{\rm mag}\,{\rm \as^{-2}}]$ 
and $\mu_{3.6}\e16\,[{\rm mag}\,{\rm \as^{-2}}]$ in dashed isophote. 
We exclude the central isophotes to better reveal the kinematic features.
Some panels display two circles projected on the disc's plane with $i\e77\degree$ at 3.2\kpc (black dashed ellipse) and 8\kpc (solid black ellipse), 
the projected disk major axis (dash black line at $\pa\e38\degree$), and the projected bar major axis (black line at $\pa\e55\degree\!.7$) and minor axis 
(black line at $\pa\e145\degree\!.7$).
The thick black lines in panels ($b2$, $c2$) mark the projected thin bar major axis, 
where the de-projected semimajor axis $r_{\rm bar}^{\rm thin}\e4.0\kpc\,\iz1000\as\de$ is in projection $R_{\rm bar}^{\rm thin}\e2.3\kpc\,\iz600\as\de$ \citepalias{Blana2017}. 
The differences between the observations and the model are shown in panel ($a$) with $\Delta\e\sigma_{\rm los}^{\rm obs}-\sigma_{\rm los}^{\rm model}$, 
($b$) with $\Delta\e||\upsilon_{\rm los}^{\rm obs}||-||\upsilon_{\rm los}^{\rm model}||$,
($c$) with $\Delta h3\e h3^{\rm obs}-h3^{\rm model}$, 
and panel ($d$) with $\Delta h4\e h4^{\rm obs}-h4^{\rm model}$.
We show the zero velocity values within a range $\upsilon_{\rm los}\e 0\pm5\kms$ (magenta) in panels $b1$ and $b2$.}
\label{fig:kinmap}
\end{center}
\end{figure*}

\begin{figure*}
\begin{center}
\includegraphics[width=18.cm]{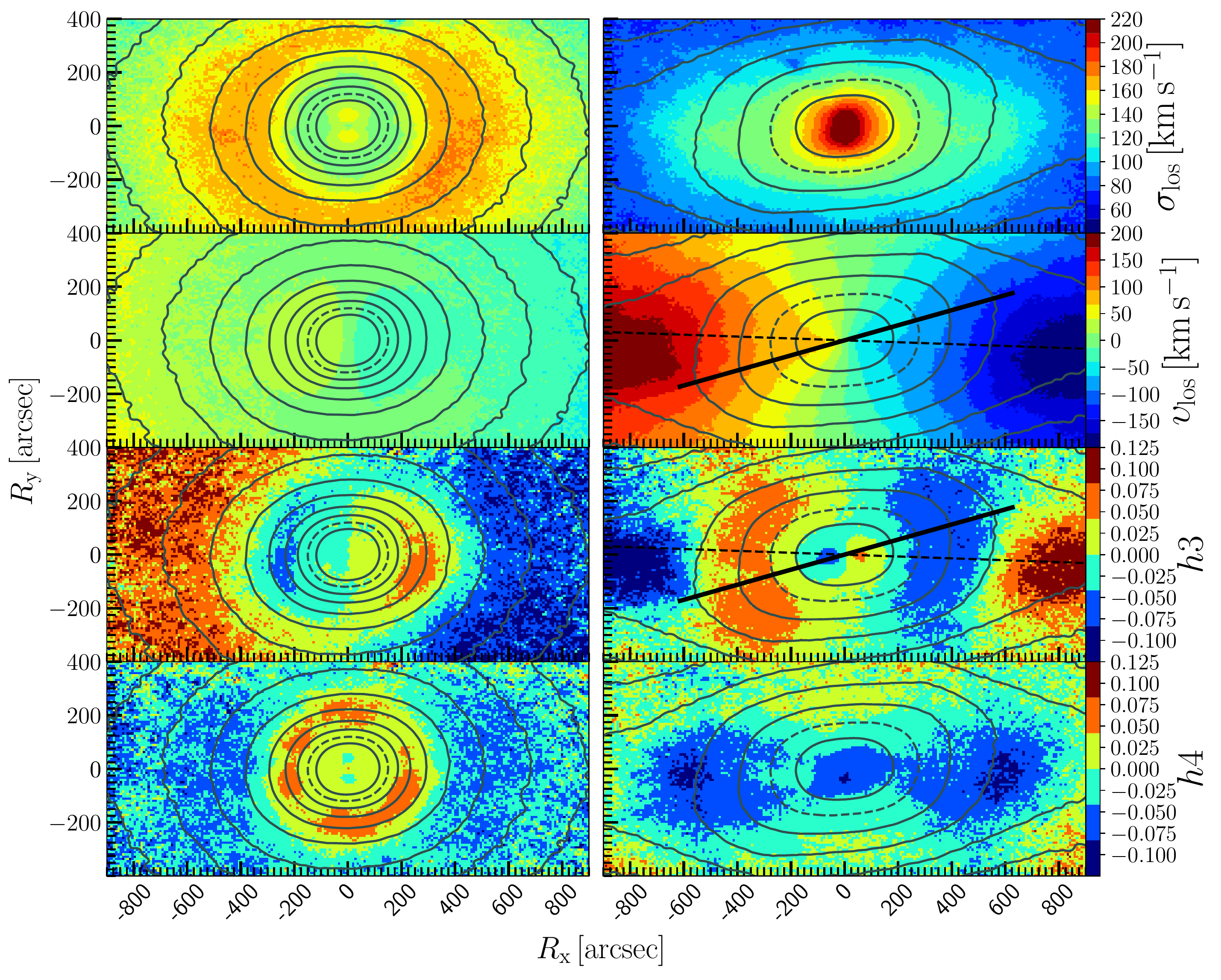}
\vspace{-0.5cm}
\caption[Kinematic maps and isophotes of the best model bulge components]{Kinematic maps and isophotes of model JR804 for 
the \CB particles (left column) and the \BPB and disc particles (right column). 
The isophotes are spaced every $\Delta \mu_{3.6}\e0.5\,[{\rm mag}\,{\rm \as^{-2}}]$ 
and the value $\mu_{3.6}\e16\,[{\rm mag}\,{\rm \as^{-2}}]$ is shown with a dashed isophote.
In the \BPB maps of $\upsilon_{\rm los}$ and $h3$ we show the projected bar major axis (solid line) and the projected disc major
axis (dashed line). 
We exclude the isophotes in the centre to better reveal the central kinematic structures of each bulge component.}
\label{fig:kinmapcomp}
\end{center}
\end{figure*}

\begin{figure}
\begin{center}
\includegraphics[width=8.9cm]{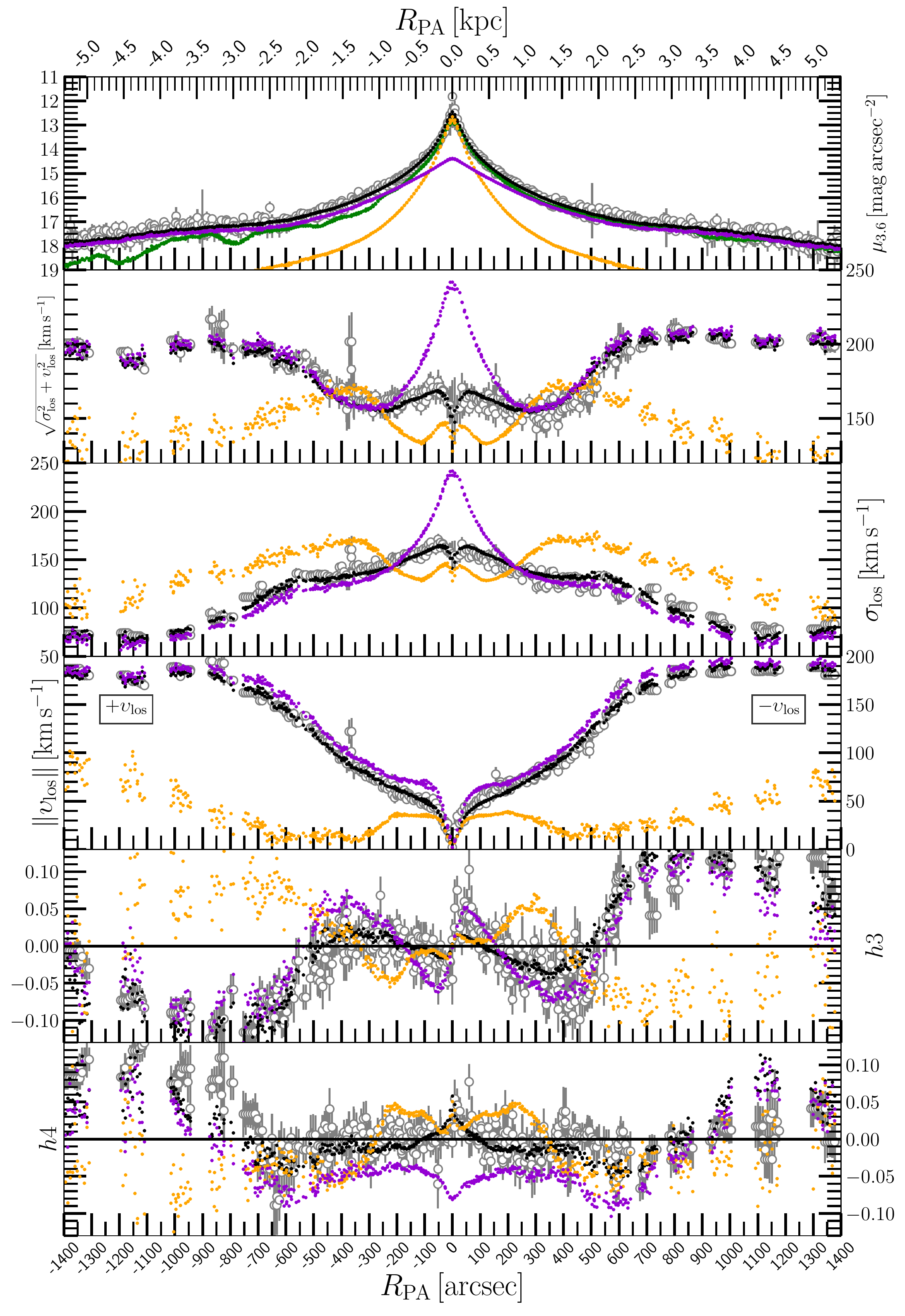}
\vspace{-0.7cm}
\caption[Surface-brightness and kinematic disc cuts]{
Surface-brightness and kinematic cuts near the disc major axis ($\pa\e33\degree$) of model JR804 (black dots) with its components, the 
\CB (orange) and the \BPB (purple), and of M31 (open circles).
We also plot the extincted surface-brightness of the model observable $L^{A{\rm V}}$ ($\mu_{A{\rm V}}$, green line).
Positive $R_{\pa}$ extends into the far side of the disc.}
\label{fig:kinprofPAD}
\end{center}
\end{figure}

\begin{figure*}
\begin{center}
\includegraphics[width=8.8cm]{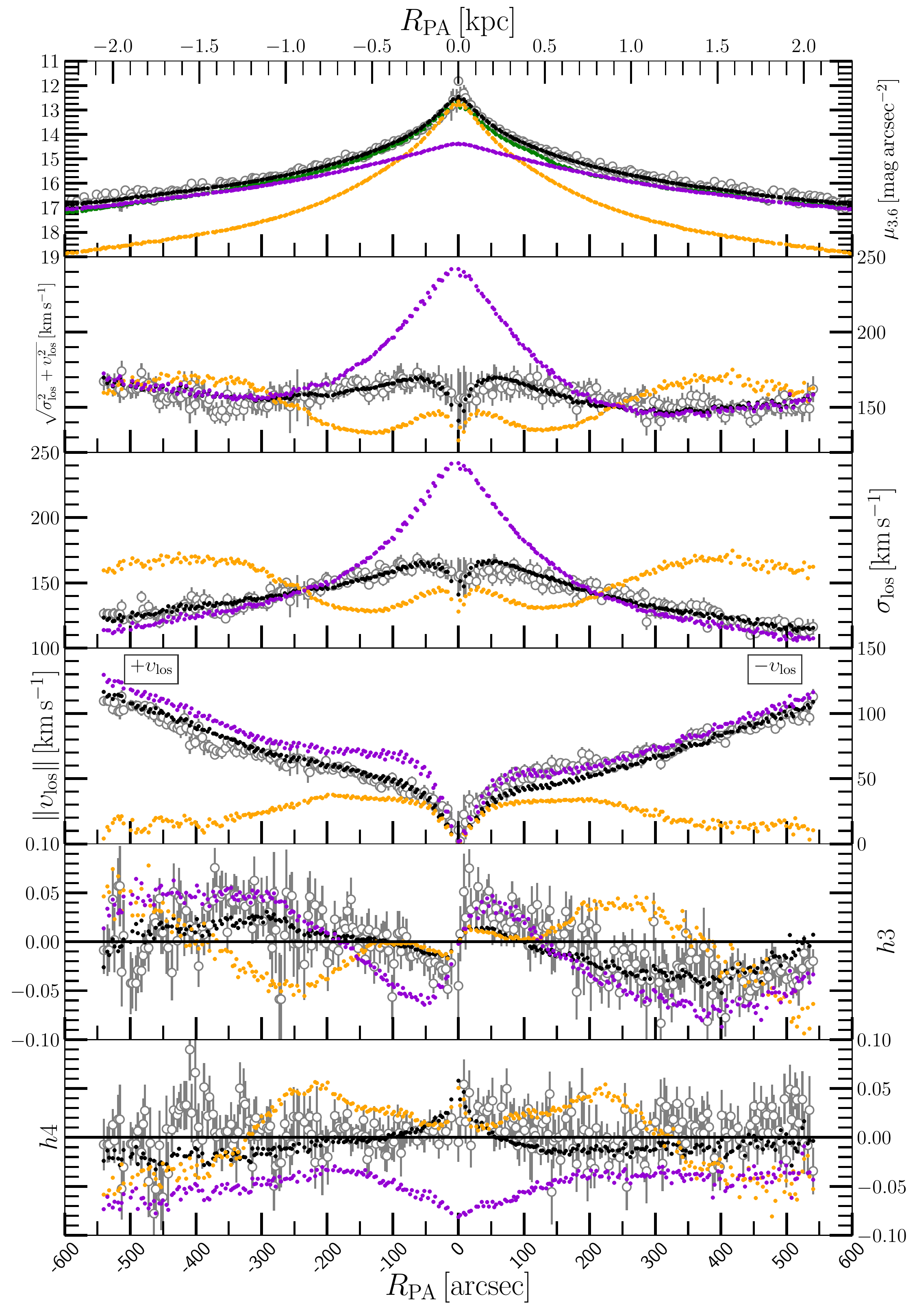}
\hspace{-0.2cm}
\includegraphics[width=8.8cm]{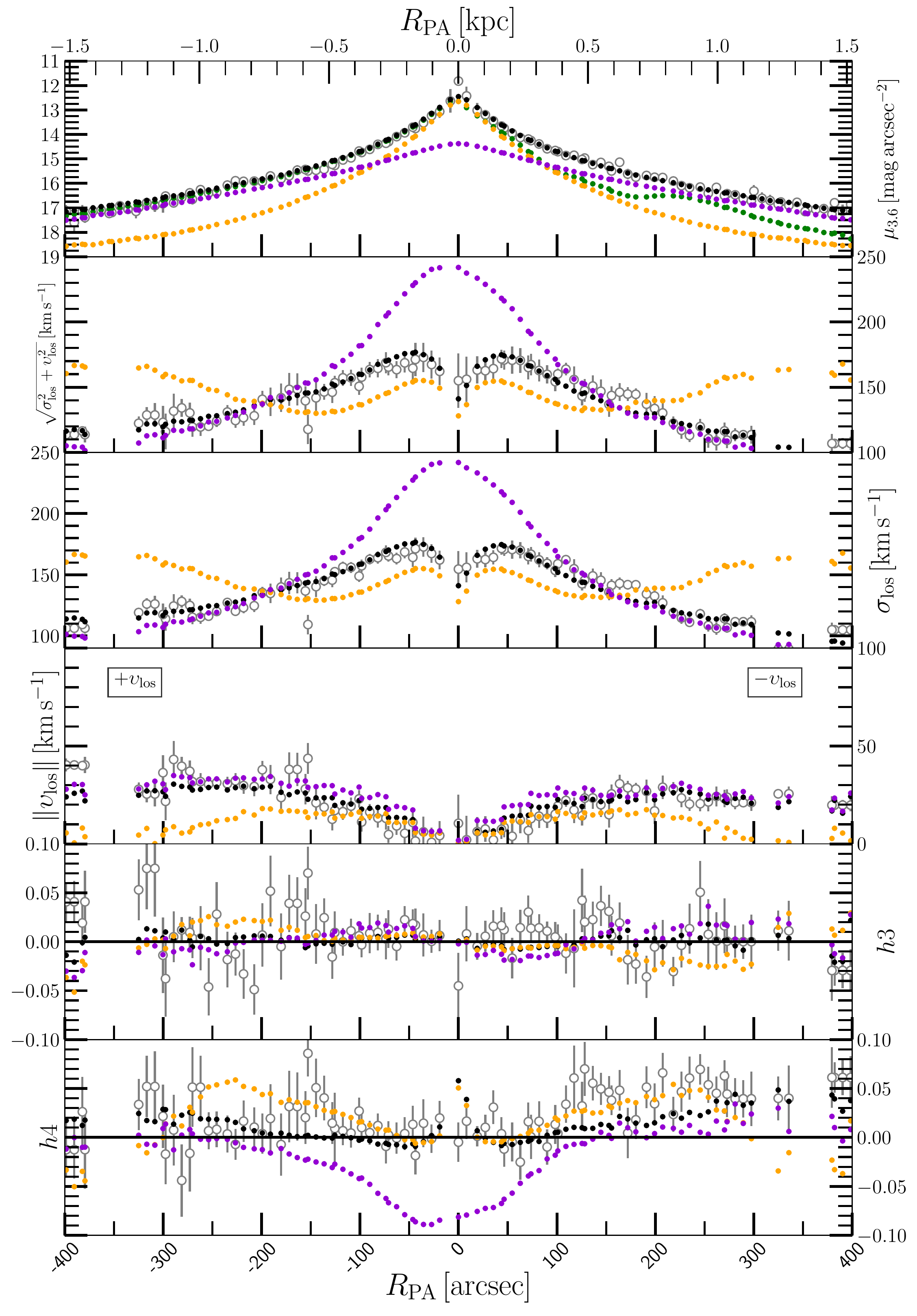}
\vspace{-0.4cm}
\caption[Surface-brightness and kinematic bulge cuts]{Surface-brightness and kinematic cuts along the projected bar major axis 
($\pa\e55.7\degree$) in the left column and the bar minor axis ($\pa\e145.7\degree$) in the right column of model JR804 (black dots) with its components, the \CB (orange) 
and the \BPB (purple), and of M31 (open circles). 
We also plot the extincted surface-brightness of the model observable $L^{A{\rm V}}$ ($\mu_{A{\rm V}}$. green line).
Positive $R_{\pa}$ extends into the near side of the disc.}
\label{fig:kinprofPA}
\end{center}
\end{figure*}

\begin{figure}
\begin{center}
\includegraphics[width=8.7cm]{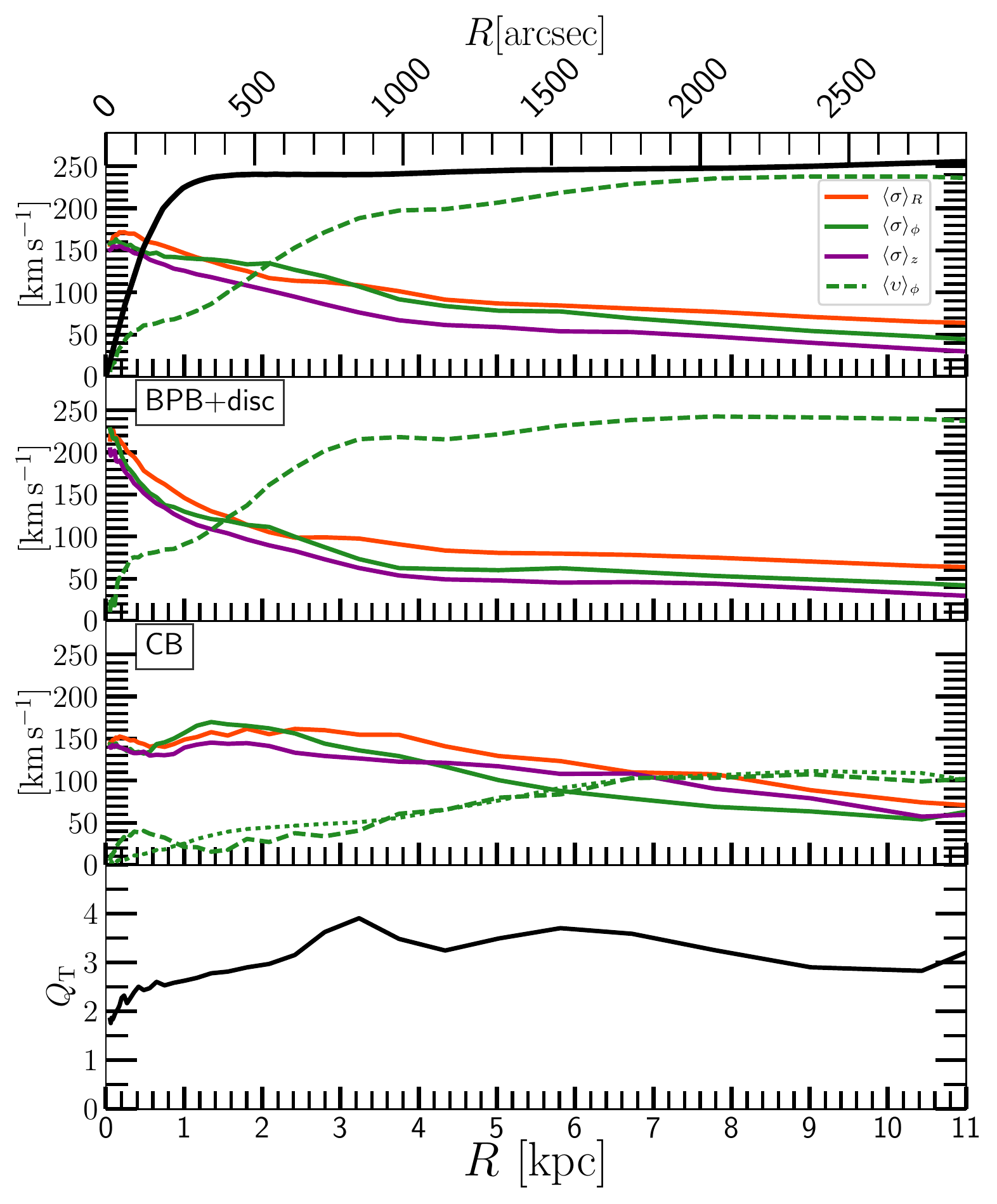}
\vspace{-0.7cm}
\caption[Intrinsic kinematic profiles]{
Intrinsic azimuthally averaged kinematic radial profiles in the disc plane for the best Einasto model for the total stellar components (top panel), 
\BPB and disc (second panel), CB component (third panel), and the Toomre parameter $Q_{\rm T}$ (bottom panel). The dispersion profiles 
are shown in the first three panels in solid line for the coordinates $\langle \sigma \rangle_{R}$ (red), $\langle \sigma \rangle_{\phi}$
(green) and $\langle \sigma \rangle_{\rm z}$ (magenta), and the streaming velocity $\langle \upsilon \rangle_{\phi}$ is shown with a 
dashed green line. We also show $\langle \upsilon \rangle_{\phi}$ of the \CB of Model 1 (green dotted line). 
The total circular velocity is shown in the first panel (black line).}
\label{fig:kinprof3D}
\end{center}
\end{figure}

In this section we present the bulge kinematics of the best model and compare them with the IFU kinematic 
measurements of \citetalias{Opitsch2017}.

For a better qualitative comparison we show kinematic maps in Figure \ref{fig:kinmap}, presenting the velocity, the dispersion, $h3$ and $h4$ 
of M31, the best model JR804 and the residuals. In Figure \ref{fig:kinmapcomp} we show separately the \CB component and the 
\BPB component of the best model.
For an easier quantitative comparison we also show kinematic profiles along the disc major axis in Figure \ref{fig:kinprofPAD}, 
and along the bar projected major and minor axis in Figure \ref{fig:kinprofPA}.

\renewcommand{\labelenumi}{\Roman{enumi})}
\renewcommand{\labelenumii}{\roman{enumii})}

\begin{enumerate}[leftmargin=4pt ,itemsep=0pt,labelsep=0pt]

\item~\textit{The line-of-sight dispersion ($\sigma_{\rm los}$)}:
it has three important features that are reproduced by the model: 
\begin{enumerate}
\item~Within $R<100\as$ the velocity dispersion of M31 shows two peaks of 
$\sigma_{\rm los}^{\rm M31, max}\si170\kms$ 
along the bulge minor axis \citepalias{Opitsch2017}, with a drop of $\sigma_{\rm los}$ in the centre. 
We find that this is produced by the concentrated \CB that dominates in the centre (also shown in \citetalias{Blana2017}) which is revealed with the dispersion maps of each component in Fig \ref{fig:kinmapcomp}, 
and in the $\sigma_{\rm los}$ profiles of Figure \ref{fig:kinprofPAD} and \ref{fig:kinprofPA}.
The \CB mass profile is similar to a Hernquist model, where the material in the centre requires lower kinetic energy to remain confined
in gravitational equilibrium, leading to a dispersion drop. 
The two dispersion peaks, and partially the dispersion drop, can also be attributed to CB particles in circular orbits near the centre, as 
\citet{Hernquist1990} shows for the Hernquist model.
When plotted separately, the maximum central velocity dispersion of the \CB alone is $\sigma_{\rm los}^{\rm CB, max}\si150\kms$,
which combined with the high maximum central dispersion of the \BPB with a peak of $\sigma_{\rm los}^{\rm BPB, max}\si240\kms$, 
reproduce the central dispersion in M31.
A very important characteristic of the \BPB is that its high central velocity dispersion is caused by the deep gravitational potential 
of the \CB component, which due to its high mass concentration increases the central circular velocity. 
This results in particles orbiting the \BPB and the thin bar that have high velocities when passing the centre. 

\item~Our model also reproduces the two elongated high $\sigma_{\rm los}$ plateaus in the bulge noted by \citetalias{Opitsch2017}
within $R_{\rm x}\e\pm600\as$ shown in Figures \ref{fig:kinmap} and \ref{fig:kinprofPA}. 
This features are reproduced in the model by the \BPB that dominates here over the classical bulge.
At the end of the \BPB along the projected bar major axis, or at the projected disc major axis at $R_{\rm x}\si-600\as$, 
the \BPB surface-brightness is $\mu_{3.6}\si17\magasq$, while the \CB is much fainter, with $\mu_{3.6}\si 18.5\magasq$.
The dispersion of the \CB component in the outer part rises again.
Further out the two $\sigma_{\rm los}$ plateaus end at $R_{\rm x}\si\pm600\as$, decreasing along the major axis (Figure \ref{fig:kinmap}) to
$\sigma_{\rm los}\si70\kms$ in the disc.

\item~Along the disc minor axis and at the near side of the disc (positive $R_{\rm y}$) the dispersion is systematically lower than the far side of the disc
(negative $R_{\rm y}$) as shown by the maps (Figure \ref{fig:kinmap}) and the profiles (Figure \ref{fig:kinprofPA}). 
This feature is also reproduced in the model, and is caused by the dust absorption. 
This can be understood from Figure \ref{fig:dust},
and the dust extinction map in Figure \ref{fig:AV}: 
the light of the near side of the bulge that is behind the dust plane is strongly extinguished by the dust, leaving 
mostly the light of the foreground disc that has a dispersion lower than the bulge. 
In contrast, at negative $R_{\rm y}$, most of the light of the bulge is transmitted, 
while part of the light of the kinematically cooler disc material, which is now behind the bulge is absorbed, resulting in dispersions slightly higher
than if the disc would be fully included.  
The dust also causes the observed asymmetry between the two $\sigma_{\rm los}$ plateaus, where for the side of $R_{\rm x}<0\as$ 
the dispersion is higher than at the side of $R_{\rm x}>0\as$. 
\end{enumerate}

\item~\textit{The line-of-sight velocity ($\upsilon_{\rm los}$)}:
We also find that the combination of both bulge components reproduces different characteristics of the M31 bulge velocity field, listing three of them below: 
 \begin{enumerate}
\item 
 In the very centre ($R<50\as$) and near the disc major axis ($\pa\e33\degree$) (Figure \ref{fig:kinprofPAD}) both bulge components show similar 
 rotation ($\upsilon_{\rm los}\si30\kms$). However, at $R\si100\as$ the \BPB rotates much faster, reaching already $\si70\kms$, while the \CB component has $\si35\kms$, which then combined reproduce the total velocity of M31 with $\si50\kms$.  
 Between 100\as and 600\as along the disc major axis, the \BPB dominates the light and the rotation increases with a constant slope, reaching a roughly
 constant value of $\upsilon_{\rm los}\si\pm200\kms$ in the disc region.

\item~Analysing the difference in velocity between the model and the observations in the panel $b$ of Figure \ref{fig:kinmap},
we find a region in the observations at $(R_{\rm x}, R_{\rm y})\e(200\as,100\as)$ that has a velocity $\si10\kms$ higher than the model.
This is an asymmetry in the M31 observations that is not reproduced by our dust modelling. 
CO observations in this region \citep{Melchior2011} indicate that the molecular gas kinematics is complex and maybe tilted in this region,
and so it may be that our dust modelling is too simple here. As the bar major axis $\upsilon_{\rm los}$ profile shows 
(Figure \ref{fig:kinprofPA}), the \BPB velocities match the observations well, suggesting that \CB light contribution
could be much weaker in this particular region.

 \item~The velocity map (Figure \ref{fig:kinmap}) shows in the centre a twist in the zero velocity values that reproduces the velocity 
 twist observed in M31. The twist is weaker in the very centre (within 100\as) due to the \CB component, which has a more oblate 
 structure (Figure \ref{fig:kinmapcomp}).

 \item~The velocity of the CB component within 50\as is higher than the initial model before being fitted. As we show later in Figure \ref{fig:kinprof3D},
the CB component of the initial model gain rotation from the bar, but mostly in the outer parts. This could mean that the observed high central 
rotation could be a legacy from the early formation of the CB.
\end{enumerate}

\item~\textit{The Gauss-Hermite coefficients $h3$ and $h4$}:
 \begin{enumerate}
\item~The $h3$ maps (Figure \ref{fig:kinmap}) and profiles (Figure \ref{fig:kinprofPA}) show that the $h3$ values in the disc region 
beyond $R_{\rm x}\!>\!700\as$ and $R_{\rm x}\!<\!-700\as$ are anti-correlated 
with the velocity $\upsilon_{\rm los}$, changing when we enter the region of the bar, with $h3$ then correlated with the velocity. 
This behaviour is expected in barred galaxies \citep{Bureau2005,Iannuzzi2015}. However, the central region of M31's bulge has a second change of sign in $h3$,
which is also reproduced by the model (\ie $h3$ and $\upsilon_{\rm los}$ are again anti-correlated for isophotes with 
$\mu_{3.6}\!\leq\!16.5\magasq$). This central $h3-\upsilon_{\rm los}$ anti-correlation feature is produced by both bulge components, and it is 
due to the near axisymmetric central density distribution, similar to the way that the asymmetric drift causes the $h3-\upsilon_{\rm los}$
anti-correlation in the axisymmetric disc.
The dust extinction also generates an asymmetry between the $h3$ at the left and the right side of the bar that is reproduced by the model (Figure \ref{fig:kinmap}).
In particular, the $h3-\upsilon_{\rm los}$ correlation in the bar region is more extended along the $R_{\rm x}$ axis in the positive side of $R_{\rm x}$.
This is because a large fraction of light from the bar at negative $R_{\rm x}$ is behind the dust and it is more strongly absorbed, leaving the foreground part of the disc 
component more visible.

  \item~The M31 $h4$ map in Figure \ref{fig:kinmap} reveals in the centre a positive region, while at the end of the \BPB 
  ($R_{\rm x}\si\pm600\as$) the $h4$ map shows negative values. In the model the \BPB $h4$ map in Figure \ref{fig:kinmapcomp} 
  shows mostly negative values, and the central positive $h4$ region is reproduced by the \CB that shows strong positive $h4$ 
  (except where the two $\sigma_{\rm los}$ peaks are detected, where the classical bulge $h4$ is negative). 
  Along the disc minor axis $h4$ has positive values at $R_{\rm y}\e\pm400\as$, with larger positive values in the near side of the disc,
  which is where the dust extinction effects are stronger. Our $h4$ maps also agree with the results for other 
  box/peanut bulge models \citep{Iannuzzi2015}, where $h4$ depends on the bar angle for bars with strong box/peanut bulges, while bars with weak or without a box/peanut bulges show a weaker dependence.  
 \end{enumerate}

\item~\textit{Stellar kinematics in the outer disc and the inner spheroid}:\\
In Figure \ref{fig:kinprof3D} we show the de-projected kinematic profiles of the best Einasto model. 
We also show the Toomre parameter $Q_{\rm T}\e \kappa\, \langle\sigma\rangle_r \left(3.36\, G\, \Sigma\left( R \right)\right)^{-1}$ \citep{Toomre1964} 
calculated with the azimuthally averaged surface mass density profile $\Sigma\left( R \right)$ and the radial velocity dispersion $\langle\sigma\rangle_r$ 
from the disc particles and the epicycle frequency $\kappa$ from the total circular velocity. The stellar disc is stable and dynamically hot 
with a radially averaged value and standard deviation of $\langle Q_{\rm T}\rangle\e 2.6\!\pm\!0.6$. 
This is consistent with \citet[][see their Figure 16]{Dorman2015} who also find that M31 has a dynamically hot stellar disc.\\

The intrinsic kinematic profiles in Figure \ref{fig:kinprof3D} also shows that in the outer parts the \CB increases its rotation to 
$\si70\kms$ at 5\kpc, similar to the values estimated for the inner spheroidal component at that radius \citep{Dorman2012}, 
reaching $\si100\kms$ at 10\kpc. 
The outer rotation of the \CB is similar to the Model 1 of \citetalias{Blana2017}, which obtained all its rotation from the angular 
momentum transfer from the bar \citep{Saha2016}. The increase of the rotation of the inner spheroid at this radius is not unexpected, 
as for example it is also observed in the Milky Way's inner stellar halo \citep{Ness2013b,Perez-Villegas2017a}.\\

 \item~\textit{Stellar and globular cluster kinematics}:\\
 Given that the kinematic properties of each bulge component are different, is there a signature or tracer which could identify each stellar component observationally?
\citet{Morrison2011} analyse velocities and metallicities of a sample of old star clusters near the M31 centre, {\bfun concluding that these clusters could be associated with the bar or the inner spheroid depending on their metallicities: the more metal-rich clusters $(\zh\!\geq\!-0.6\dex)$} near the disc major axis beyond $R\si4\kpc$ have velocities similar to the M31 surrounding field stars ($\si200\kms$), but within the \BPB region ($R<2\kpc$), the clusters reach higher velocities ($\si300\kms$) similar to the \BPB velocity dispersion presented here.
 Given that they assumed that the bar is roughly edge on ($\theta_{\rm bar}\si20\degree$) they associate the metal-rich component with 
the $x_2$ orbits that are perpendicular to the bar. However, in \citetalias{Blana2017} and here we find $\theta_{\rm bar}\si55\degree$ with the end of the thin bar 
in projection at $R\e2.3\kpc\,(600\as)$, approximately the location where the metal-rich star clusters velocity changes.
{\bfun The more metal-poor clusters $(\zh\!\leq\!-0.8\dex)$} show less co-rotation with the field stars with a broader velocity distribution, similar to the {\bfun outer} classical bulge kinematic properties, {\bfun or to a spheroid component like the inner stellar halo.}
Therefore, the bulge models presented here support the scenario where the star clusters in the centre could be associated with different stellar components.
\end{enumerate}

\subsubsection{Kinematics: circular velocity \& the Lindblad resonances.}
\label{sec:res:bm:kin:vcps}
\begin{figure}
\begin{center}
\includegraphics[width=8.8cm]{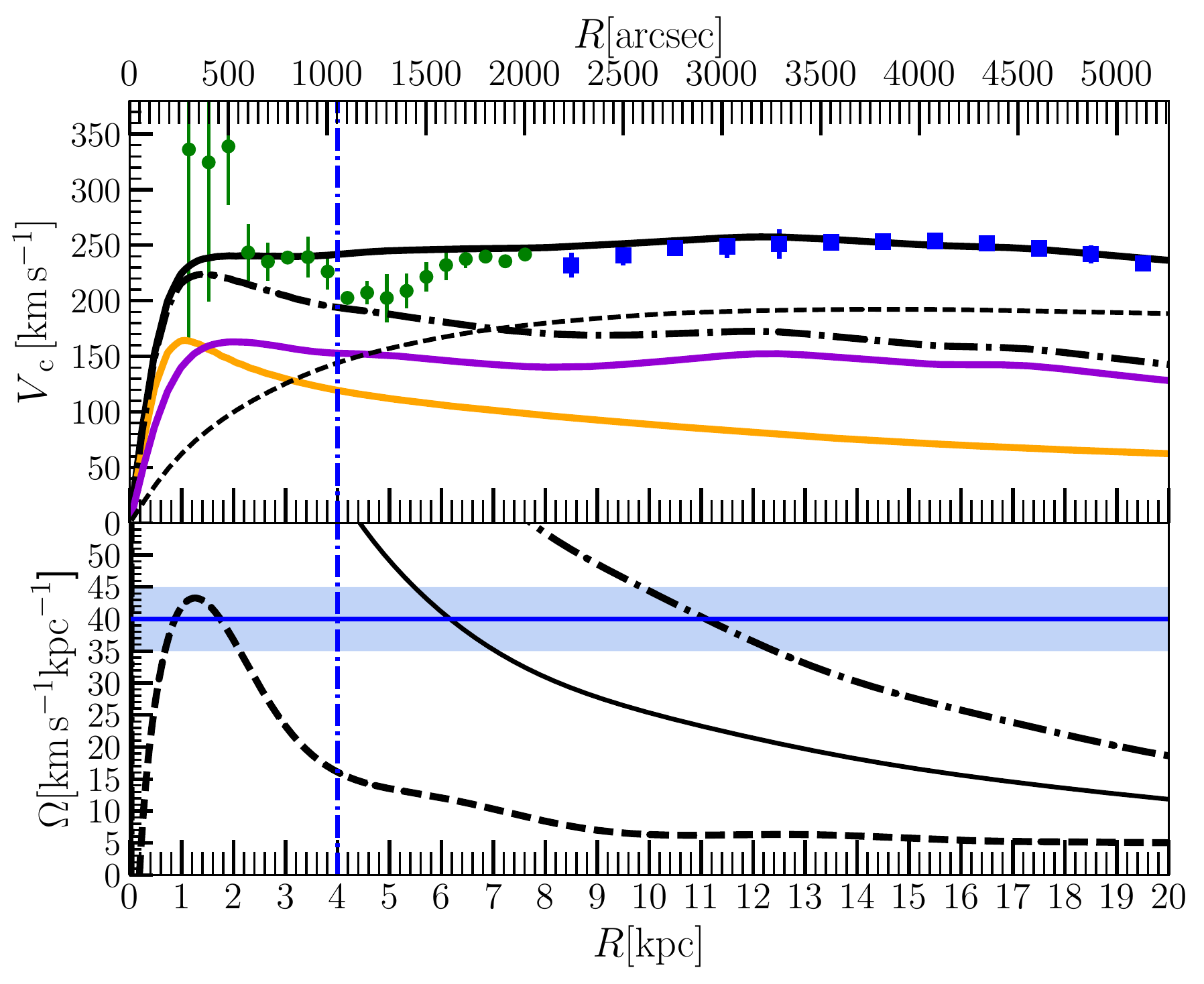}
\vspace{-0.7cm}
\caption[Best model circular velocity profiles, angular frequency profiles and resonances]{Top panel: azimuthally averaged circular velocity 
in the plane of the disc of the model JR804 with the different components: 
the \CB (orange), the \BPB and the stellar disc (purple), the total stellar component (dot dashed line), the dark matter (dash) 
and the total circular velocity (solid black). The \HI data of \citet{Corbelli2010} is shown out to 20\kpc (blue squares). 
We also show the \HI data of \citet{Chemin2009} within 8\kpc (green circles).
Bottom panel: 
the angular frequency profile $\Omega$ (solid curve), $\Omega_{\rm ILR}\e\Omega-\kappa/2$ (dash curve), 
and $\Omega_{\rm OLR}\e\Omega+\kappa/2$ of the model JR804 with a bar pattern speed \bps (blue  horizontal line and shaded region). 
The corotation radius, the Lindblad resonances inner inner, outer inner, and outer are located at $r_{\rm cor}\e6.5\pm1.0\kpc$, 
$r_{\rm IILR}\e1.0\kpc$, $r_{\rm OILR}\e1.8\kpc$ and $r_{\rm OLR}\e11.2\pm1.0\kpc$. 
The end of the thin bar \citepalias{Blana2017} is shown in both panels (blue vertical dotted dashed line).}
\label{fig:kinvc}
\end{center}
\end{figure}

In Figure \ref{fig:kinvc} we show the total circular velocity $V_{\rm c}$ profile of the best model JR804 with its different components. 
The \CB component reaches a maximum circular velocity of $V_{\rm c}^{\rm CB, max}\e165\kms$ at 1.0\kpc, dominating over 
the \BPB component within $R\!\leq\!0.5\kpc$, and then drops nearly Keplerianly. 
The \BPB component reaches a maximum of $V_{\rm c}^{\rm BPB, max}\e160\kms$ at 2.0\kpc, where it dominates over the classical bulge, which 
has 140\kms at that radius. 
The total circular velocity increases fast due to the \CB contribution, reaching $V_{\rm c, o}\e235\kms$ at 1.6\kpc where 
it stays roughly flat reaching a maximum of $V_{\rm c, max}\e255\kms$ at \si12.5\kpc.
We also show the \HI rotation curve of \citet{Corbelli2010} that is used to fit the dark matter density profile, which is in general well fitted.
The authors neglect the inner $R<8.5\kpc$, arguing the presence of an inner warp and the non-circular motion of the gas. 
For comparison we also show the inner $R<8\kpc$ of the \HI rotation curve from \citet{Chemin2009} although we similarly caution about the non-axisymetric motion of the gas here.

In the bottom panel of Figure \ref{fig:kinvc} we show the angular frequency profile ($\Omega$) of the best model, 
with the range of best bar pattern speed \bps. 
The corotation radius, where $\pps\e\Omega$, is located at $r_{\rm cor}\e6.5\pm1.0\kpc$. 
The isophotal comparison of the M31 bulge with N-body models in \citetalias{Blana2017} suggests that the thin bar length of M31 is 
$r_{\rm bar}^{\rm thin}\si4.0\kpc$, which would classify M31's bar as a slow bar with $\mathcal{R}\e1.6\pm0.2$, where \citet{Debattista2000}
define slow bars to be when $\mathcal{R}\e r_{\rm cor}/r_{\rm bar}^{\rm thin}\!\leq\!1.4$ .
The inner inner and the outer inner Lindblad resonances  \citep[$\Omega_{\rm ILR}\e\Omega-\kappa/2$][]{Lindblad1956},
in this model are located at $r_{\rm IILR}\e1.0\kpc$ and $r_{\rm OILR}\e1.8\kpc$. 
The outer Lindblad resonance $\Omega_{\rm OLR}\e\Omega+\kappa/2$ is then at $r_{\rm OLR}\e11\pm1\kpc$.

The gas kinematics and its distribution in M31 shows many substructures that are consistent with 
the typical properties observed in other barred galaxies.
In the centre of the bulge between \si1\kpc (260\as) and \si2\kpc (500\as) the gas velocity measured by \citet{Chemin2009} reaches 
$\si340\kms$, higher than the circular velocity $V_{\rm c}\si230\kms$.
However, this difference is expected in barred galaxies where the gas has a non-circular motion with
in-falling streams of gas, as shown by \citet[][see their Figure 5, see also \citealt{Gerhard1986, Binney1991, Li2015, Chemin2015}]{Kim2012}.
Such streams are typically located near the inner Lindblad resonances, which in this model are at 
$r_{\rm IILR}\e1.0\kpc\,(260\as)$ and $r_{\rm OILR}\e1.8\kpc\,(470\as)$, almost exactly where \citetalias{Opitsch2017} also detects 
the presence of high velocity streams of gas with $\si\pm300\kms$.

A second signature is that the \HI gas velocity drops in the transition between the bar and the disc, as observed between 4\kpc and 6\kpc.
This again is typically produced in barred galaxy simulations due to the non circular motion of the gas in the non axisymmetric potential produced by the bar.

Finally, there is the 10\kpc ring-like substructure \citep{Habing1984, Gordon2006, Barmby2006}. This is made of stars, 
gas and dust and it is where most of the current star formation occurs \citep{Ford2013, Rahmani2016}, with a 
star formation timescale longer than 500\Myr \citep{Lewis2015}. This is longer than the
dynamical time scale at this radius, making an ephemeral collision origin unlikely, as proposed by \citet{Block2006, Dierickx2014} 
\citep[see however][]{Hammer2018}. Assuming that this structure is located at 10\kpc and that it is related to a resonance with the bar,
\citetalias{Blana2017} predict a bar pattern speed of $\pps\e41\psu$. Here we use the bulge stellar kinematics as fitting constraints, 
finding \bps, placing the outer Lindblad resonance at $r_{\rm OLR}\e11\pm1\kpc$
near the ring structure. This suggests that the OLR could be related to the formation of the ring, as also observed in other galaxies
\citep{Buta1991, Buta2017b}.

\section{Summary and discussion}
\label{sec:conc}
We have presented here dynamical models for M31 built with a classical bulge component (\CB) 
and a box/peanut bulge component (\BPB).
We use the M2M method to measure the main properties of M31's bulge:
the IRAC 3.6\mum mass-to-light ratio \pml, the dark matter mass within 3.2\kpc of the bulge \pdm,
and the pattern speed of the \BPB and the thin bar \pps. For this we directly fit 
simultaneously new IFU VIRUS-W bulge stellar kinematic observations \citep{Opitsch2017},
and the 3.6\mum IRAC photometric data \citep{Barmby2006}, with the following main results:

\begin{enumerate}[label=\arabic*),leftmargin=*,itemsep=0pt,labelsep=5pt]
 \item The range of parameters that best reproduce all the observations simultaneously are 
\bml, \bps and \bdm, using an Einasto dark matter profile.
These models have a total dynamical mass within the composite 
bulge of $M_{\rm dyn}^{\rm B}\e4.25^{+0.10}_{-0.29}\times10^{10}\sm$ with a stellar mass and percentage of 
$M_{\star}^{\rm B}\e3.09^{+0.10}_{-0.12}\times10^{10}\sm$(73\%).
The CB has $M_{\star}^{\rm CB}\e1.18^{+0.06}_{-0.07}\times10^{10}\sm$(28\%)
and the \BPB has $M_{\star}^{\rm BPB}\e1.91\pm0.06\times10^{10}\sm$(45\%).
We also obtain similar values within the errors for our grid of models with NFW dark matter haloes.
The bulge dark matter mass agrees with the expected values for an adiabatically contracted NFW halo with M31's virial mass.
However, the best Einasto models fit the bulge stellar kinematics generally better than the models 
with NFW haloes, favouring a shallow central dark matter halo distribution,
similar to that found in the Milky Way \citep{Portail2017a}.
This also reveals the importance of kinematic data with high spectral and spatial resolution, and the appropriate 
modelling to accurately determine the central dark matter mass distribution in galaxies.

\item How does the model of \citetalias{Blana2017} compare with the best M2M models?
They explored N-body simulations build with \CB components of different masses and sizes, 
where the \BPB formed from the bar instabilities of the initial disc during the simulations. 
They find a best model selected from photometric comparisons with M31's bulge. 
Here we improve the models of B17 by fitting directly the data using the M2M method.
The main properties of their \CB are similar to the ones found here. 
The B17 \BPB luminosity is also similar to the value presented here, however, 
given their slightly larger mass-to-light ratio their \BPB mass is 15 per cent higher.
Their kinematic maps qualitatively match several features observed in M31, however the M2M model
highly improves the match quantitatively. For example, the M2M model presented here reproduces 
now the velocity dispersion in the outer parts of the \BPB due to a more massive dark matter halo.

\item Our best model has two bulge components with completely different kinematics
that only together successfully reproduce the detailed properties of the kinematic and the photometric maps.
Furthermore, our modelling includes dust absorption effects that can approximately reproduce the kinematic asymmetries in the observations.
The model, for example, reproduces the higher dispersion of the far side of the galaxy compared to the near side.

\item Our results present new constraints on the early formation of M31 given 
the lower mass found for the \CB component compared to previous estimations in the literature.
An implication is on the relation between bulges and central super massive black holes (SMBH).
SMBH masses show correlations with classical bulges and not pseudobulges \citep{Hu2008,Kormendy2013a,Saglia2016}.

Using the $M_{\bullet}\!-\!M_{\rm bulge}\!-\!\sigma$ relation\footnote{for the sample CorePowerEClassPC} from \citet{Saglia2016} for 
the \CB component alone with a mass of $M_{\star}^{\rm CB,10\kpc}\e1.71\times10^{10}\sm$ with $\sigma^{\rm CB, max}\si130$ - $150\kms$  
predicts a SMBH mass of $M_{\bullet}\e0.4^{+0.4}_{-0.2}\, -\, 0.6^{+0.7}_{-0.3}\times10^8\sm$, where the errors are the instrinsic scatter in the relation. 
This is somewhat lower than the measured $M_{\bullet}\e1.4^{+0.9}_{-0.3}\times10^8\sm$ \citep{Bender2005},
but lies within the scatter. Using the $M_{\bullet}\!-\!M_{\rm bulge}\,{}^3$  relation 
predicts a mass of $M_{\bullet}\e0.9^{+1.5}_{-0.5}\times10^8\sm$, that is closer to the measured value in M31.

\item The tightly constrained stellar mass-to-light ratio value of \bml is in agreement with the expected values from 
stellar populations with a Chabrier IMF \citep{Meidt2014}, with the metallicities and ages measured in M31's bulge and bar \citep{Opitsch2016, Saglia2018}. 
Considering the \CB alone a Chabrier IMF would be consistent with \citet{Cappellari2012} (using $\Upsilon_{\rm r}\si4\ml$ 
and $\sigma_{\rm CB}\si150\kms$).
It is however inconsistent with the Salpeter IMF found for more massive classical bulges measured by \citet[][SWELLS survey]{Dutton2013}.

\item Our findings agree with the photometric \citep{Fisher2008} and kinematic \citep{Fabricius2012} bulge classification
criteria using the S\'ersic index ($n$) and the central kinematics to distinguish classical bulges ($n>2$) from pseudobulges ($n<2$).
As \citet{Fisher2008} mention, and \citet{Erwin2015} investigate further, composite bulges can have an effect on the bulge selection 
criteria, and they can manifest both bulge type properties. Here we find that M31's composite bulge S\'ersic index is at the boundary with 
$n_{\rm M31}\si2$ and it shows kinematic properties of both bulge types.
Moreover, considering the \CB alone we find $n_{\rm CB}\si4$, a \CB to total mass ratio ${\rm B}/{\rm T}\e0.21$, 
effective radius $R_{\rm e}^{\CB}\si1\kpc$, and central dispersion $\sigma_{\rm CB}\si150\kms$, which also agree with the criteria for 
classical bulge types. 

Here we present two properties of a composite bulge that could improve the selection criteria: i) a composite bulge 
with $n\!\approx\!2$ can host a classical bulge with a high S\'ersic index, where the composite bulge has a value lowered by the presence of a box/peanut bulge, 
and ii) the presence of a classical bulge component can increase the total central dispersion by increasing the dispersion of the box/peanut bulge that lives within the classical 
bulge potential. This suggests that other observed bulges that show low S\'ersic values ($n\lesssim2$), but with high central dispersion and a large dispersion 
gradient ($\nabla\sigma$) \citep{Neumann2017}, could be hosting a compact classical bulge.

\item Our best M31 bar pattern speed value is $\bps$ which results in $\mathcal{R}\e1.6\pm0.2$, placing this bar 
among the slow bars. This is within the range of recent measurements of $\mathcal{R}$ of barred galaxies, finding $\mathcal{R}\e1.41\pm0.26$
\citep[{\it Spitzer} with gas kinematics][]{Font2017} and  $\mathcal{R}\e1.0^{+0.7}_{-0.4}$ \citep[CALIFA survey][]{Aguerri2015}.
Furthermore, our pattern speed measurement places the inner Lindblad resonances near the inner gas rings and streams observed 
within the bulge \citep{Opitsch2017}, and the outer Lindblad resonance near the 10\kpc ring, which could 
explain its origin and persisting star forming activity \citep{Lewis2015}.

\end{enumerate}

Finally, the M2M models presented here have many possible uses. They can be applied to investigate further the early formation and the secular evolution of 
M31. For example, including gas to reproduce the outer 10\kpc ring-like substructure, or the bulge gas distribution.
Also, from stellar population and chemodynamical galaxy formation simulations \citep{Kobayashi2011} it is expected that 
the stars with different chemical elements have different spatial distributions. 
The M31 bulge metallicity maps \citep{Saglia2018} could be used to dissect the galaxy's orbital structure using a chemodynamical modelling, 
as similarly done for the MW \citep{Portail2017b}.
Other applications of our model involve the interpretation of pixel micro-lensing events in M31's halo for the
observational campaigns PAndromeda \citep{Lee2012} and WeCAPP \citep{Lee2015}. 
For the pixel lensing modelling an important ingredient are accurate dynamical models of the stellar mass distribution
to take into account the self-lensing events, and thereby better constrain the lensing events in the halo \citep{Riffeser2006}.
The models presented here are the most appropriate as these include the barred nature of the Andromeda galaxy.\\
These models may be available upon request to the authors.

\section*{Acknowledgements}
\label{sec:acknow}
Mat\'ias Bla\~{n}a (MB) would like to thank Carolina Agurto, Fabrizio Finozzi, Laura Morselli and Mar\'ia de los Angeles P\'erez Villegas 
for several insightful and delightful scientific discussions, and also to thank Achim Bohnet for the technical support.
We thank Pauline Barmby for providing the Spitzer 3.6\mum IRAC1 data. 
We also thank Jerry Sellwood and Monica Valluri for making their potential solver code available to us.
We are grateful to the anonymous referee for constructive comments that improved the manuscript.
MB would also like to thank the Deutscher Akademischer Austauschdienst (DAAD) for the 
grant supporting this doctoral project with a Research Grant 
for Doctoral Candidates and Young Academics and Scientists (57076385). 
MB also thanks the powerful python and its wonderful community.
\bibliographystyle{mnras}

\label{lastpage}
\appendix
\section{NFW grid of parameters and Einasto cube of parameters}
\label{sec:appA}

\begin{figure}
\begin{center}
\includegraphics[width=8.cm]{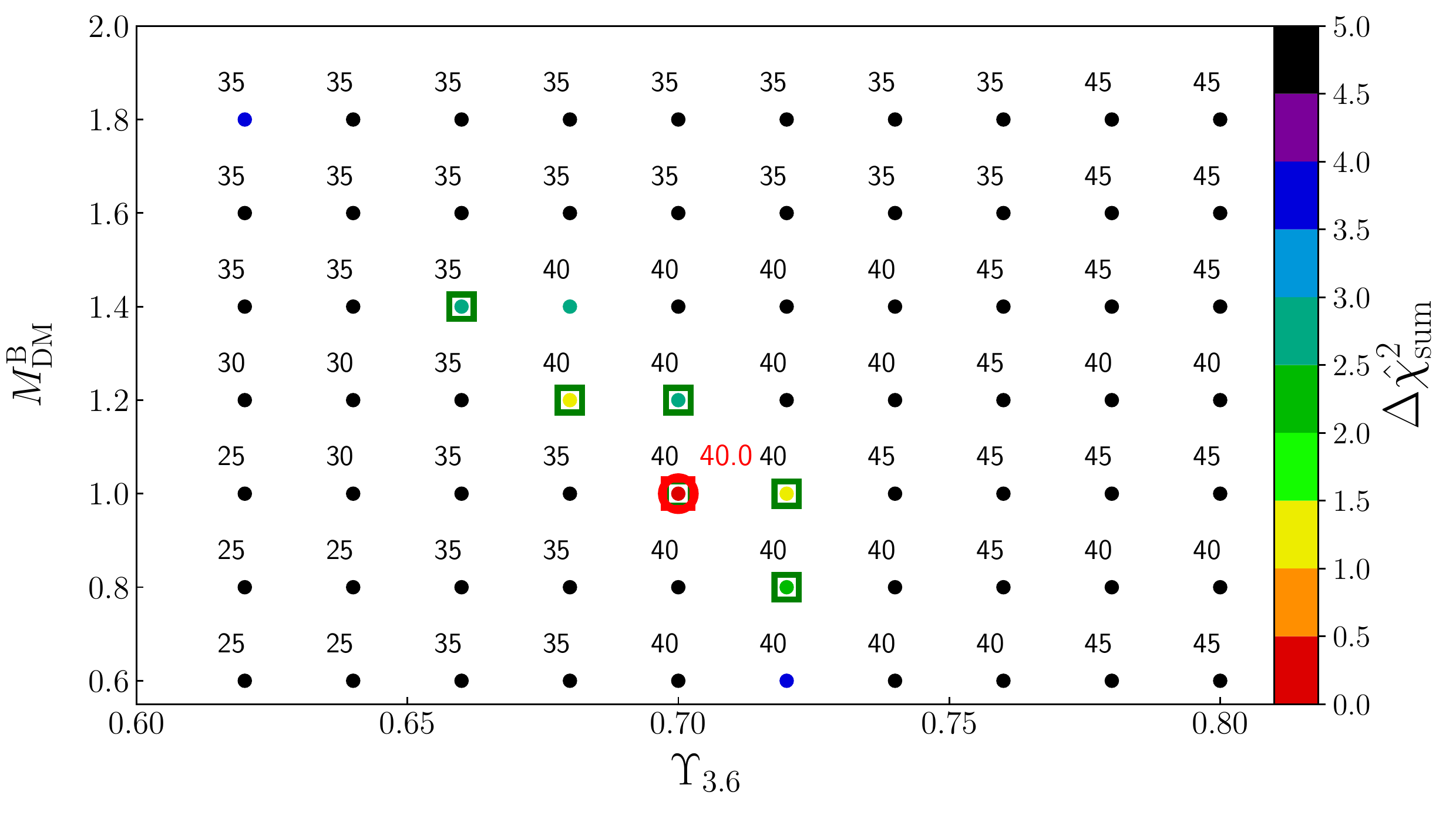}\\ 
\vspace{-0.1cm}
\includegraphics[width=8.cm]{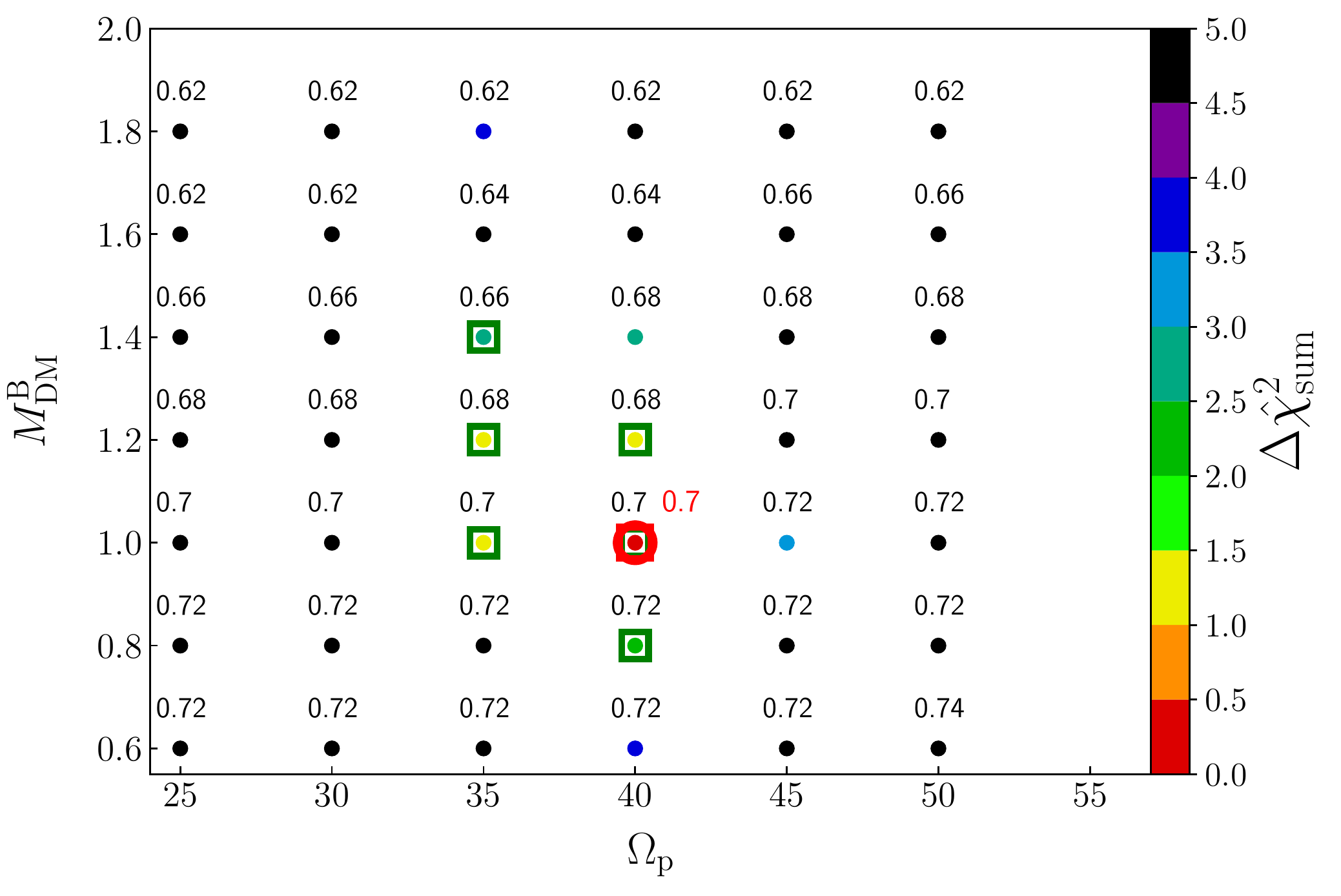}
\vspace{-0.2cm}
\caption[Parameter \setdsum: best models]{Range of acceptable models defined by the total
  goodness-of-fit \setdsum, for an NFW halo.  The green squares mark the range of acceptable
  models, with the red circles marking the overall best matching model KR241.  Top panel: \setdsum
  in the \pml, \pdm plane, always selecting the minimum value along the parameter \pps axis.  Bottom
  panel: \setdsum as function of \pps and \pdm, selecting the minimum value along the parameter \pml
  axis.  In black are shown models with $\setdsum\ge5$).}
\label{fig:XsumNFW}
\end{center}
\end{figure}

\begin{figure}
\begin{center}
\includegraphics[width=8cm]{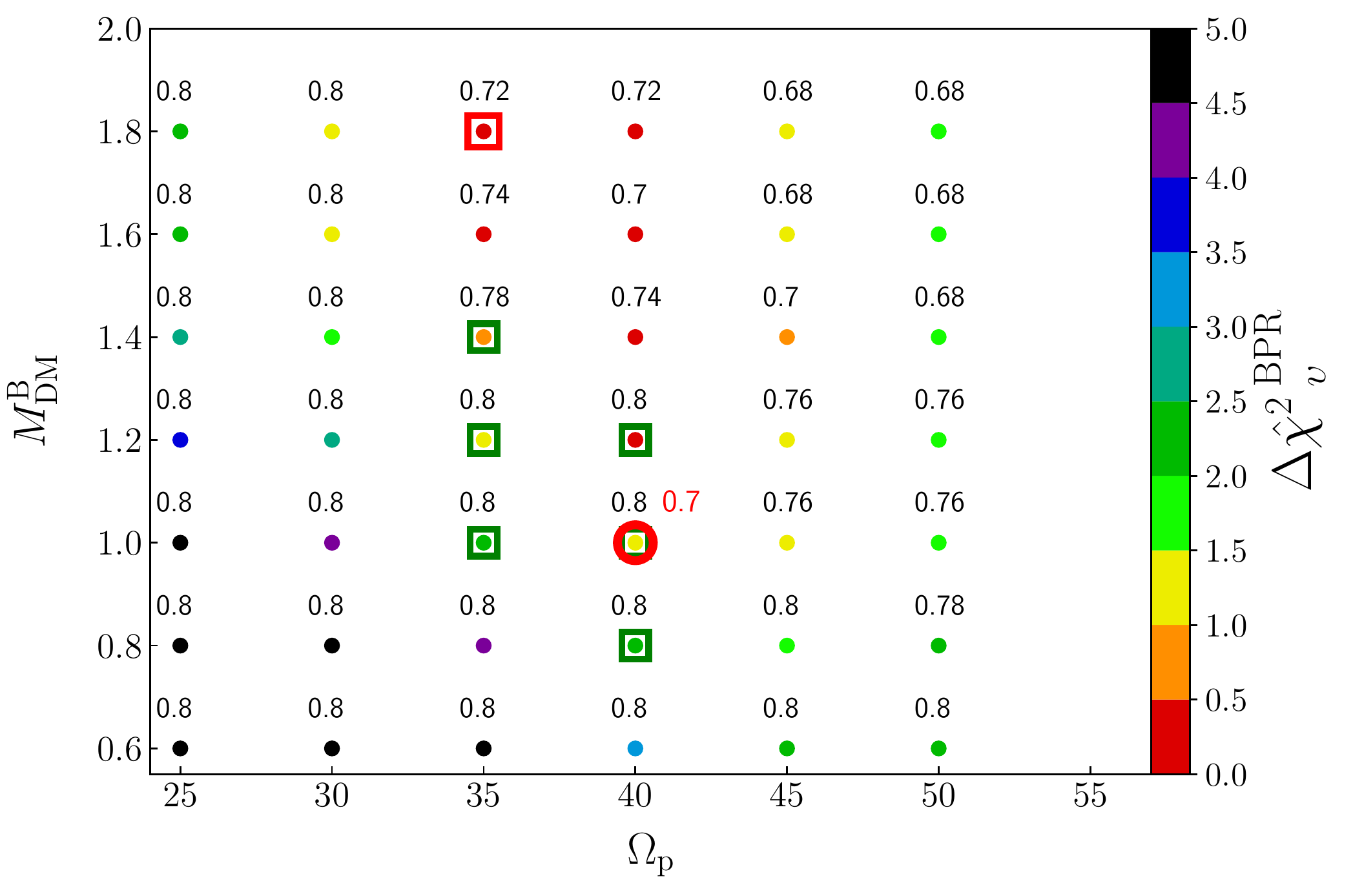}\\
\vspace{-0.1cm}
\includegraphics[width=8cm]{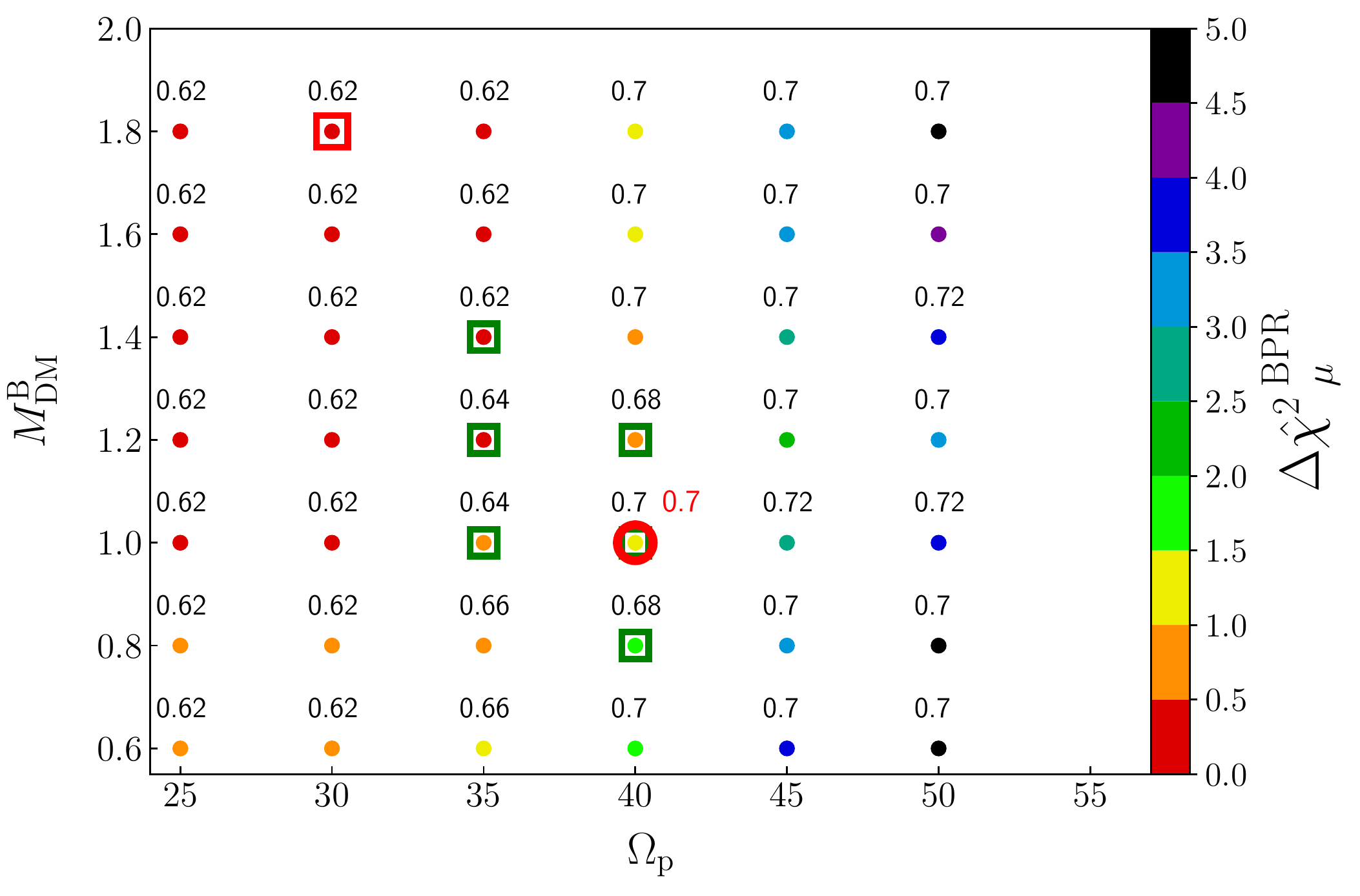}\\
\vspace{-0.1cm}
\includegraphics[width=8cm]{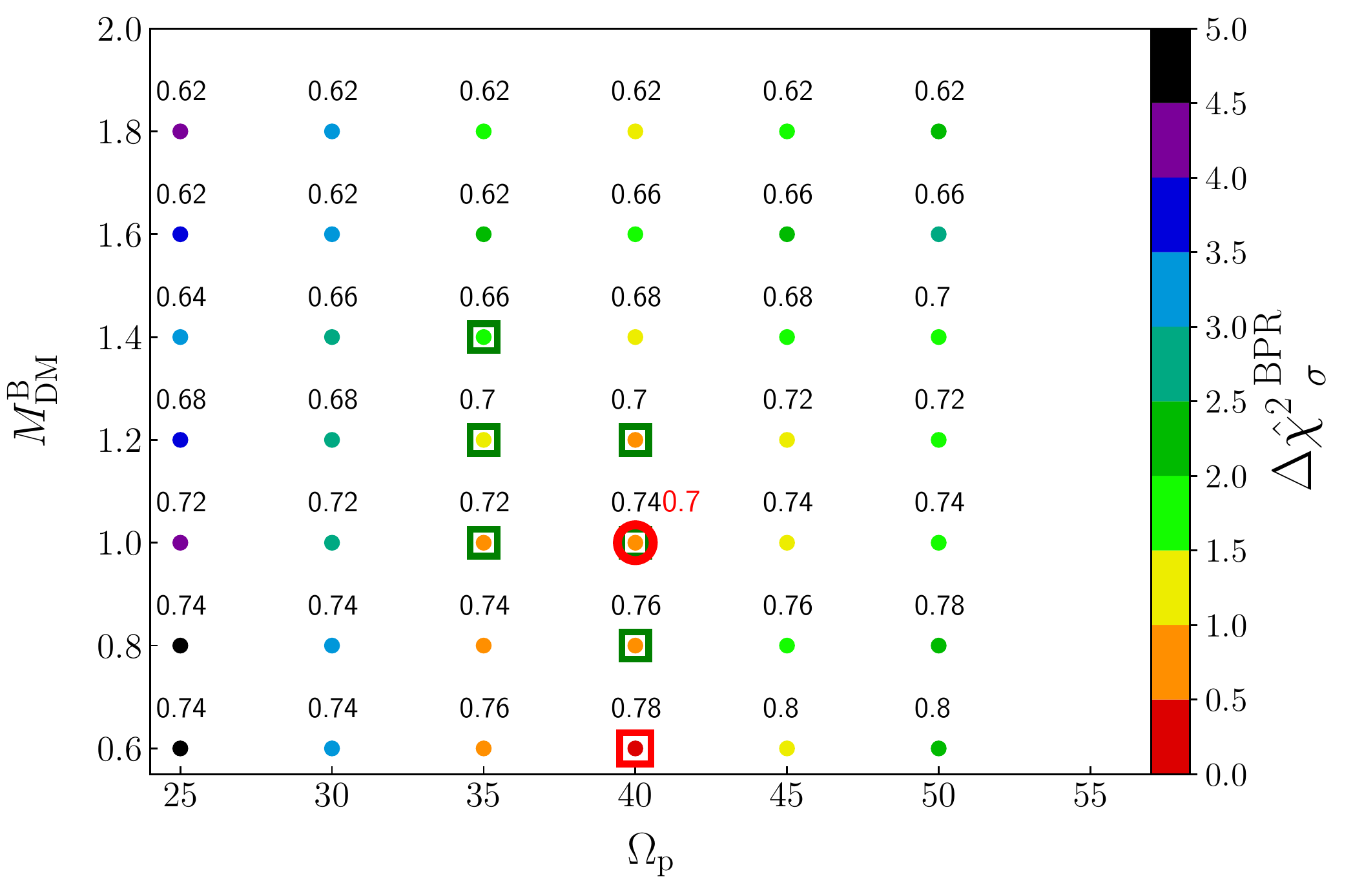}
\vspace{-0.4cm}
\caption[Subset results for the NFW grid of model, part 2]{Results of the grid of models for the NFW dark matter halo for
\setE, \setC and \setD as function of the parameters 
$\pps$ and $\pdm$ selecting the lowest value along the axis of the parameter $\pml$.
The values of each subset are the points that are coded in the coloured bar, and the number corresponds to the selected $\pml$. 
We mark the best model KR241 (red circle), the models with the minimum values in each subset (red squares), 
and the range of the acceptable models \bmnfw (green squares). 
The green squares do not necessarily agree with the shown \pml.}
\label{fig:sspart2NFW}
\end{center}
\end{figure}

\begin{figure}
\begin{center}
\includegraphics[width=8.cm]{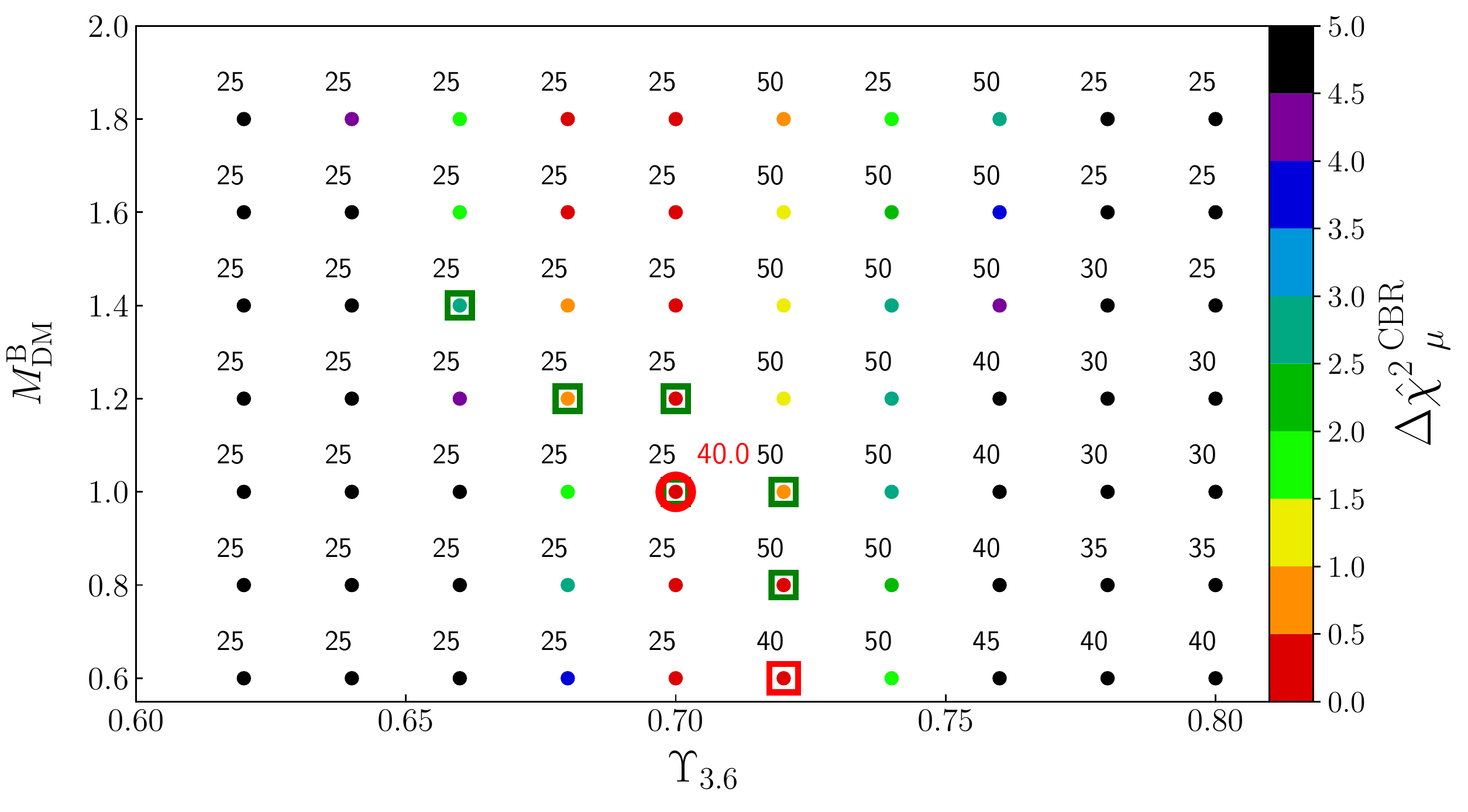}\\
\includegraphics[width=8.cm]{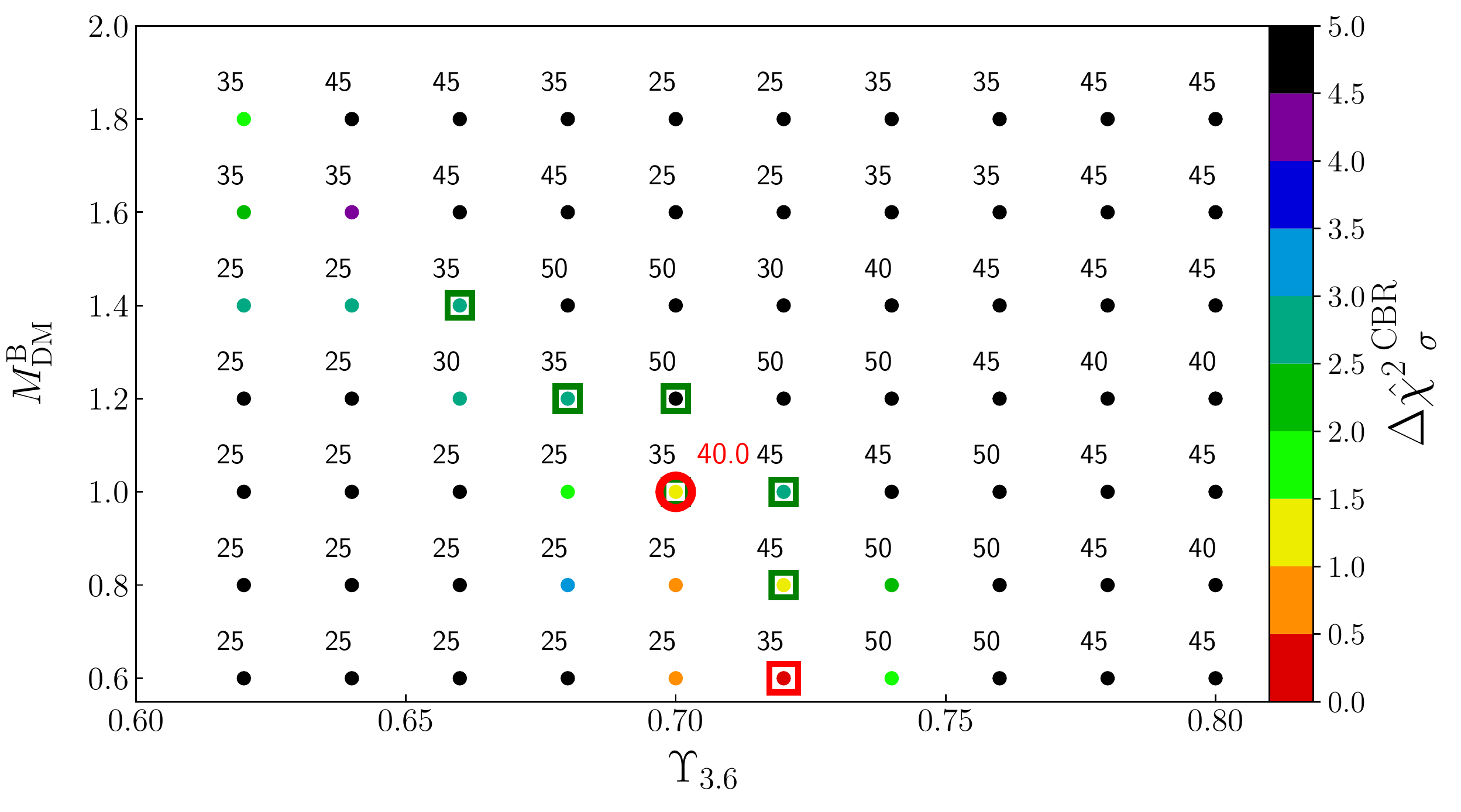}\\
\includegraphics[width=8.cm]{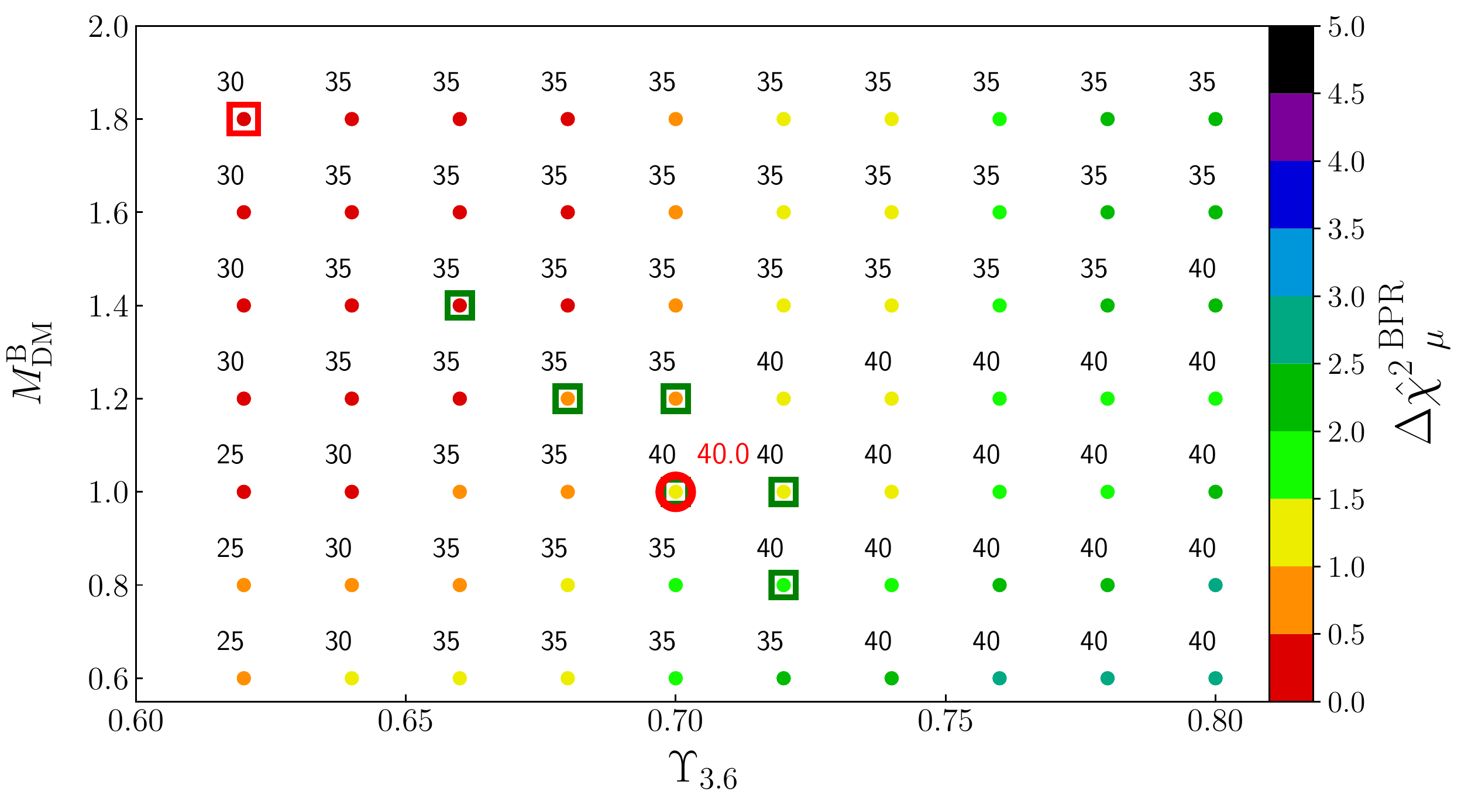}\\
\includegraphics[width=8.cm]{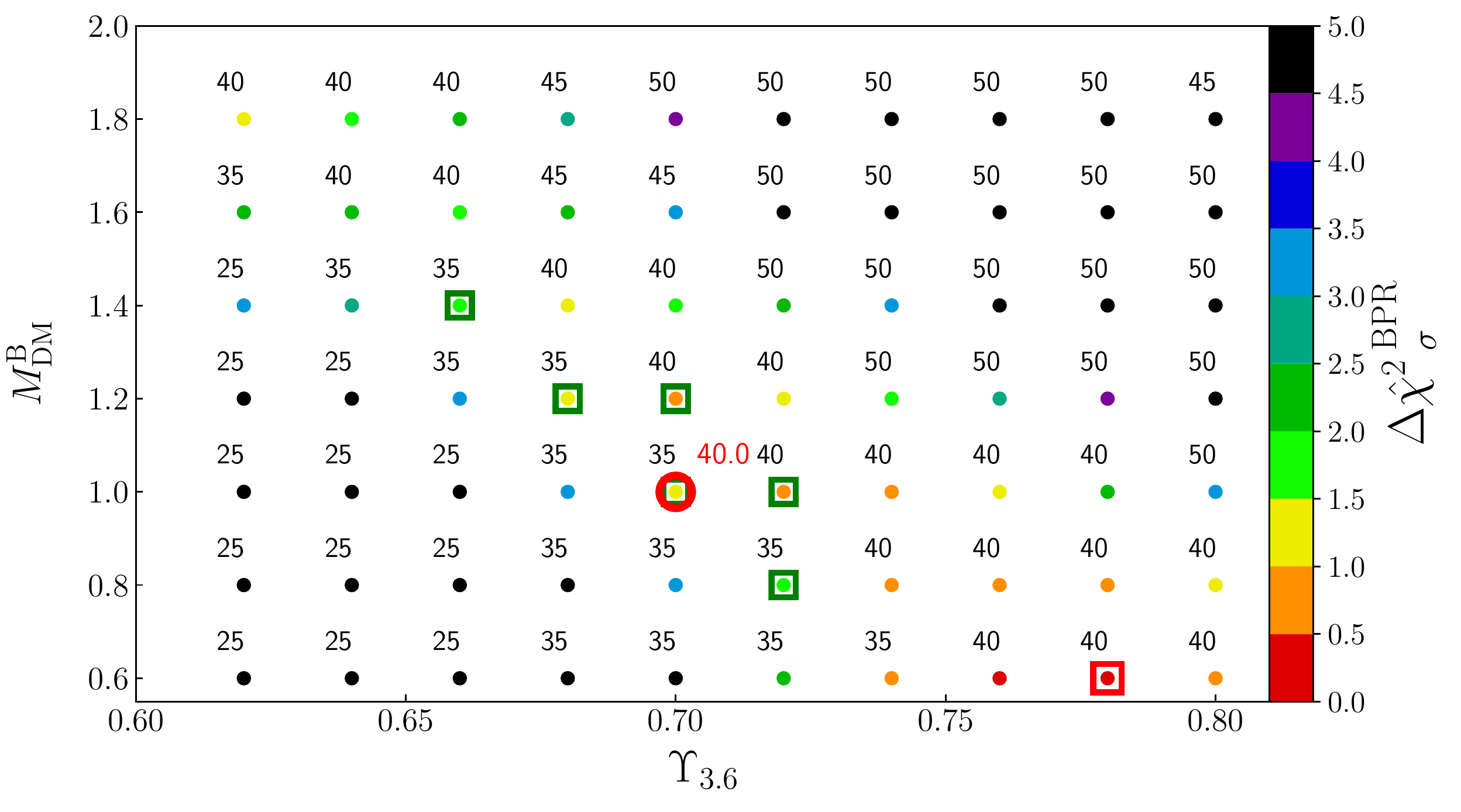}
\vspace{-0.3cm}
\caption[Subset results for the NFW grid of model, part 1]{Results of the grid of models for the NFW dark matter halo for
\setA, \setB, \setC and \setD as function of the parameters 
$\pml$ and $\pdm$ selecting the lowest value along the axis of the parameter $\pps$.
The values of each subset are he points that are coded in the coloured bar, and the number corresponds to the selected $\pps$. 
We mark the best model KR241 (red circle), the models with the minimum values in each subset (red squares), 
and the range of the acceptable models \bmnfw (green squares). The green squares do not necessarily agree with the pattern speed shown.}
\label{fig:sspart1NFW}
\end{center}
\end{figure}

\begin{figure*}
\begin{center}
\includegraphics[width=18cm]{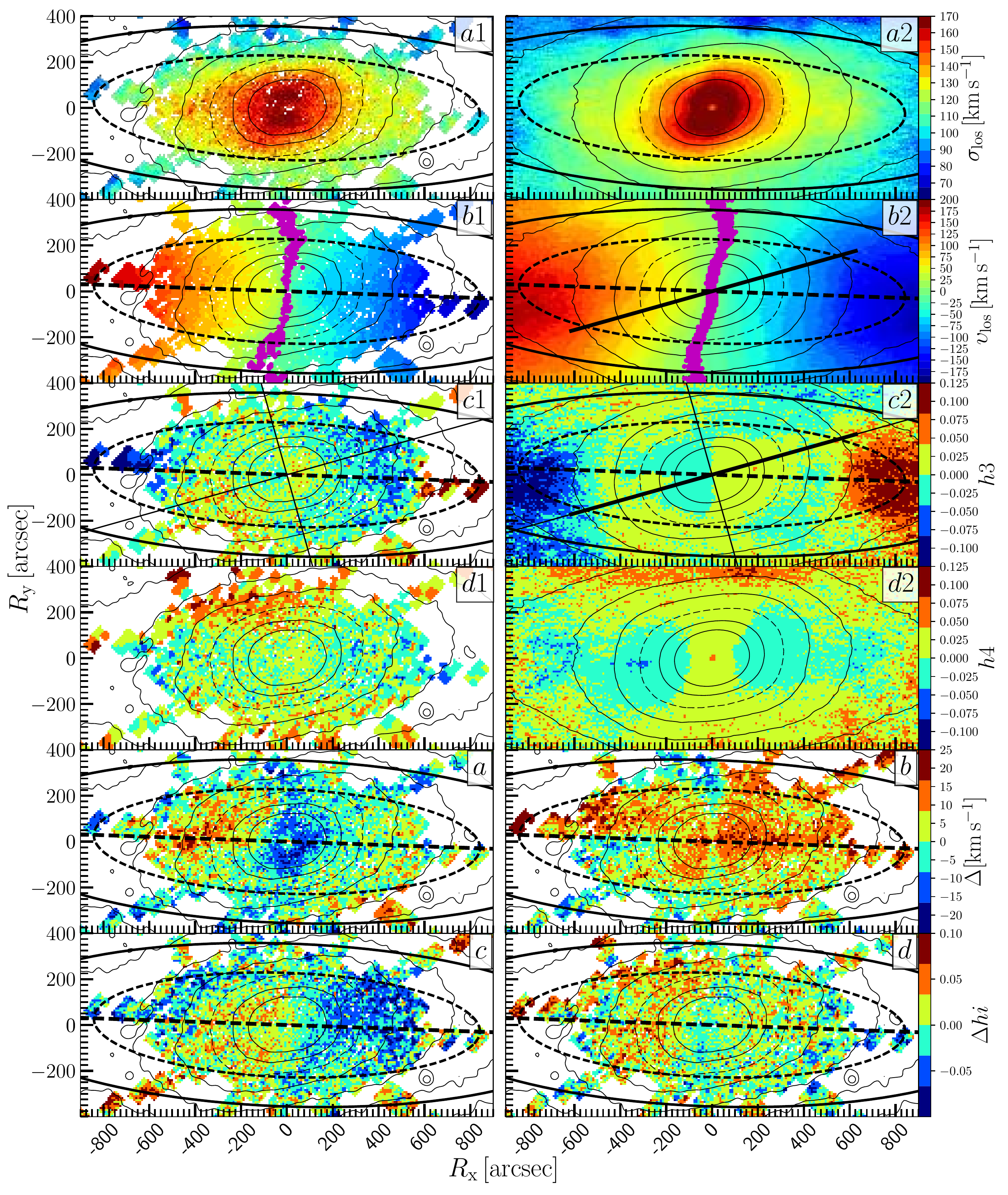}
\vspace{-0.55cm}
\caption[Kinematic maps and isophotes of M31 and the non-rotating model]{Surface-brightness isophotes and kinematic maps of M31 (left column) and a {\it triaxial elliptical galaxy bulge model} \ie a model with the same parameters as model JR804, but with no pattern speed ($\pps\e0\psu$) (right column). Labels similar as Figure \ref{fig:kinmap}. 
The differences between the observations and the model are shown in panel ($a$) with $\Delta\e\sigma_{\rm los}^{\rm obs}-\sigma_{\rm los}^{\rm model}$, 
($b$) with $\Delta\e||\upsilon_{\rm los}^{\rm obs}||-||\upsilon_{\rm los}^{\rm model}||$,
($c$) with $\Delta h3\e h3^{\rm obs}-h3^{\rm model}$, 
and panel ($d$) with $\Delta h4\e h4^{\rm obs}-h4^{\rm model}$. }
\label{fig:kinmap_PS0}
\end{center}
\end{figure*}

\end{document}